\documentclass[12pt]{iopart}
\usepackage{iopams}
\usepackage{amssymb}
\usepackage{graphicx}
\usepackage{xcolor,hyperref}
\usepackage{etoolbox}
\usepackage{braket}
\usepackage[utf8]{inputenc}
\usepackage[T1]{fontenc}
\usepackage{bm}% bold math

\makeatletter
\newrobustcmd{\fixappendix}{%
  \patchcmd{\l@section}{1.5em}{7em}{}{}%
  \patchcmd{\l@subsection}{2.3em}{7em}{}{}%
}
\makeatother

\usepackage{har2nat}
\hypersetup{breaklinks=true,colorlinks=true,linkcolor=blue,citecolor=blue,filecolor=magenta,urlcolor=cyan}

\renewcommand{\harvardurl}[1]{\url{#1}}

\newcommand{\iso}[2]{\ensuremath{^{#2}\mathrm{#1}}}
\newcommand{\thetaQCD}{$\bar{\theta}_\textrm{\tiny QCD}$ }

\begin{document}

\title[Opportunities for Fundamental Physics Research with Radioactive Molecules]{Opportunities for Fundamental Physics Research with Radioactive Molecules}

\vspace{10pt}
\begin{indented}
\item[]\today
\end{indented}

\begin{abstract}

Molecules containing short-lived, radioactive nuclei are uniquely positioned to enable a wide range of scientific discoveries in the areas of fundamental symmetries, astrophysics, nuclear structure, and chemistry.  Recent advances in the ability to create, cool, and control complex molecules down to the quantum level, along with recent and upcoming advances in radioactive species production at several facilities around the world, create a compelling opportunity to coordinate and combine these efforts to bring precision measurement and control to molecules containing extreme nuclei.  In this manuscript, we review the scientific case for studying radioactive molecules, discuss recent atomic, molecular, nuclear, astrophysical, and chemical advances which provide the foundation for their study, describe the facilities where these species are and will be produced, and provide an outlook for the future of this nascent field.

\end{abstract}

\newpage

\makeatletter
\setcounter{tocdepth}{2}
\tableofcontents
\newpage

{ \footnotesize
\author{	Gordon Arrowsmith-Kron	}	\address{	Facility for Rare Isotope Beams, Michigan State University, East Lansing, Michigan 48824, USA	}
\author{	Michail Athanasakis-Kaklamanakis	}	\address{	Experimental Physics Department, CERN, CH-1211 Geneva 23, Switzerland	}
			\address{	KU Leuven, Department of Physics and Astronomy, Instituut voor Kern- en Stralingsfysica, B-3001 Leuven, Belgium	}
\author{	Mia Au	}	\address{	CERN, Geneva, Switzerland	}
			\address{	 Johannes Gutenberg-Universit{\"a}t Mainz, Mainz, Germany	}
\author{	Jochen Ballof	}	\address{	Accelerator Systems Department, CERN, 1211 Geneva 23, Switzerland	}
			\address{	Facility for Rare Isotope Beams, Michigan State University, East Lansing, Michigan 48824, USA	}
\author{	Robert Berger	}	\address{	Fachbereich Chemie, Philipps-Universität Marburg, Hans-Meerwein-Straße 4, 35032 Marburg, Germany	}
\author{	Anastasia Borschevsky	}	\address{	Van Swinderen Institute for Particle Physics and Gravity, University of Groningen, The Netherlands	}
\author{	Alexander A. Breier	}	\address{	Institute of Physics, University of Kassel, Heinrich-Plett-Str. 40, 34132 Kassel, Germany	}
\author{	Fritz Buchinger	}	\address{	McGill University, Canada 	}
\author{	Dmitry Budker	}	\address{	Helmholtz-Institut, GSI Helmholtzzentrum fur Schwerionenforschung and Johannes Gutenberg University, Mainz 55128, Germany, Mainz 55128, Germany	}
			\address{	Department of Physics, University of California at Berkeley, Berkeley, California 94720-7300, USA	}
\author{	Luke Caldwell	}	\address{	JILA, NIST and University of Colorado, Boulder, Colorado 80309, USA	}
			\address{	Department of Physics, University of Colorado, Boulder, Colorado 80309, USA	}
\author{	Christopher Charles	}	\address{	TRIUMF, 4004 Wesbrook Mall, Vancouver, BC V6T 2A3, Canada	}
			\address{	University of Western Ontario, 1151 Richmond St. N., London, Ontario, Canada, N6A 5B7	}
\author{	Nike Dattani	}	\address{	HPQC Labs, Waterloo, Ontario, Canada	}
			\address{	HPQC College, Waterloo, Ontario, Canada	}
\author{Ruben P. de Groote}
\address{Instituut voor Kern- en Stralingsfysica, KU Leuven, Leuven, Belgium}
\address{Department of Physics, University of Jyväskylä, Jyväskylä, Finland}
\author{	David DeMille	}	\address{	University of Chicago, Chicago, IL, USA	}
			\address{	Argonne National Laboratory, Lemont, IL, USA	}
\author{	Timo Dickel	}	\address{	GSI Helmholtzzentrum für Schwerionenforschung GmbH, 64291 Darmstadt, Germany	}
			\address{	II. Physikalisches Institut, Justus-Liebig-Universität Gießen, 35392 Gießen, Germany	}
\author{	Jacek Dobaczewski$^*$	}	\address{	School of Physics, Engineering and Technology, University of York, Heslington, York YO10 5DD, UK	}
			\address{	Institute of Theoretical Physics, Faculty of Physics, University of Warsaw, ul. Pasteura 5, PL-02-093 Warsaw, Poland	} \ead{jacek.dobaczewski@york.ac.uk}
\author{Christoph E. Düllmann}
\address{Department of Chemistry - TRIGA Site, Johannes Gutenberg University,
Fritz-Strassmann-Weg 2, 55128 Mainz, Germany}
\address{GSI Helmholtzzentrum für Schwerionenforschung, Planckstr. 1, 64291
Darmstadt, Germany
Helmholtz Institute Mainz, Staudingerweg 18, 55128 Mainz, Germany}
\author{	Ephraim Eliav	}	\address{	School of Chemistry, Tel Aviv University, Ramat Aviv, Tel Aviv 69978, Israel	}
\author{	Jonathan Engel	}	\address{	Department of Physics and Astronomy, University of North Carolina,
Chapel Hill, North Carolina  27599-3255,  USA	}
\author{	Mingyu Fan	}	\address{	Department of Physics, University of California, Santa Barbara, California 93106, USA	}
\author{	Victor Flambaum	}	\address{	University of New South Wales, Sydney 2052, Australia	}
\author{	Kieran T. Flanagan	}	\address{	Photon Science Institute, Department of Physics and Astronomy, University of Manchester, Manchester, M13 9PL, UK	}
\author{	Alyssa N. Gaiser	}	\address{	Facility for Rare Isotope Beams, Michigan State University, East Lansing, Michigan 48824, USA	}
\author{	Ronald F. Garcia Ruiz$^*$	}	\address{	Massachusetts Institute of Technology, Cambridge, MA 02139, USA	} \ead{rgarciar@mit.edu}
\author{	Konstantin Gaul	}	\address{	Fachbereich Chemie, Philipps-Universität Marburg, Hans-Meerwein-Straße 4, 35032 Marburg, Germany	}
\author{	Thomas F. Giesen	}	\address{	Institute of Physics, University of Kassel, Heinrich-Plett-Str. 40, 34132 Kassel, Germany	}
\newpage
\author{	Jacinda S. M. Ginges	}	\address{	School of Mathematics and Physics, The University of Queensland, Brisbane QLD 4072, Australia	}
\author{	Alexander Gottberg	}	\address{	TRIUMF, 4004 Wesbrook Mall, Vancouver, BC V6T 2A3, Canada	}
\author{	Gerald Gwinner	}	\address{	Department of Physics and Astronomy, University of Manitoba, Winnipeg, MB R3T 3M9, Canada	}
\author{	Reinhard Heinke	}	\address{	CERN, Geneva, Switzerland	}
\author{	Steven Hoekstra	}	\address{	Van Swinderen Institute for Particle Physics and Gravity, University of Groningen, The Netherlands	}
\address{	Nikhef, National Institute for Subatomic Physics, Amsterdam, The Netherlands	}
\author{	Jason D. Holt	}	\address{	TRIUMF, 4004 Wesbrook Mall, Vancouver, BC V6T 2A3, Canada	}
			\address{	Department of Physics, McGill University, Montreal, QC H3A 2T8, Canada	}
\author{	Nicholas R. Hutzler$^*$	}	\address{	California Institute of Technology, Pasadena, CA 91125, USA	} \ead{hutzler@caltech.edu}
\author{	Andrew Jayich$^*$	}	\address{	Department of Physics, University of California, Santa Barbara, California 93106, USA	} \ead{jayich@physics.ucsb.edu}
\author{	Jonas Karthein	}	\address{	Massachusetts Institute of Technology, Cambridge, MA 02139, USA	}
\author{	Kyle G. Leach	}	\address{	Facility for Rare Isotope Beams, Michigan State University, East Lansing, Michigan 48824, USA	}
			\address{	Colorado School of Mines, Golden, CO 80401, USA	}
\author{	Kirk W. Madison	}	\address{	Department of Physics and Astronomy, University of British Columbia, Vancouver, BC V6T1Z1, Canada	}
\author{	Stephan Malbrunot-Ettenauer	}	\address{	TRIUMF, 4004 Wesbrook Mall, Vancouver, BC V6T 2A3, Canada	}
			\address{	Department of Physics, University of Toronto, 60 St. George St., Toronto, Ontario, Canada	}
\author{	Takayuki Miyagi	}	\address{	TRIUMF, 4004 Wesbrook Mall, Vancouver, BC V6T 2A3, Canada	}
\author{	Iain D. Moore	}	\address{	Accelerator Laboratory, Department of Physics, University of Jyväskylä, Jyväskylä, 40014, Finland.	}

\author{	Scott Moroch	}	\address{	Massachusetts Institute of Technology, Cambridge, MA 02139, USA	}

\author{	Petr Navratil	}	\address{	TRIUMF, 4004 Wesbrook Mall, Vancouver, BC V6T 2A3, Canada	}
\author{	Witold Nazarewicz$^*$	}	\address{	Facility for Rare Isotope Beams and Department of Physics and Astronomy, Michigan State University, East Lansing, Michigan 48824, USA	} \ead{witek@frib.msu.edu}
\author{	Gerda Neyens	}	\address{	KU Leuven, Department of Physics and Astronomy, Instituut voor Kern- en Stralingsfysica, B-3001 Leuven, Belgium	}
\author{	Eric B. Norrgard	}	\address{	Sensor Science Division, National Institute of Standards and Technology, Gaithersburg, Maryland 20899, USA	}
\author{	Nicholas Nusgart	}	\address{	Facility for Rare Isotope Beams, Michigan State University, East Lansing, Michigan 48824, USA	}
\author{	Lukáš F. Pašteka	}	\address{	Van Swinderen Institute for Particle Physics and Gravity, University of Groningen, The Netherlands	}
			\address{	Department of Physical and Theoretical Chemistry, Faculty of Natural Sciences, Comenius University, Bratislava, Slovakia	}
\author{	Alexander N Petrov	}	\address{	Petersburg Nuclear Physics Institute named by B.P.\ Konstantinov of National Research Center ``Kurchatov Institute'' (NRC ``Kurchatov Institute'' - PNPI), 1 Orlova roscha mcr., Gatchina, 188300 	}
			\address{	Saint Petersburg State University, 7/9 Universitetskaya nab., St. Petersburg, 199034 Russia	}
\author{	Wolfgang R. Plaß	}	\address{	GSI Helmholtzzentrum für Schwerionenforschung GmbH, 64291 Darmstadt, Germany	}
			\address{	II. Physikalisches Institut, Justus-Liebig-Universität Gießen, 35392 Gießen, Germany	}
\author{	Roy A. Ready	}	\address{	Department of Physics, University of California, Santa Barbara, California 93106, USA	}
\author{	Moritz Pascal Reiter	}	\address{	School of Physics \& Astronomy, The University of Edinburgh, Peter Guthrie Tait Road, EH9 3FD Edinburgh, United Kingdom	}
\author{	Mikael Reponen	}	\address{	Accelerator Laboratory, Department of Physics, University of Jyväskylä, Jyväskylä, 40014, Finland.	}
\author{	Sebastian Rothe	}	\address{	CERN, Geneva, Switzerland	}
\newpage
\author{	Marianna S. Safronova	}	\address{	Department of Physics and Astronomy, University of Delaware, Newark, Delaware 19716, USA	}
			\address{	Joint Quantum Institute, National Institute of Standards and Technology and the University of Maryland, Gaithersburg, Maryland 20742, USA	}
\author{	Christoph Scheidenerger	}	\address{	GSI Helmholtzzentrum für Schwerionenforschung GmbH, 64291 Darmstadt, Germany	}
			\address{	II. Physikalisches Institut, Justus-Liebig-Universität Gießen, 35392 Gießen, Germany	}
			\address{	Helmholtz Forschungsakademie Hessen für FAIR (HFHF), Campus Gießen, Gießen, Germany	}
\author{	Andrea Shindler	}	\address{	Facility for Rare Isotope Beams \& Physics Department, Michigan State University, East Lansing, Michigan 48824, USA	}
\author{	Jaideep T. Singh$^*$	}	\address{	Facility for Rare Isotope Beams, Michigan State University, East Lansing, MI, USA	} \ead{singhj@frib.msu.edu}
\author{	Leonid V. Skripnikov	}	\address{	Petersburg Nuclear Physics Institute named by B.P.\ Konstantinov of National Research Center ``Kurchatov Institute'' (NRC ``Kurchatov Institute'' - PNPI), 1 Orlova roscha mcr., Gatchina, 188300 Leningrad region, Russia	}
			\address{	Saint Petersburg State University, 7/9 Universitetskaya nab., St. Petersburg, 199034 Russia	}
\author{	Anatoly V. Titov	}	\address{	Petersburg Nuclear Physics Institute named by B.P.\ Konstantinov of National Research Center ``Kurchatov Institute'' (NRC ``Kurchatov Institute'' - PNPI), 1 Orlova roscha mcr., Gatchina, 188300 Leningrad region, Russia	}
			\address{	Saint Petersburg State University, 7/9 Universitetskaya nab., St. Petersburg, 199034 Russia	}
\author{	Silviu-Marian Udrescu	}	\address{	Massachusetts Institute of Technology, Cambridge, MA 02139, USA	}
\author{	Shane G. Wilkins	}	\address{	Massachusetts Institute of Technology, Cambridge, MA 02139, USA	}
\author{	Xiaofei Yang	}	\address{	School of Physics and State Key Laboratory of Nuclear Physics and Technology, Peking  University, Beijing 100871, China	}
}
\bigskip
$^*$ Corresponding authors

\newpage

\section{Executive Summary}

Radioactive molecules hold great promise for their discovery potential in diverse fields.  The extreme nuclear properties of heavy, short-lived nuclei and the intrinsic sensitivity, flexibility, and quantum control opportunities available to molecules make them a competitive platform for advancing high-energy particle physics, cosmology, nuclear physics, astrophysics, and  chemistry -- a breadth which is reflected in the corresponding research community.  In addition to their discovery potential, the common threads which tie the community together are the two challenges of working with radioactive molecules: controlling heavy and complex molecules at the level of single quantum states, and the creation and handling of radioactive isotopes.  

Fortunately, these two challenges are being rapidly overcome across many fronts.    Diatomic and even polyatomic molecules have been trapped at ultracold temperatures \cite{Fitch2021Review,Vilas2021}, yielding new opportunities for quantum information science, quantum chemistry, many-body physics, precision measurement, and quantum sensing~\cite{Bohn2017,Safronova2018,Isaev2018,Chupp2019,Hutzler2020Review,Fitch2021Review}.  Both the internal and motional states of trapped molecular ions have been controlled at the single quantum level \cite{Chou2017, Chou2020}.  Molecules hold the record for the most sensitive measurement of the electron's electric dipole moment by orders of magnitude \cite{Roussy2022,ACME2018}.  Note that while these advances have been primarily with stable species, the efforts are highly synergistic with work to control radioactive molecules, especially with high efficiency, and stable molecules will continue to serve as a testbed for new techniques.  

In addition to advances in controlling molecules, the capabilities for working with radioisotopes continues to expand, including recent landmark spectroscopy of radioactive molecules~\cite{GarciaRuiz2020,Udrescu2021}.  University research groups have effectively used radioisotopes in small quantities \cite{Hucul2017, Fan2019}, with efficiency gains expected \cite{Hunter}.  Meanwhile, nuclear facilities continue expanding their radioactive molecule capabilities, and their ability to generate usable quantities of promising short-lived radioisotopes.\footnote{Here we consider an isotope to be ``short-lived'' when its lifetime is shorter than $\sim$10$^6$ years, meaning that it must be created artificially to be studied. Long-lived, naturally occurring species such as \iso{Th}{232} are radioactive, but can be available in macroscopic quantities as their extremely long lifetimes ($\tau=1.4$ $\times 10^{10}$ years) make them stable for practical purposes.}  Extending state-of-the-art molecular measurement and control methods to radioactive species will be a non-trivial endeavor.

Importantly, molecular experiments have the potential to continue offering major technological improvements.  Many advances, including upgrades to existing experiments, new experiments, and new theoretical approaches provide a pathway to significant improvements in this decade and those to come. Implementing advances from the quantum information science (QIS) community will extend coherence and improve control, leading to improved experimental sensitivity for a wide range of precision measurements in fundamental physics and chemistry.  Heavy, deformed, exotic nuclei offer additional significant enhancements in sensitivity to hadronic symmetry violations and opportunities for nuclear structure studies.  Expanding molecular spectroscopy of short-lived species will advance \textit{ab~initio} molecular theory, and provide the ground-work for studies of fundamental symmetries. Precision spectroscopy of radioactive molecules is critical for their identification in astrophysical studies.

Radioactive molecules will play an important role in advancing science in these areas.  For example, sensitivity to violations of fundamental symmetries tend to scale very rapidly with proton number, typically as $Z^{2}$ to $Z^{5}$.  Therefore, the heaviest nuclei, which are radioactive, will also provide the greatest sensitivities.  Furthermore, nuclear deformations provide additional dramatic enhancements.  Nuclei with an octupole $(\beta_3)$ deformation enhance sensitivity to charge-parity (CP) violating hadronic physics via a nuclear Schiff moment (NSM) by a factor of 100 to 1,000, in addition to molecular enhancements.  For example, suitable radium-bearing molecules have around $10^{5}$ to $10^6$ more intrinsic sensitivity to hadronic CP-violation than \iso{Hg}{199} ~\cite{Kudashov:13}, which currently sets the  best limits on  multiple hadronic CP-violating sources \cite{Graner2016}.  

\begin{figure}
    \centering
    \includegraphics[width=\textwidth]{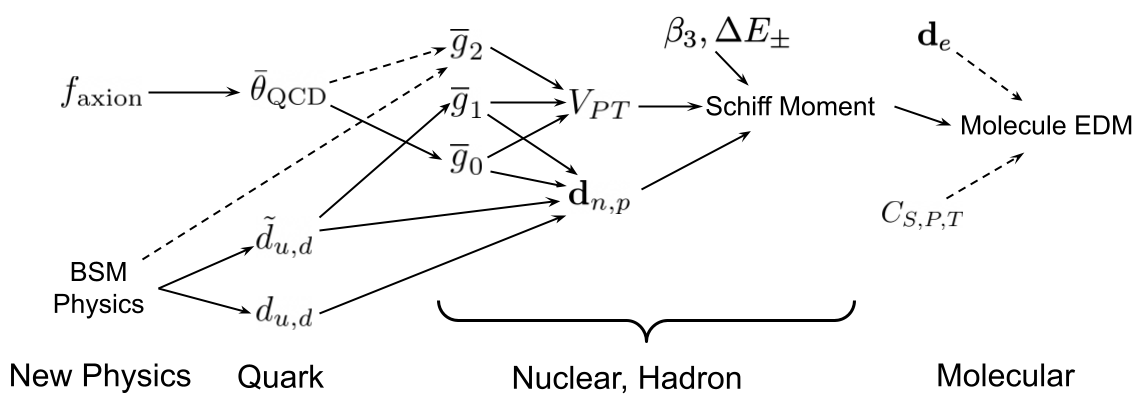}
    \caption{ Sources of BSM physics, such as supersymmetric particles couple to SM particles through quark electric EDMs ($d_{u,d}$) and chromo EDMs ($\tilde{d}_{u,d}$), or via an axion field ($f_\mathrm{axion}$) which couples to the QCD \thetaQCD parameter.  These sources then propagate through hadronic and nuclear physics to time symmetry violating moments of radioactive molecules which can be measured.  The isoscalar, $\bar{g}_0$, and isovector, $\bar{g}_1$, pion-nucleon-nucleon ($\pi NN$) coupling parameters result in a $P, T$-odd nuclear Hamiltonian term, $V_{PT}$, as well as proton and neutron EDMs, $\boldsymbol{\mathrm{d}}_{n,p}$.  Whereas the isotensor parameter, $\bar{g}_2$, only couples to $V_{PT}$.  The Schiff moment is determined by $V_{PT}$, the nucleon EDMs, and is significantly enhanced in octupole deformed nuclei by $\Delta E_\pm$, see Eq. \ref{SchiffOctupole}.  Both $P,T$-odd electron-nucleon couplings, $C_{S,P,T}$, and the electron EDM, $\mathbf{d}_e$, will also couple weakly into an overall molecular EDM.  But these contributions will be small in molecules with paired valence electrons, e.g. $^1\Sigma$, which are well-suited to a hadronic CP violation measurement due to their magnetic field insensitivity.  Figure inspiration from \cite{Graner2017a}. }
    \label{fig:new-physics-to-edm}
\end{figure}

Realizing discoveries with radioactive molecules will require a very broad base of theory support.  A wide range of atomic, molecular and nuclear theory are critical for identifying molecules for study, designing experiments, and interpreting results. The impact of certain measurements relies on understanding signals that originate at the nucleon level but are manifested in molecular spectra.  Thus theory is needed to connect physics across multiple fields. Atomic and molecular theory is needed for translating laboratory measurements to CP-violating observables such as electric dipole moments (EDMs) and nuclear symmetry violating moments. Nuclear theory establishes the connection between nuclear properties and underlying physical sources such as quark EDMs and strong force CP-violation. High-energy theory is essential to connect these sources to fundamental physics, including Standard Model extensions, see Fig. \ref{fig:new-physics-to-edm}.

This manuscript reviews the scientific motivation for radioactive molecules and the relevant experimental and theoretical techniques, it discusses existing and upcoming opportunities at facilities which can produce these species, and it provides an outlook for the field's future.  While combining the challenges of molecular experiments and radioactive nuclei is daunting, there are important recent advances, such as spectroscopy of RaF \cite{GarciaRuiz2020,Udrescu2021}, and trapping and cooling of radium-containing molecular ions \cite{Fan2021,Yu2021}. The young field of radioactive molecules is full of opportunities for a wide range of scientists, it is exciting to consider the developments that will take place over the next decade, especially the inevitable unforeseen advances and ideas that will drive discoveries.

\section{Science Motivation}\label{Science_Motivation}
  
In this section we focus on a few areas where radioactive molecules offer unique opportunities for major advances.  First, radioactive molecules can combine molecular and nuclear enhancements to dramatically amplify signatures of symmetry violations, both within and beyond the Standard Model.  Second, the electromagnetic environment experienced by nuclei in molecules provides pathways to study detailed nuclear structure, such as higher-order parity-violating effects and distributions of nuclear charge and weak currents.  Third, radioactive molecules are important astrophysical probes, and  laboratory-based spectroscopy is critical to reveal detailed information about stellar processes.  Fourth, the availability of isotopes with extreme mass and charge will enable the study of a largely unexplored and chemically distinct area of the periodic table.

\subsection{Charge-parity violation}\label{sec:bsm}

Molecules are sensitive to a wide variety of physics, both within and beyond the Standard Model.  Fundamental symmetry violations are of particular interest and relevance since molecules are uniquely sensitive to the signatures of these symmetry violations, enabling sensitive probes of physics at high energies \cite{Engel2013,Roberts2015,DeMille2017,Safronova2018,Isaev2018,Cesarotti2019,Chupp2019,Cairncross2019,Skripnikov:2020c,Hutzler2020Review,Snowmass2021EDM}. 

The motivation to study fundamental symmetry violations is quite broad.  First, there is specific motivation provided by the observed imbalance between matter and anti-matter in the universe, which cannot be explained within the Standard Model, and suggests the existence of new CP-violating physics~\cite{Huet1995,Dine2003}.  Second, since CP is not a symmetry of the universe~\cite{Christenson1964,Fanti1999,Abe2001,Aubert2001,Aubert2004}, there is no reason to expect that new particles and forces will preserve CP, making symmetry-violation searches a generically powerful probe for new physics.  Third, the Strong CP Problem~\cite{Kim2010StrongCPReview} motivates measurement of strong CP violation, which manifests in low-energy observables, but also hints at the existence of new CP-violating particles.  Finally, parity violation in atomic systems gives critical information about electroweak physics at low momentum transfer, where such precision information is lacking \cite{Safronova2018,Roberts2015}.

In the low energy regime, CP-violation manifests as symmetry-violating electromagnetic moments, such as electric dipole moments, nuclear Schiff moments, and magnetic quadrupole moments (MQM).  Observable effects of these moments are enhanced by orders of magnitude in molecules relative to other experimental platforms. This is due to the unique combination of extreme internal electromagnetic fields present in molecules and the ability to fully orient these fields with modest externally applied fields \cite{Roberts2015,Safronova2018,DeMille2017,Chupp2019,Cairncross2019,Hutzler2020Review,Snowmass2021EDM}. Despite the considerable challenges of working with molecules, their increased intrinsic sensitivity through this ``molecular enhancement,’’ combined with a variety of new experimental techniques that are discussed in this manuscript, has resulted in molecules now starting to overtake atomic experiments and provide the most sensitive searches for CP-symmetry-violating effects.  Furthermore, all of these searches will benefit from the availability of heavy radioactive nuclei; the intrinsic sensitivity of molecules to CP-violation scales roughly as $Z^{2}$ to $Z^3$.

The most sensitive limits on the electron EDM (eEDM) come from experiments using HfF$^+$ \cite{Cairncross2017,Roussy2022} and ThO \cite{ACME2018} molecules.  These experiments, along with one using YbF \cite{Hudson2011}, provide more sensitivity  than the best limits obtained from atomic experiments \cite{Regan2002}.  Molecular experiments have improved the eEDM limit by a factor of around 100 in around 10 years, and are already probing parameter space outside of the reach of direct searches at high-energy colliders -- up to $\sim$50~TeV for new CP-violating particles which couple to the electron~\cite{ACME2018,Cesarotti2019,Snowmass2021EDM}.  Since the CP-violation in the Standard Model is still many orders of magnitude smaller than the uncertainty of these experiments~\cite{Yamaguchi2020,Ema2022}, they provide a background-free probe for new physics beyond the Standard Model.

An important theoretical consideration is that molecular CP-violation is sensitive to a wide range of observables, including the electron EDM, \thetaQCD, quark EDMs, chromo-EDMs, CP-odd four-quark operators, and the CP-violating electron-nucleus interactions \cite{Engel2013,Yamanaka2017,Safronova2018,Cesarotti2019,Dekens:2013zca,Kley:2021yhn,Snowmass2021EDM}.  The measurement of CP-violation in any given molecular species is sensitive to a combination of these effects, and thus cannot by itself isolate and probe a single underlying source~\cite{Chupp2015,Chupp2019}; global constraints from multiple experiments provide the best bounds on CP-violating parameters.  In order to robustly identify sources of new physics, we require measurements in multiple species, as well as molecular and nuclear theory to understand the sensitivity of a molecule to the different sources.  This motivates the continued development of the experimental and theoretical tools discussed in section \ref{sec:developments}.

Nuclear electric  dipole moments are screened by electrons in neutral atoms and molecules, as the charged nuclei will move to an equilibrium position with zero average electric field~\cite{Schiff1963}. However, an atomic or molecular EDM can still be detected through the nuclear Schiff moment (NSM)~\cite{Flambaum2002}. Octupole deformed nuclei have significantly enhanced nuclear Schiff moments \cite{Auerbach1996} resulting from  three factors: the collective nature of the NSM in the nuclear reference  frame, a small energy interval $\Delta E_\pm$ between the doublets of opposite-parity nuclear states, and a large T,P-violating interaction between these doublet states.  

The NSM of a nucleus with quadrupole and octupole deformation parameters $\beta_2$ and  $ \beta_3$, respectively, scales roughly as~\cite{Spevak1997,Flambaum2002}
\begin{equation}\label{SchiffOctupole}  
    S \propto \frac{\beta_2\beta_3^2ZA^{2/3}}{\Delta E_\pm}.
\end{equation}
Radioactive isotopes of Fr, Ra, Th, and Pa\footnote{Section \ref{sec:pa} deals with the peculiar case of \iso{Pa}{229} \cite{Haxton1983}, which is expected to have an anomalously small parity doublet splitting.} can have NSM enhancements of $10^{2-6}$ compared to spherical nuclei such as Hg~\cite{Haxton1983,Auerbach1996,Spevak1997, Dobaczewski2005,2015Ra,Dobaczewski2018,Flambaum2019Schiff}. 

Another advantage of octupole deformation is the stability of the results of calculations. Indeed, the expression for the Schiff moment $\vec{S}$\ has two terms of a comparable size  and opposite sign: the main term and screening term the latter of which is proportional to the nuclear EDM~\cite{Chupp2019}:
\begin{equation}\label{eqn:schiffmoment}
\vec{S} = \frac{1}{10}\int r^2\ \vec{r}\rho_Q\ d^3r - \frac{1}{6Z}\int r^2d^3r \int \vec{r}\ \rho_Q \ d^3r,
\end{equation}
\noindent where $\rho_Q$~is the nuclear charge distribution and $\vec{d}_N = \int \vec{r}\rho_Q d^3r$\ is the nuclear EDM.  That is, the NSM arises from a difference between the mass and charge distribution in the nucleus.
This difference of terms with similar size produces some instability of the results in nuclei like \iso{Hg}{199} and \iso{Xe}{129}. However, in nuclei with an octupole deformation, where the Schiff moment and EDM have a collective nature, the contribution of the screening term is relatively small since the collective EDM appears only due to the difference between the proton and neutron distributions, which is small. 
The expression for the Schiff moment in Eq. (\ref{SchiffOctupole}) is proportional to  $\beta_3^2$. The value of $\langle\beta_3^2\rangle$ may be large in the nuclei with soft octupole vibration mode even if the static deformation $\langle\beta_3\rangle=0$. This may significantly extend the list of nuclear candidates \cite{Engel2000,Flambaum2003}.

Note that the magnetic quadrupole moment of nuclei also provides access to hadronic CP-violation through atomic and molecular measurements. The MQM has a collective nature in nuclei with a quadrupole deformation due to the so called spin-hedgehog mechanism, which gives about an order of magnitude enhancement \cite{Flambaum1994}.  Since there are many heavy nuclei with large quadrupole deformations, there is a correspondingly large number of options for molecules with significant sensitivity, including many which are not radioactive~\cite{Flambaum2014,Maison2019,Denis2020,Fleig2017,Maison:20a}

\subsection{Parity violation}\label{sec:pv}

 Concurrent with searches for physics beyond the Standard Model, it is important to test predictions to ensure the model's validity in the realm of ``ordinary'' matter. Measurements of parity violation (PV) are powerful tests of the electroweak sector of the Standard Model \cite{Safronova2018}.  PV measurements are sensitive to several Standard Model electroweak and nuclear parameters which are currently poorly-known, including electron-quark electroweak neutral current couplings, nucleon-nucleus couplings, nuclear anapole moments, the nuclear weak charge, and the weak quadrupole moment.  Precise values of these parameters are vital to understanding how nucleons combine to form nuclei,  describing processes governing nuclear decay,  and providing a description of nuclear electroweak structure across the nuclear chart.  Simultaneously, PV measurements may be used to search for physics beyond the Standard Model, such as a leptophobic $Z^\prime$ boson \cite{Langacker1992, NovSusFla77-1,Gonzalez2013}. Searches for oscillating PV signals have been proposed as a means to detect axion-like particles, another leading dark matter candidate \cite{Stadnik2014,gaul:2020c,gaul:2020d}.

Nuclear spin-independent parity violation (NSI-PV) has been measured in 
PV electron scattering on 
protons and in a number of heavy atoms \cite{Qweak2018,Macpherson1991,Meekhof1993,Vetter1995,Nguyen1997,Wood1997,Antypas2019} and found to be in agreement with Standard Model PV predictions  due to the weak charge,
$Q_W$, at 0.5 to 1.5 $\sigma$ with 0.5\% precision in \iso{Cs}{133}, pending resolution of discrepancies in auxiliary experiments~\cite{Toh2019,Bennett1999} and many-body atomic theory~\cite{Porsev2009,Dzuba2002,Dzuba2012,Tran2022}.  Conversely, while measurements of PV are on-going in several heavy atomic species \cite{Toh2014,Antypas2019}, including radioactive atoms \cite{Fr2}, 
and molecular \cite{Altuntas2018} species, the only non-zero measurement (14\,\% fractional uncertainty) of atomic nuclear spin-dependent parity violation (NSD-PV)  comes from \iso{Cs}{133} \cite{Wood1997}, and this result implies constraints on Standard Model meson-nucleon couplings which disagree with other 
nuclear PV measurements \cite{Haxton2001,Johnson2003}. Remarkably, the proximity of states of different parity in certain molecules can provide an enhancement of more than eleven orders of magnitude in sensitivity to NSD-PV with respect to state-of-the-art experiments with atoms \cite{Sushkov1978,Flambaum1985,Altuntas2018}.

Nuclear spin-dependent parity violation (NSD-PV) in chiral molecules is predicted to lead predominantly to parity-violating splittings in nuclear magnetic resonance (NMR) \cite{barra:1988a,laubender:2003,soncini:2003,weijo:2005,weijo:2007,nahrwold:09,Eills2017}, electron paramagnetic resonance (EPR) \cite{khriplovich:1985} and M{\"o}{\ss}bauer \cite{khriplovich:1985} spectra of enantiomers, the two non-identical mirror images of a chiral molecule. Whereas NSI-PV is expected to cause a parity-violating energy difference between enantiomers \cite{yamagata:1966,letokhov:1975} as well as parity-violating resonance frequency differences in high-resolution rovibronic spectra \cite{letokhov:1975,quack:1989,berger:2004a,crassous:2005,quack:2008,schwerdtfeger:2010}. None of these predicted tiny splittings in spectra of chiral molecules could be resolved so far despite several decades of experimental attempts, but a comparatively tight upper bound to PV effects in chiral molecules stems from high-resolution infrared spectroscopy of enantiomerically enriched CHBrClF \cite{daussy:1999,ziskind:2002}; the bound on the fractional vibrational frequency difference between enantiomers  $\Delta\nu/\nu \le 10^{-13}$  falls short of the predicted effect size by about three orders of magnitude. As NSI-PV effects in chiral molecules scale steeply with $Z^5$, due to the essential role of spin-orbit coupling for parity-violating energy shifts, specifically tailored molecules with heavy radioactive nuclei such as astatine \cite{berger:2007} have the potential for a first successful measurement of electroweak PV in chiral molecules and to serve as sensitive probes for parity violation due to pseudovector cosmic fields \cite{gaul:2020c,gaul:2020d}. Such pseudovector fields are considered as signatures of local Lorentz invariance violations \cite{colladay:1998} or as candidates for dark matter \cite{an:2015,graham:2016}.

\subsection{Nuclear Structure}\label{Nuclear_Structure}

Atomic and molecular spectra can reveal detailed information about the nucleus, including information about the distributions of charge, mass, and magnetization, as well as the weak nuclear force and its higher-order effects.  Large and heavy nuclei can push the magnitude of these effects to their limits, and offer opportunities to access this information in new ways and with higher resolution.

Most of the spectroscopic data that we have on nuclear moments comes from the measurement of the hyperfine splitting of atomic lines \cite{Yan2022}. These measurements performed over long isotopic chains can set stringent tests on different nuclear models \cite{Gar2016,deGroote2020b,(Ver22b)}. However, for certain elements the atomic or ionic structure may not offer sensitivity to higher-order nuclear moments \cite{Koszorus2021}. For atoms in a given electronic level, the specific moments are identically zero due to angular momentum selection rules, or the levels are too short lived to allow for the extraction of the moments. On the other hand, in the electronic ground state of molecules, the expectation value of these moments is non-zero, while the existence of long-lived rotational levels allow their measurement with high accuracy \cite{Kel1998}. Hence molecules can offer a significant advantage over their atomic counterpart.

Experiments to search for PV effects in atoms can be used to measure the quadrupole distribution of the neutrons inside the nucleus~\cite{Flambaum:17}. For this, one can consider the tensor part of the PV Hamiltonian and introduce the weak quadrupole moment~\cite{Flambaum:17}. Such an interaction is dominated by the PV interaction of the neutrons with electrons due to much larger weak charge of neutrons in comparison to protons. Molecules with a $^3\Delta_1$ electronic state are interesting systems for such a problem: $^3\Delta_1$ and $^3\Delta_{-1}$ states can be directly mixed by such PV tensor interaction. It has been shown for the \iso{Hf}{177}F$^+$ molecule~\cite{Skripnikov:2019a} that the tensor weak interaction induced by the weak quadrupole moment gives the dominating contribution to the PV effects and exceeds contributions of the vector anapole moment and the scalar weak charge. The latter effects can contribute only via the interference with non-adiabatic effects and are therefore suppressed.

Nuclear magnetic moments play an essential role in the search for P-violating and T,P-violating effects in molecules. Predictions of the atomic and molecular enhancement factors of such effects are usually probed by comparing the theoretical prediction of the hyperfine structure (HFS) constants with the experimental data. Furthermore, real nuclei have a finite size, which leads to the finite nuclear magnetization distribution effect known as the Bohr-Weisskopf (BW) effect~\cite{bohr1950influence,bohr1951bohr,sliv1951uchet}. This effect is hard to predict accurately~\cite{Senkov:02} and in most cases approximate nuclear models are used, which may lead to uncertainties in HFS constants that amount to several percent~\cite{Ginges2017,Roberts2021}. Methods to determine this effect from a combination of theory and experiment have been proposed for atoms~\cite{Ginges2018,Roberts2022,Sanamyan2022,Prosnyak:2020} and molecules~\cite{Skripnikov:2020e} (see below). Studies of the BW effect may also allow one to probe the distribution of neutrons in atomic nuclei~\cite{Zhang2015}, needed for the accurate evaluation of the atomic parity violation amplitude~\cite{Safronova2018}. 
(See section~\ref{sec:DFT} for more discussion of nuclear theory.)

Another problem for the accurate prediction of the magnetic dipole hyperfine structure (HFS) constants for atomic and molecular systems is the uncertainty of the nuclear magnetic moments for some heavy nuclei. Recently, it resulted in the Bi hyperfine ``puzzle'' problem \cite{Ullmann:17}  -- the experimental result~\cite{Ullmann:17} exhibited a 7$\sigma$ deviation from the theoretical prediction~\cite{Shabaev:01a} carried within the bound-state QED in strong fields. As was shown~\cite{Skripnikov:18a} the source of the puzzle was due to the inaccurate tabulated value of the nuclear magnetic moment of \iso{Bi}{209}. The reason for the inaccuracy of the tabulated value was caused by the use of an approximate theory to calculate the shielding constant, which is required to extract the nuclear magnetic value from the experimental nuclear magnetic resonance data. A new approach, based on the relativistic coupled cluster theory, has been suggested to solve this problem~\cite{Skripnikov:18a}. Consideration of the BW effect and hyperfine magnetic anomaly in neutral atoms are essential to predict nuclear magnetic dipole moments of short-lived isotopes~\cite{Persson1998,Schmidt:2018,Barzakh2020,Prosnyak:2021,konovalova2017calculation,Prosnyak:2020}. Here, one usually needs both experimental and theoretical input for atomic HFS constants of stable and short-lived isotopes for two electronic states. If accurate atomic theory is available, the nuclear magnetic moments may be determined from comparison of theoretical and experimental HFS constants for a single electronic state, as was carried out for isotopes of francium~\cite{Roberts2020}.

It was shown that the consideration of the BW effect can be important for the prediction of the HFS constant in heavy-atom molecules. For example, this effect contributes approximately 4\% of the hyperfine shift in the ground electronic state of \iso{Ra}{225}F~\cite{Skripnikov:2020e}. There it was shown that one can factorize the BW effect contribution to the HFS constant for heavy-atom molecules into a pure electronic part and one universal parameter, which depends on the nuclear magnetization distribution. This important relation allows one to combine experimental and theoretical data to extract this nuclear magnetization distribution parameter for an atom and use it in molecular predictions or vice versa. It means that it is possible to avoid direct nuclear structure calculation of the magnetization distribution if an accurate electronic structure prediction is possible. It also means that one can use experiments and theoretical input for radioactive molecules to obtain the value of the nuclear magnetic moments of short-lived isotopes similar to the atomic case~\cite{Schmidt:2018}.

\subsection{Astrophysics}

\begin{figure}[ht]
\centering
\includegraphics[width=1.0\linewidth]{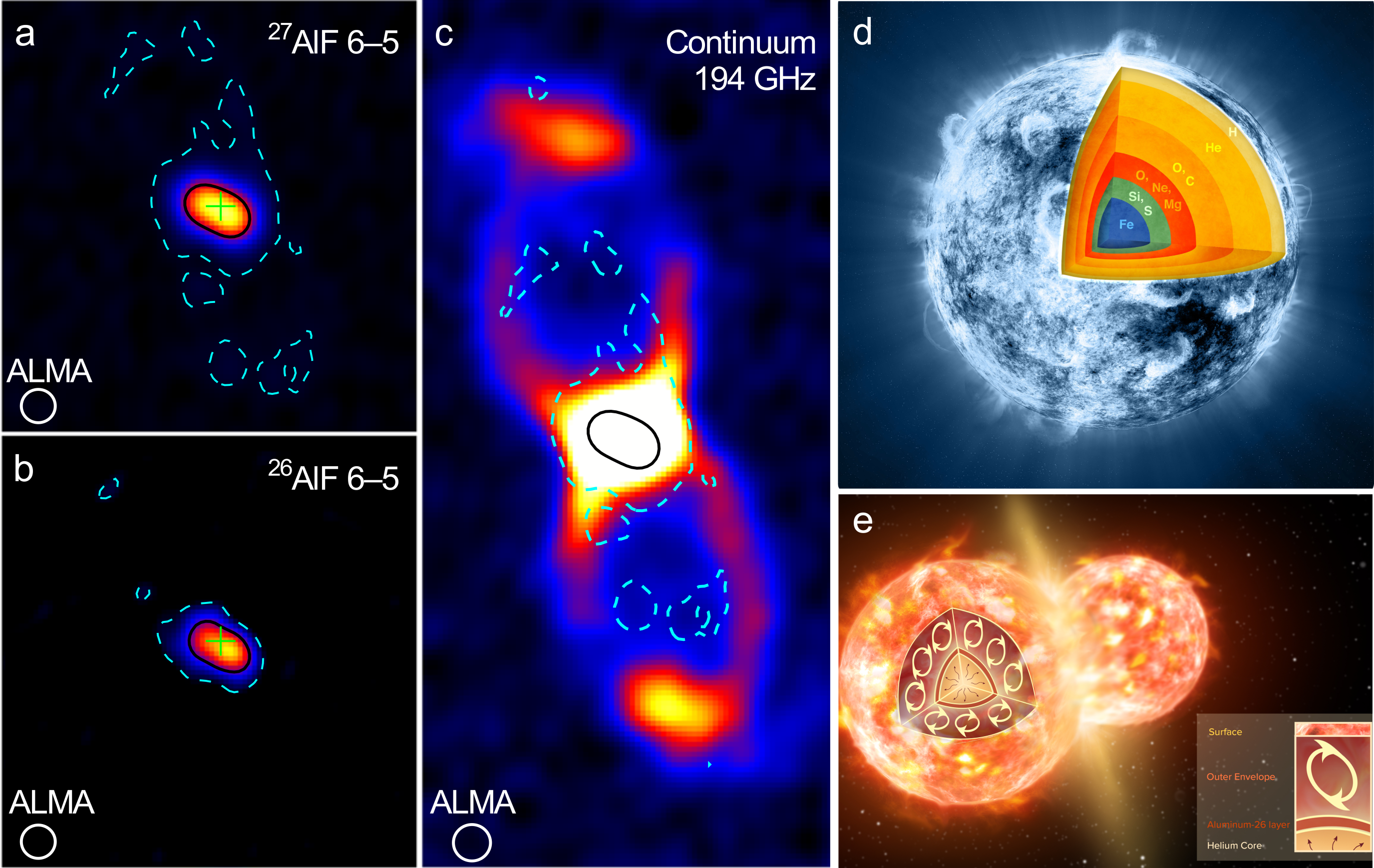}
\caption{Maps of molecular emission of \iso{Al}{27}F [a], \iso{Al}{26}F [b] and dust continuum [c] around stellar remnant \textit{CK Vul} \cite{Kaminski2018}. Crosses indicate the position of the radio source. Cross section of a supergiant showing nucleosynthesis and elements formed [d] Credit: NASA/CXC/M. Weiss. An artist impression of the collision of two stars, like the ones that formed \textit{CK Vul} [e] Credit: NRAO/AUI/NSF; S. Dagnello.}
\label{fig:astro} 
\end{figure}

Over millions of years, generations of stars have enriched primordial matter with heavy chemical elements through nuclear fusion processes \cite{Asplund:2009}. Even the heaviest quasi-stable species, Th and U, have been observed in stellar atmospheres \cite{Frebel:2007,Yong:2021}. The stable nuclei resulting from stellar nucleosynthesis give information about the long-term evolution of stars over large, extended regions of space; in contrast, unstable or radioactive nuclei with short to intermediate lifetimes reflect the dynamics of current stellar evolution \cite{Langer:2012,Tur:2007,Tur:2010,Kaminski2018,Brinkman:2019,Brinkman:2021}. Until recently, radioactive nuclei of the present stellar population are astronomically detected by their highly energetic $\gamma$-ray decay signal \cite{Diehl:2021}. However, as recently demonstrated in the case of aluminum monofluoride, \iso{Al}{26}F, radioactive nuclei can be detected via low-energy photon emission in the millimeter wavelength region once they are incorporated into molecules. These radioactive molecules can then be observed with ultra-high sensitivity and with unprecedented spatial resolution, offered by  the current generation of telescope facilities, i.e., ALMA~\cite{ALMA:2015}, TEXES~\cite{Lacy:2002}, SOFIA~\cite{Young:2012SOFIA}. Before a \iso{Al}{26} nucleus releases a high-energy photon, its  molecular counterpart,  i.e., \iso{Al}{26}F, has emitted one billion low-energy photons. Fig. \ref{fig:astro} shows the spatial resolution of \iso{Al}{26}F around the merger \textit{CK Vul}, observed by the ALMA large array telescope facility.

Stellar objects eject atomic and ionic matter during unstable stellar phases which then can form small- to medium-size molecules in the expanding circumstellar envelope of cooling gas \cite{Campbell:2015,Endres:2016}. After one year of expansion, the expelled stellar matter reaches moderate temperatures that are suitable to form stable molecules. The first of these molecules condense from refractory elements, e.g., metals and heavy atoms, also including short-lived radioactive species with lifetimes of several tens to thousands of years. The local physical and chemical conditions of the expanding stellar shells govern the molecular formation processes in accordance with reaction enthalpies \cite{Gail:2014}, which may also enable the formation of larger radioactive molecular agglomerates and dust grains \cite{Wallner:2016,Groopman:2015,Diehl:2021}. 
 In general, the stellar evolutionary phases are intimately linked to the occurrence of specific radioactive tracer molecules. In the final stage of a supernova event, most of the heavy atoms are locked in dust grains; however, there is a substantial amount of thermal energy released by radioactive decay that can heat and partially vaporize the material of the dust cocoon surrounding supernova remnants like \textit{Cas A} \cite{Grefenstette:2016} and SN1987A \cite{Boggs:2015}. Of particular interest, besides \iso{Al}{26} and \iso{Fe}{60}, are radioactive nuclei like \iso{Ni}{56}, \iso{Co}{60}, \iso{Ti}{44} and \iso{Si}{32}, that decay during the early dense phase of dust and gas expansion.    
 Several attempts to observe radioactive molecules in the circumstellar region of the ageing carbon-rich star, IRC 10216 \cite{Guelin:1995,Forestini:1997}, before they are locked up in interstellar dust particles, have not been successful \cite{Cernicharo:2000}. However,  heavier stellar objects, like luminous blue variables and Wolf-Rayet stars, should produce a higher yield of long-term stable radioactive nuclei, which condense in their hydrogen-poor, slowly expanding, detached outer stellar layer into molecular compounds, enabling their astronomical identification.
 
The observation of the low-energetic rovibrational motions of radioactive molecules in stellar sources requires very precise knowledge of their specific characteristic frequencies, demanding accurate laboratory measurements. While accurate spectra of diatomic molecules can be derived from laboratory measurements of their stable isotopologues \cite{Breier:2018,Breier:2019,Wassmuth:2020}, this indirect method fails for triatomic species such as \iso{Al}{26}OH and for all larger species due to the increasing complexity of the molecular potential. In most stellar environments larger species are expected to be more abundant~\cite{Agundez:2020}. This requires \textit{in situ} spectroscopic measurements of radioactive molecules. Facilities such as ISOLDE-CERN, TRIUMF, and FRIB present excellent opportunities to perform this molecular spectroscopy, as discussed in section~\ref{sec:facilities}. Spectroscopic studies of radioactive species will enable future astronomical observations that will provide critical insights about the stellar nucleosynthesis processes at play in the interiors of massive stars. 

\subsection{Radiochemistry}

Radiochemistry is a vastly understudied field. Within recent decades, technological advancements have occurred to allow safe experimentation with radioactive elements in addition to optimizing small scale experiments. Different radionuclides are often generated together, so it is essential to optimize isotope
separation for studying high-purity samples. The separation of elements can be dependent on their concentrations, thus further complicating separation~\cite{Nash2006}. Moreover, many of the isotopes of interest, like \iso{Pa}{229}, are understudied due to their relative scarcity and difficulty in handling. 

For future work with \iso{Pa}{229} the best protactinium isotope for optimizing protactinium chemistry is \iso{Pa}{231}~(32,760 year half-life). Samples of \iso{Pa}{231}, primarily obtained through extraction from uranium ores~\cite{Kirby1960}, have generated significant quantities of radioactive daughter products such as \iso{Ac}{227} over several decades. These high specific activity daughter products, despite having little impact on the overall chemistry as they are by mass a minor impurity, make the dose of the \iso{Pa}{231} significantly higher and thus more difficult to work with safely. Protactinium is also known to deposit onto glass surfaces, complicating research, particularly with small quantities.

Despite the obstacles inherent in elements with exclusively radioactive isotopes, there have been astounding results and progress with \iso{Ra}{225}. Radium poses a unique challenge in that it decays into a gaseous daughter, radon, in addition to having daughter isotopes with high specific activities~\cite{Failla1932}. This has made the safe handling and experimentation with radium exceptionally difficult, particularly on a bulk scale where fundamental chemistry can be established. Regardless of this hindrance, spectroscopic studies of \iso{Ra}{225} atoms and molecules containing it have shown great promise for studying radioactive isotopes of interest~\cite{2015Ra,GarciaRuiz2020}. 

\section{Recent Developments}
\label{sec:developments}

There has been recent and rapid progress in a wide range of techniques which will be necessary to realize the major scientific advances mentioned above.  In this section, we review a selection of relevant experimental and theoretical advances.   Many of these techniques do not directly concern radioactive molecules currently, but represent a firm foundation which will enable advances with radioactive species.  Furthermore, radioactive molecules provide a unique opportunity to bring together these very diverse fields in the solution of a common challenge.

\subsection{Spectroscopy of radioactive molecules} 

The study of radioactive molecules requires overcoming major challenges.  These molecules can only be produced or handled in minuscule quantities, often $<$ 1 nanogram. The efficient use of these 
synthetic molecules is therefore critical.  They also have short lifetimes requiring rapid experimental work and important safety considerations. 

Recently, sensitive studies of the rotational and hyperfine structure of radium monofluoride (RaF) were performed at room temperature by leveraging ion traps, fast beams, and collinear resonance ionization \cite{GarciaRuiz2020,Udrescu2021}. Predictions of the suitability for laser cooling of RaF \cite{Isaev2010} are supported by these experimental results. To overcome challenges inherent with short-lived molecules, collinear resonance ionization spectroscopy was used in combination with bunched ion beams. A subsequent interpretation of radium molecule isotope shift measurements suggest high RaF sensitivity to nuclear size effects~\cite{Udrescu2021}. These results provide experimental bounds to further develop quantum chemical theory, an important ingredient for interpreting molecular spectroscopy in terms of nuclear structure. 

Recently, a new experimental campaign achieved $\sim 50$ MHz resolution and high sensitivity (less than 100 molecules in a given rotational state) spectra of RaF molecules using the collinear resonance ionization spectroscopy technique \cite{Udrescu2022}. These
measurements will allow high-precision isotope shifts, rotational constants, and hyperfine structure
parameters of different RaF molecules to be extracted.  As discussed in section \ref{sec:facilities}, the study of RaF and other radioactive molecules requires access to these exotic species made available by specialized facilities.

\subsection{CP-violation searches in molecules}

Molecules have enhanced sensitivity to a variety of fundamental symmetry violations~\cite{Safronova2018,Chupp2019,Hutzler2020Review,Snowmass2021EDM}, and offer several key advantages over atoms. The large electromagnetic fields in molecules leads to amplification of the effects of CP-violating electromagnetic moments, including the electron EDM, nuclear Schiff moments, nuclear magnetic quadrupole moments, as well as CP-violating interactions between electrons and nucleons.  Since molecules can be efficiently polarized in laboratory fields, whereas atoms generally cannot, the achievable sensitivity with molecules is typically $\gtrsim 10^{2}$ to $10^3$ times larger.  Furthermore, certain molecules offer an ``internal co-magnetometer'' scheme for robust rejection of systematic errors~\cite{Baron2014,Cairncross2017,Kozyryev2017PolyEDM,ACME2018}.
 
There are now three molecular experiments with electron EDM sensitivity greater than that of the most sensitive atomic experiment (Tl~\cite{Regan2002}): YbF~\cite{Hudson2011}, HfF$^+$\cite{Cairncross2017,Roussy2022}, and ThO~\cite{Baron2014,ACME2018}.  Each of these experiments is being actively upgraded with the goal of improving the current limit by at least an order of magnitude in the next few years~\cite{Panda2019,Wu2020,Ho2020,Alauze2021,Masuda2021,Ng2022}, with pathways for improvements in future generations.  These experiments are already probing up to the $\sim 50$~TeV scale for generic CP-violating physics~\cite{ACME2018,Cesarotti2019,Snowmass2021EDM}, and offer opportunities for significant improvements using many of the techniques discussed in this manuscript.  There are many promising candidates for new molecular eEDM searches, including BaF~\cite{Aggarwal2018}, $^{174}$YbOH,~\cite{Kozyryev2017PolyEDM,Denis2019,Prasannaa2019,Gaul2020,Augenbraun2020YbOH,Zakharova:2021a,petrov2021sensitivity}, \iso{Ra}{226}OH~\cite{Isaev2017RaOH,Kozyryev2017PolyEDM,Gaul2020,Zakharova:2021b}, BaOH~\cite{Denis2019,Gaul2020}, SrOH~\cite{Kozyryev2017SrOH,Gaul2020,Lasner2022SrOH}, and species isolated in matrices (see section \ref{sec:matrix}).
Experiments with all these systems can be also used to probe the effect induced by the exchange of axionlike particles between electrons and nucleons~\cite{Stadnik2018,Maison:2021,maison2021axion} as such interactions can induce static EDMs of atoms and molecules. Data from the HfF$^+$ experiment, for example, has been used to constrain masses of axionlike particles over several orders of magnitude~\cite{Roussy2021}.

Experiments are also under underway to search for hadronic CP-violation in molecules.  Nuclear Schiff moments and magnetic quadrupole moments give access to sources of hadronic parameters such as those discussed in section \ref{sec:bsm}~\cite{Ginges2004,Engel2013,Safronova2018,Snowmass2021EDM}.  These experiments include CENTReX \iso{Tl}{205}F~\cite{Hunter2012,Grasdijk2021},  \iso{Yb}{173}OH~\cite{Kozyryev2017PolyEDM,Maison2019,Denis2020,Pilgram2021}, \iso{Ta}{181}O$^+$~\cite{Fleig2017,Chung2021}, \iso{Lu}{175}OH$^+$~\cite{Maison:20a,Maison2022} and experiments with short-lived radioactive species such as RaF~\cite{Kudashov2014,GarciaRuiz2020,gaul:2020,Udrescu2021} and radium-bearing molecular ions, e.g. RaOCH$^+_3$\cite{Fan2021,Yu2021}. A compilation of expected energy shifts for some of these molecules in terms of the QCD parameter \thetaQCD and quark chromo-EDMs is given in~\cite{Maison:20a}. Note that there is motivation to search for oscillating CP-violating observables as well, in particular hadronic sources, as they can probe the axion and axion-like fields~\cite{Graham2011,Graham2013,Stadnik2014,Budker2014,Flambaum2020SpinRotation,Arvanitaki2021}.  Additionally, the paramagnetic molecules traditionally used in electron EDM experiments using heavy nuclei without spin, such as $^{232}$ThO, are in fact sensitive to hadronic and nuclear CP violation through higher-order effects~\cite{Flambaum2020Hadronic,Flambaum2020Internucleon,Flambaum2020pEDM}.

\subsection{P-violation in molecules}
\label{sec:pvmol}

The weak interaction between electrons and nucleons produces several parity-violating effects which mix states of opposite parity in atoms and molecules. PV effects are enhanced in systems with close-lying opposite-parity states, making molecules an ideal platform for PV measurements. For example, neighboring molecular rotational levels have opposite parity and may be tuned near to degeneracy in a magnetic field on the order of $0.1$ T to 1 T \cite{Kozlov1985,Flambaum1985,DeMille2008Anapole}. In addition to rotational structure, linear polyatomic molecules have opposite-parity $\ell$-doublets which are roughly 100 times closer in energy than rotational levels, and can be tuned near to degeneracy using correspondingly smaller magnetic fields \cite{Norrgard2019}. In chiral molecules, i.e. molecules which are not superimposable with their mirror image and possess a high barrier for stereomutation, the degeneracy of states of opposite parity becomes lifted only by quantum mechanical tunnelling \cite{hund:1927}.
This tunnelling induces a splitting which is expected to be orders of magnitude smaller than parity-violating matrix elements in configurationally stable chiral molecules~\cite{berger:2001a,quack:2008,sahu:2021}.
This leads to profound consequences for the structure and dynamics of chiral molecules \cite{quack:1989,berger:2001a,quack:2002,berger:2004a,berger2019review} (see below). 
PV may in principle be measured in each constituent nucleus of the molecule to provide strong consistency checks of both systematic errors and theory.
As with atoms, many molecular species may be laser cooled and trapped in order to extend the interaction time and increase the sensitivity compared to beam-based PV measurements \cite{Isaev2010,Isaev2016Poly,Isaev2018}.  

Precision measurements of NSD-PV of multiple nuclei are needed, ideally in both very heavy and light nuclei to distinguish between the two leading contributions: the nuclear anapole moment and $Z_0$ boson exchange between an electron and individual nucleons.  Promising candidates for future measurements include RaF \cite{Isaev2010} and RaOH molecules \cite{Isaev2017RaOH,Norrgard2019}.  NSD-PV matrix elements are predicted to be an order of magnitude larger in RaF and RaOH compared to $^{137}$BaF, primarily due the relativistic enhancement $\propto Z^2$ and the $\propto A^{2/3}$ scaling of the anapole moment. RaF and RaOH molecules have two favorable properties for laser cooling: highly-closed optical transitions \cite{Isaev2010,Isaev2017RaOH,GarciaRuiz2020}, and  (for the \iso{Ra}{213} and \iso{Ra}{225}  isotopologues), the minimal nuclear spin $I= 1/2$ necessary for NSD-PV. A challenge in using RaF or RaOH to study PV is that both molecular- and nuclear-structure calculations, for example to determine molecular sensitivity to PV signals and the magnitude of different sources of PV, are difficult in heavy systems.  However, radium has nine isotopes with $I>0$ and half-lives  $\tau > 10$ s. The isotope chain presents an opportunity to  isolate nuclear- and molecular-structure effects, and thereby enable robust tests of each. Complimentary  NSD-PV measurements which are more sensitive to $Z_0$ boson exchange are possible in a number of light molecules \cite{DeMille2008Anapole,Norrgard2019, Hao2020}, where both the molecular and nuclear calculations are most accurate.  

\subsection{Molecular beams}

Collimated beams of atoms and molecules have been at the heart of atomic physics for almost a century and are the most mature technique for molecular spectroscopy~\cite{Scoles1988}.  It is often desirable to study molecules at a temperature of a few Kelvin or below to collapse population into the lowest few rotational states; however, this presents a challenge as molecules at this temperature will freeze onto surfaces.  The most common techniques to circumvent this issue rely on producing beams of molecules cooled by inert gases which are themselves cooled, for example supersonic beams~\cite{Scoles1988} or cryogenic buffer gas beams (CBGB)~\cite{Hutzler2012}.  These beam techniques have been used to perform spectroscopy on an extremely wide range of species, from diatomics to large organics, open and closed shell, and reactive, refractory or otherwise challenging species. The techniques used to produce these species offer tremendous flexibility, and include ablation of solid precursors, injection of hot vapors, reaction of metals with gases, and optically-driven chemical production~\cite{Hutzler2012,Patterson2009,Jadbabaie2020}.

Molecular beams were used in two of the three molecular eEDM experiments which surpassed the sensitivity of the Tl experiment: ThO~\cite{ACME2018} and YbF~\cite{Hudson2011}.  Interrogation times can be increased using CBGB techniques designed to create slow beams~\cite{Lu2011,Hutzler2012}, and the flux at the detector can be enhanced using magnetic or electrostatic lenses, or transverse laser cooling~\cite{DeMille2013,Ho2020,Alauze2021}, all of which are being implemented to improve molecular beam-based EDM searches. Molecular beams also serve as an important starting point for many other techniques, such as laser cooling and trapping (sec. \ref{sec:lc}), which aim to trap molecules to achieve still longer coherence times.

\subsection{Direct laser cooling}\label{sec:lc}

One method for improved precision is to directly laser cool molecules~\cite{DiRosa2004, Shuman2010, Fitch2021Review}. Laser cooling is a well-established technology for atoms~\cite{Metcalf1999}, and is a major driver of advances in atom-based quantum science including atomic clocks, quantum information processing, and quantum simulation, in both neutrals and ions.  The technique relies on repeated absorption-spontaneous emission cycles to transfer momentum from a laser field to an atom or molecule, thus applying a force.  Combining multiple laser fields with control over frequencies, polarization, and external DC fields can result in cooling and trapping in optical fields down to microKelvin temperatures and high densities.  The motivation for implementing laser cooling in molecular precision measurements is strong; increases in coherent interrogation time, state preparation and readout efficiency, and better control over electromagnetic fields in small volumes could lead to orders-of-magnitude increases in sensitivity to fundamental symmetry violations~\cite{Tarbutt2013,Kozyryev2017PolyEDM}, searches for dark matter~\cite{Kozyryev2021}, and much more~\cite{Safronova2018,Hutzler2020Review}.

\begin{figure}
    \centering
    \includegraphics[width=0.4\textwidth]{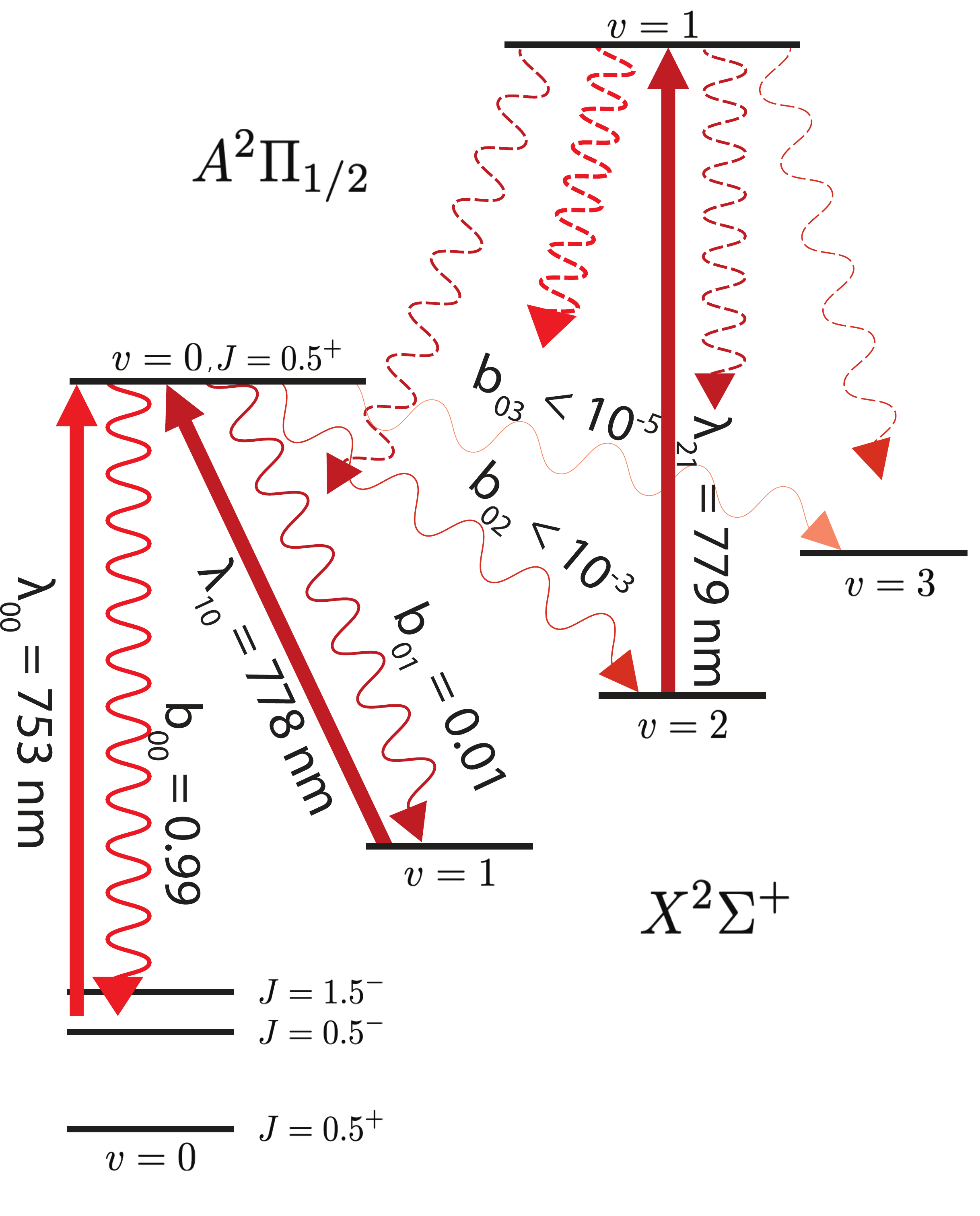}
    \caption{Proposed laser cooling scheme for RaF~\cite{Isaev2010,GarciaRuiz2020,Udrescu2022} with vibrational branching ratios based on experimental measurements and theoretical predictions.}
    \label{fig:LaserCooling}
\end{figure}

The primary challenge with laser cooling molecules are their rotational and vibrational degrees of freedom, which can become excited via spontaneous emission.  Certain molecules exhibit optical transitions with nearly closed decay paths, enabling many thousands of photons to be cycled with just a few lasers~\cite{DiRosa2004,Shuman2010,Isaev2016Poly,Fitch2021Review}.  These molecules generally feature a decoupling of electronic and vibrational degrees of freedom; thus the molecule can absorb a photon, receive a momentum kick, and then spontaneously decay back to the same starting state without vibrational excitation, as shown in Fig. \ref{fig:LaserCooling}. Recently, light diatomic molecules have been confined in magneto-optical traps \cite{Barry2014, Truppe2017SubDoppler, Anderegg2017, Collopy2018}, cooled to microkelvin temperatures \cite{Cheuk2018, Caldwell2019} and transferred to conservative magnetic \cite{Williams2018, McCarron2018SrF} or optical \cite{Anderegg2019, Langin2021} traps suitable for precision measurements with long coherence times. A number of species containing heavy atoms have been identified as suitable for laser cooling, including RaF~\cite{Isaev2010} (see \ref{fig:LaserCooling}), YbF~\cite{Tarbutt2013,Lim2018}, BaF~\cite{Aggarwal2018}, RaOH~\cite{Isaev2017RaOH,Kozyryev2017PolyEDM}, YbOH,~\cite{Kozyryev2017PolyEDM,Augenbraun2020YbOH}, HgF~\cite{Prasannaa2015HgX}, TlF~\cite{Cho1991,Hunter2012,Grasdijk2021}, and SrOH~\cite{Kozyryev2017SrOH,Gaul2020,Lasner2022SrOH}. Recently, it has been shown that some molecular cations such as AcOH$^+$ may be considered for laser-cooling~\cite{Oleynichenko:2021AcOH}. Due to the need to scatter more photons to slow heavy molecules experiments with heavier and more complex species are more challenging but advancing rapidly with demonstrations of photon cycling and laser-cooling in one and two dimensions \cite{Kozyryev2017SrOH, Mitra2020, Augenbraun2020YbOH, Lim2018}, and recent magneto-optical trapping of polyatomic molecules~\cite{Vilas2021}.

\subsection{Assembly from laser cooled atoms}

A powerful technique to realize large numbers of ultracold, trapped molecules is to assemble them \emph{in situ} from their ultracold atom constituents.  Compared to direct laser cooling of molecules, this has the significant advantage of needing to cool only atoms to the desired ultracold temperatures, techniques for which are very well-established for many atomic species.  Unbound atoms can be coherently driven, via a combination of Feshbach resonances and/or laser transitions, to bound states---including the absolute ground state. This enables the conversion of ultracold trapped atoms to ultracold trapped molecules, with efficiency of up to $\approx 50\%$ reported~\cite{duda2021transition}. This process has been demonstrated with many different diatomic species consisting of two alkali atoms ~\cite{Ni2008,RvachovKetterleNaLi2017,HuNiKRbReaction2019,ParkZwierleinNaK2015,SeebelbergBlochNaKCoherenceTime2018,YangJWPNaK2019,LiuNiNaCsTweezer2019,GuoWangNaRb2016,MolonyCornishRbCs2014,TakekoshiNagerlRbCs2014,VogesOspelkausNaK2020}, which are the easiest to cool and trap. Such experiments have created large ensembles of ultracold, optically trapped molecules at temperatures of $\sim 100$ nK and with (nuclear) spin coherence times approaching 10 s \cite{Gregory2021}. However, use of this molecular assembly technique for symmetry-violation experiments is limited to molecules that can be formed with laser-coolable atoms. 

Until recently, molecules consisting of two laser-coolable atoms that also have high sensitivity to EDMs or NSMs had not been identified \cite{Meyer2009}. Also, to date, only molecules with closed electronic shells have been assembled, though there is significant effort towards extending the techniques to species with an unpaired electron as well \cite{BarbeSchreckRbSr2018,WilsonCornishCsYb2021}. In a recent proposal, Ag was identified as a promising alkali-like atom for pairing with heavy laser-coolable radioisotopes such Fr or Ra \cite{Fleig2021,Klos2022}. Such an experiment could realize exquisite sensitivity to CP violating physics via the long spin coherence times enabled by trapping at ultracold temperatures and the large number of molecules that can be created.  In order to realize systems of this type, there is much important preliminary work to be done, such as realizing near-degenerate gases of co-trapped Ag atoms with similarly near-degenerate Fr or Ra, and predicting and finding the necessary Feshbach resonances~\cite{Chin2010} and stimulated Raman transitions~\cite{BergmannSTIRAProadmap2019} to place the molecules in the science state of interest. Finally, the platform of optically trapped ultracold molecules has the potential to utilize advanced techniques such as spin squeezing to further enhance sensitivity~\cite{Hosten2016,Fr1,Takano2009,TscherbulDipolarSqueezing2022,BilitewskiDynamicalSqueezing2021}, or to apply promising new CPV measurement schemes \cite{Verma2020}.

\subsection{Ion Trapping}

\begin{figure}
    \centering
    \includegraphics[width=0.4\textwidth]{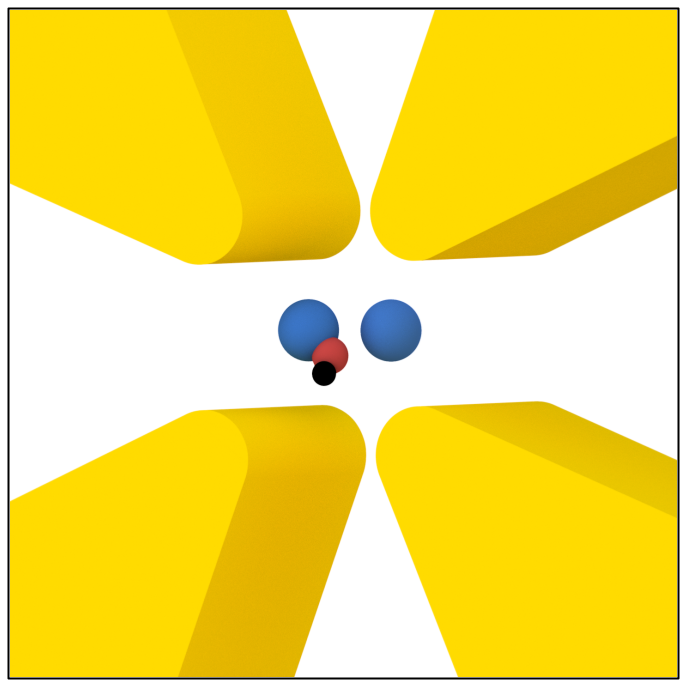}
    \caption{The essence of ion trapping: a small number of particles can be held for long periods of time without the need for laser interrogation.  The figure depicts a charged atom, e.g. Ra$^+$, and a molecular ion held in a Paul trap.  The trap's radiofrequency (RF) electrodes (gold) confine the ions radially, operating on the same principal as a quadrupole mass spectrometer, and axial confinement is provided by DC electrodes (not shown).  The trapping mechanism is only sensitive to a particle's charge and mass, and is therefore independent of internal structure.  The traps are very deep and can be generated with robust RF and DC sources, making it possible to hold ions continuously for weeks \cite{Olmschenk2009b}.  Radioactive molecules may be synthesized in the trap by starting with trapped radioactive ions (e.g. Ra$^+$, Ac$^{2+}$, Pa$^+$) and then introducing neutral gas reagents.  To an extent the loss in number sensitivity compared to a neutral atom or molecule experiment can be made up for with long measurement coherence times.}
    \label{fig:trapped-ions}
\end{figure}

Ion traps hold charged particles with various electromagnetic field configurations, where the two most common types are the Penning trap and the Paul trap.  An ion trap acts on a particle's charge, and the trap's stability depends only on the particle's charge-to-mass ratio.  The trapping mechanism's insensitivity to an atom or molecule's internal structure gives flexibility to work with a diverse range of species.  Typical traps are deep enough to hold warm atoms or molecules, as well as making them robust to elastic collisions between the trapped ions and neutral background gas particles. The Coulomb interaction limits the density of trapped ions, but measurement coherence times~\cite{Zhou2020,Wang2021} can be long and with repeated measurements (often of the same trapped ions) low statistical uncertainty can be realized.  Additionally, ion-trapping systems can be characterized to achieve high measurement precision \cite{Brewer2019}.  The ability to realize high-accuracy measurements with small numbers of atoms or molecules makes ion traps appealing for efficiently working with radioactive species.

There have been recent advances with stable molecular ions that are relevant to their radioactive counterparts.  In an experiment at JILA a relatively large ion trap was used to hold HfF$^+$ molecules in a measurement that set a limit on the size of the electron's EDM \cite{Cairncross2017,Roussy2022}.  In this experiment a rotating electric field polarized the trapped molecules, and then state-selective photodissociation is used for read out \cite{Zhou2020}.

Cold samples of molecules can be produced by using co-trapped and laser-cooled atomic ions to sympathetically cool the molecular ions. For atomic and molecular ions of comparable charge-to-mass ratios, the motion of the atoms and molecules are strongly coupled via the Coulomb interaction, thereby linking the molecule motional temperature to that of the laser-cooled atoms.  With this technique as many as 1,000 molecular ions in large Coulomb crystals have been cooled to temperatures of ~100 mK  \cite{Moelhave2000}.  
These trapped, cold molecules are a good starting point for spectroscopy, as line-broadening effects such as the Doppler effect have been reduced. 
 In the limit of single molecular ions co-trapped with single atomic ions, sympathetic cooling has been used to cool molecules to their motional ground state \cite{Wolf2016, Chou2017}.  For example, CaH$^+$ was sympathetically cooled to its motional ground state, then using a technique known as quantum logic spectroscopy (QLS) \cite{Schmidt2005}, CaH$^+$ was prepared in a single state and then coherently controlled with a Ca$^+$ ``logic'' ion \cite{Chou2017}.  The stimulated Raman transitions used in \cite{Chou2017} to drive transitions between molecular states avoids spontaneous emission which stochastically populates ro-vibrational levels.  With the addition of a frequency comb, quantum logic spectroscopy was used to coherently drive between different rotational states and measure rotational transition frequencies at the ppm level of precision \cite{Chou2020}.  The coherent control afforded by QLS was used to prepare entangled states between molecular rotational levels and internal states of the co-trapped atom \cite{Lin2020}.

Radioactive molecular ions, including RaOH$^+$ \cite{Kozyryev2017PolyEDM}, RaOCH$_3^+$~\cite{Yu2021}, and  PaF$^{3+}$~\cite{Zulch2022}, have been proposed as species to search for CP violation, yet have no experimental spectroscopic information. The QLS techniques developed with stable molecules and stable atoms should be directly applicable to such heavy radioactive molecules, where Yb$^+$ or Ra$^+$ may serve as the logic ion for singly charged molecules.  RaOCH$_3^+$, along with its isotopologue RaOCD$_3^+$, and two other candidate molecules for CP violation (RaOH$^+$ and RaOD$^+$), were also recently synthesized, trapped and cooled \cite{Fan2021}.  Similar results in unpublished work have since been realized with RaSH$^+$~\cite{Hutzler2021DAMOP, Jayich2021}.  However, in order to apply QLS to these molecules, or essentially any non-hydride with small rotational constants, a cryogenic environment is needed to suppress black-body radiation excitation rates for rotational transitions.  At these low temperatures the initial state occupancy of molecules of interest for CP violation, such as the radium-bearing molecules, is similar to CaH$^+$ at room temperature.

\begin{figure}
    \centering
    \includegraphics[width=\linewidth]{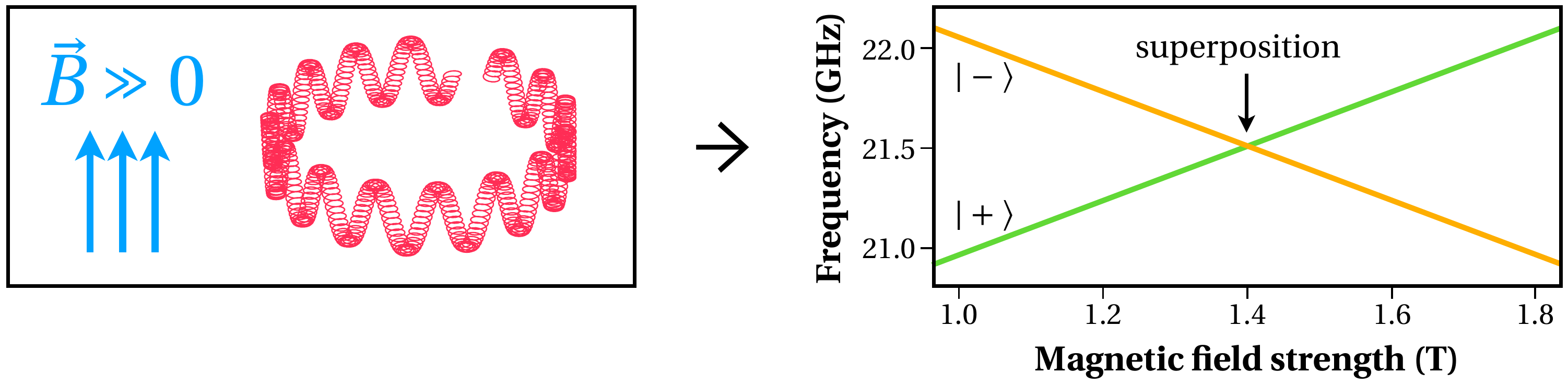}
    \caption{Sketch of the eigenmotion (curly red) of a molecular ion trapped in the strong magnetic field of a Penning trap (left) and the resulting shift of molecular opposite parity levels into superposition (right), thereby strongly enhancing hadronic parity violation.}
    \label{fig:PenningTrap}
\end{figure}

Penning traps operate by applying a weak electrostatic field and a strong magnetic field of several Tesla to trap ions. These traps have been used successfully \cite{Mougeot2021} in precision mass spectrometry of radioisotopes as they allow for direct access to the ion's cyclotron frequency \cite{Brown1986}. There are ongoing developments~\cite{Karthein2021DNP} to use the strong magnetic fields of a Penning trap to Zeeman-shift opposite-parity molecular states into superposition (see Fig.$\,$\ref{fig:PenningTrap}). As recently demonstrated in molecular beam experiments \cite{Altuntas2018}, this superposition state can vastly amplify hadronic parity violation effects, such as the nuclear anapole moment and nuclear spin dependent $Z^0$-boson exchange between the electrons and the nucleus  \cite{Sushkov1978,Flambaum1985,Safronova2018} (see section \ref{sec:pv}). Penning traps, thereby, allow for a substantial increase in the coherence time to measure parity violation at high precision. However, a systematic study of different molecules is necessary to disentangle the different sources of parity violating through their different scaling with mass (see sec. \ref{sec:pvmol}).

\subsection{Contemporary Solid State Approaches}
\label{sec:matrix}

Searching for T-violation in the solid state is thought to hold great promise because of the potentially high number density of sensitive particles. Traditionally, these types of experiments have been formulated as measurements of bulk linear magnetoelectric effects \cite{shapiro68,ignat69,bmw86} in a variety of carefully chosen materials: either (1) the material is magnetized and the resulting electrical polarization due to non-zero atomic EDMs is sensed or (2) the material is exposed to an electric field causing the non-zero atomic EDMs to be aligned and the resulting magnetization is detected using sensitive magnetometry. It has been suggested to apply this approach to matrix-isolated atoms \cite{pryor1987} and molecules \cite{kd2006}.

Matrix isolation is the technique of isolating atoms and molecules within a solid formed from an inert gas such as parahydrogen or noble gases. The number density can be tuned to be as large as $10^{16}/\mathrm{cm^3}$ (1 ppm dopant fraction) \cite{xuthesis} while still keeping the guest species very well isolated from one another.  Spin relaxation times are ultimately limited by long-range dipolar interactions among neighboring guests, which at these concentrations, can be as long as 1000 s for nuclear spins and 1 s for electronic spins \cite{vanvleck}. Large (100 mm$^3$) optically transparent polycrystalline samples are relatively straightforward to grow, making the possibility of laser manipulation and fluorescence-based readout feasible. The dopant capture efficiency is roughly 0.3 \cite{loseththesis}, which is attractive when small dopant numbers are involved as would be the case for rare isotopes.

\begin{figure}
    \centering
    \includegraphics[width=0.25\linewidth]{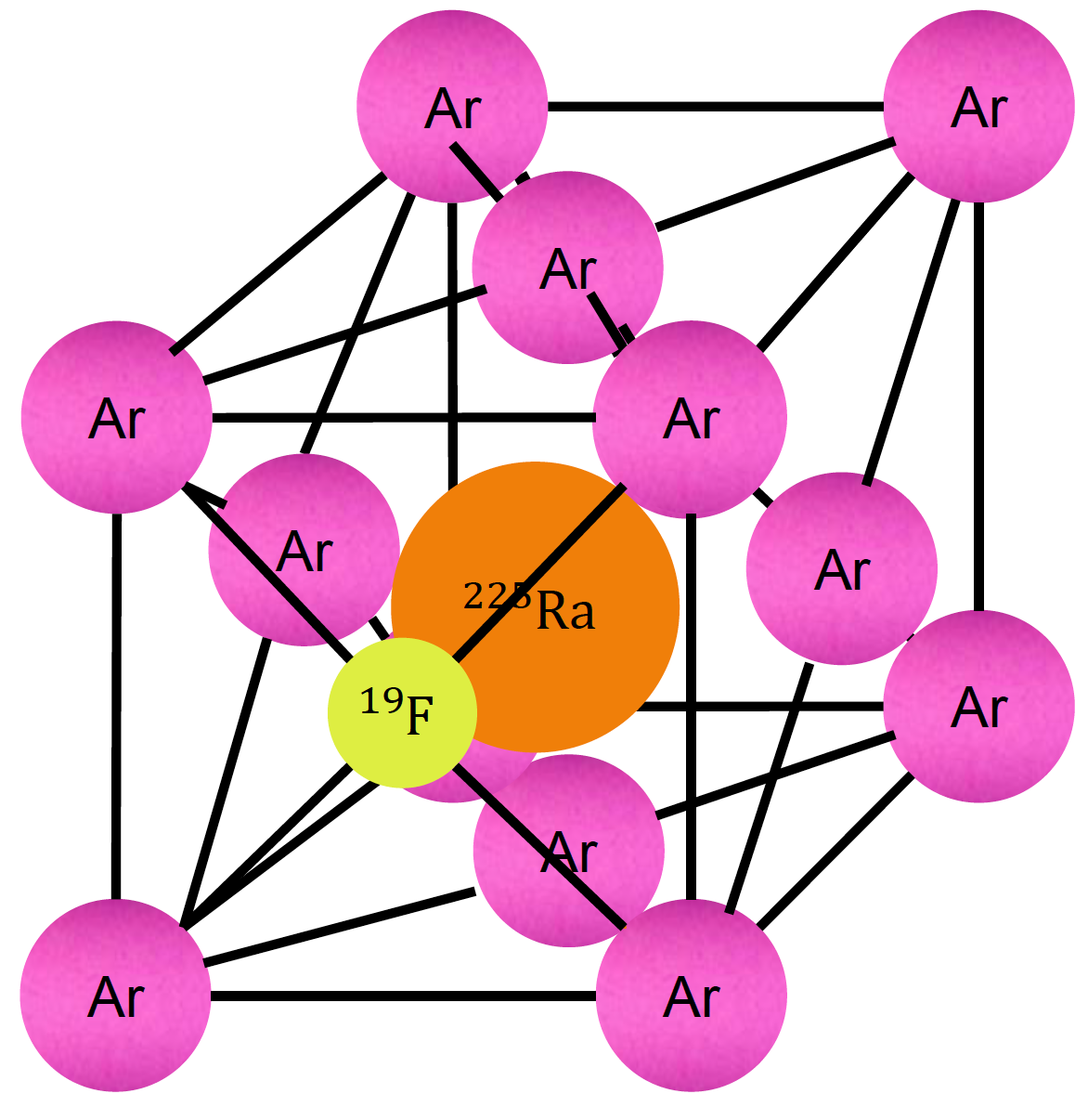}
    \caption{RaF molecules oriented in solid Argon.}
    \label{fig:NNRaF}
\end{figure}

The idea of using lasers and RF radiation to manipulate atoms \cite{Arndt1993} and molecules \cite{Hinds1997} in the condensed phase is not new \cite{mhw2008}, but it has gained renewed interest \cite{Xu2011,Vutha2018atoms} in recent years. This is due in part to the demonstration of optical pumping \cite{Kanagin2013, Upadhyay2016} with long (0.1 s) spin coherence times \cite{Upadhyay2020} in matrix-isolated alkali atoms. In addition, it has been posited that molecules orient themselves along the local crystal axes of rare gas solids, see Fig.~(\ref{fig:NNRaF}), which, if realized, could eliminate many of the systematic effects associated with applying an electric field during the EDM measurement cycle \cite{Vutha2018pra}. Coupled with the ability to grow single crystals which could suppress inhomogeneous broadening effects, matrix-isolated molecules are a promising platform for CP-violation searches that offer statistical sensitivity potentially orders of magnitude better than current EDM limits. Target radioactive molecules include \iso{Ra}{225}F, \iso{Ra}{225}O, \iso{Pa}{229}O, and \iso{Pa}{229}N which could be trapped and isolated in solid Ar or isotopically enriched solid \iso{Ne}{20,22} ($I=0$). Major challenges include the growth of doped single crystals with minimal impurities such as $\mathrm{O_2}$, the demonstration of optical pumping of molecules in solids, the realization of shot-noise limited laser-induced fluorescence readout of quantum states, and a co-magnetometry scheme for closed shell molecules to search for hadronic sources of CP-violation analogous to a co-magnetometry scheme for open shell molecules \cite{Vutha2018pra}.) Finally, the opportunities and challenges of an alternative laser and RF-based scheme using actinide ions, such as \iso{Pa}{229}, implanted in optical crystals was presented in \cite{Singh2019}.

\subsection{Ab initio molecular theory }

Electronic structure theory plays 
many crucial roles
in fundamental physics research with radioactive molecules.
It can be used to identify molecular candidate systems in various electronic states suitable for a given experiment, and theoretical understanding has been critical in guiding both ongoing and planned experiments. For this, favourably scaling electronic structure approaches on the independent particle level,  Hartree-Fock (HF) or density functional theory (DFT), can be used efficiently
\cite{isaev:2014,gaul:2017,gaul:2019,gaul:2020}. These methods allow predictions of most molecular properties of heavy-atom containing compounds with uncertainties below 20\,\% within an hour for diatomic molecules and are applicable to systems with up to 100 atoms \cite{gaul:2017,gaul:2020,gaul:2020b,Skripnikov:16a}. 

Theory is also needed to extract properties of the electron and the nuclei from the experimental measurements. Among such characteristics are the electron electric dipole moment,
nuclear magnetic dipole and electric quadrupole moments, nuclear Schiff moment, magnetic quadrupole moment as well as other symmetry-violating effects. One of the most powerful methods for treatment of heavy atoms and molecules is the relativistic coupled cluster approach \cite{Visscher:1996,EliBorKal15}. It allows high-accuracy calculations of energies and various parameters needed for interpretation of spectroscopic experiments \cite{Skripnikov2016,Sudip:2016b,Abe:2018,KahBerLaa19,BarAndRai21,LeiKarGuo20,KanYanBis20,GusRicRei20,Skripnikov:2021b,Zhang:2021}, and has also shown great predictive power. This method also lends itself to systematic and realistic uncertainty estimates \cite{Skripnikov2015ThO,Skripnikov2016,LeiKarGuo20,HaaEliIli20,HaaDoeBoe21}. Significant efforts have been undertaken for the development of multireference configuration interaction methods~\cite{Fleig:2017}.

In recent years, the accuracy of molecular transition energy predictions has become rather high. For example, it is now possible to take into account the effects of quantum electrodynamics in the four-component fully relativistic molecular calculations~\cite{Skripnikov:2021b,Skripnikov:2021c} at the same level of accuracy as was available for atoms~\cite{Shabaev:13,Flambaum:2005}. Furthermore, new developments in the relativistic Fock-space coupled cluster approach have improved its accuracy and extended its applicability to systems with complex electronic structure \cite{Oleynichenko:EXPT:20}. 

Hamiltonians such as Dirac-Coulomb are now routinely used to study the properties of small molecules. Formally approximate Hamiltonians such as the relativistic effective core potentials~\cite{Titov:99,Petrov2002,Titov:06amin,Titov:05a,Mosyagin:2016} can be successfully applied as well~\cite{Zakharova:2021a,Zakharova:2021b}. Both approaches can be combined to provide 
accurate theoretical predictions~\cite{Skripnikov2016}. 
One can also directly use the two-step approach to predict such properties of crystals as the enhancement of the oscillating nuclear Schiff moment~\cite{Skripnikov:16a}, which is relevant for systems such as crystals proposed for dark matter searches~\cite{Aybas:2021}, 

Theoretical methods can be also used to analyze systematic effects and predict properties of molecules placed in external fields. For example, it is possible to predict the dependence of the $g$-factor of a molecule on the external electric field~\cite{Petrov2014}. In Ref.~\cite{Petrov:2020} a state with zero sensitivity to the electron electric dipole moment suitable to test systematic effects in RaF was found. Also, the values of magnetic fields corresponding to the crossing of the levels of opposite parity were found. Such data can be used to plan experiments to search for the T,P- and P-violating effects using RaF. In Ref.~\cite{petrov2021sensitivity} the dependence of the molecular sensitivity to P,T-violating effects on the external electric field value was studied for linear tri-atomic molecules.

\subsection{Nuclear theory}
\label{ss:theory}

CP violation, as we have seen, can result in a nuclear Schiff moment, which  induces a molecular EDM.  Consequently, interpreting the results of molecular experiments 
requires a calculation of the dependence of Schiff moments on the underlying parameters of the CP-violating fundamental theory.  Such calculations are difficult. Heavy nuclei of interest, viewed as  proton-neutron clusters, are strongly-interacting systems with  200-250 particles.  Our ability to accurately compute their properties  is limited, and Schiff moments are particularly difficult to predict as discussed in Sec. \ref{sec:bsm}.  The CP-violating nucleon-nucleon potential that induces those moments is not well known, and unlike other nuclear quantities Schiff moments have never been observed; we thus have no direct tests of our ability to compute them.   In what follows we briefly present the current state of the art.

\subsubsection{Density Functional Theory}\label{sec:DFT}

For calculating the properties of heavy complex nuclei, the tool of choice is nuclear DFT   \cite{Bender2003}; the  validated global  energy density functionals (EDFs), modelling the effective in-medium nuclear interaction,  often provide a level of accuracy
typical of phenomenological approaches based on parameters locally
optimized to experiment, and enable extrapolations into nuclear {\it terra incognita} \cite{Neufcourt2020}. 
Realistic nuclear charge densities and currents computed by nuclear DFT \cite{Reinhard2021a}  can be used to provide quantitative  predictions for the charge radii, higher-order radial moments \cite{Reinhardr4}, quadrupole moments, hyperfine interaction constants, and other matrix elements needed for extracting  BSM physics \cite{Allehabi2021,Hur2022}.
A key challenge is the ability to assess 
uncertainties of the theoretical predictions using advanced tools of uncertainty quantification  \cite{Dobaczewski2014,Neufcourt2018,Neufcourt2020}.

\begin{figure*}[!htb]
\centering
\includegraphics[width=1.0\linewidth]{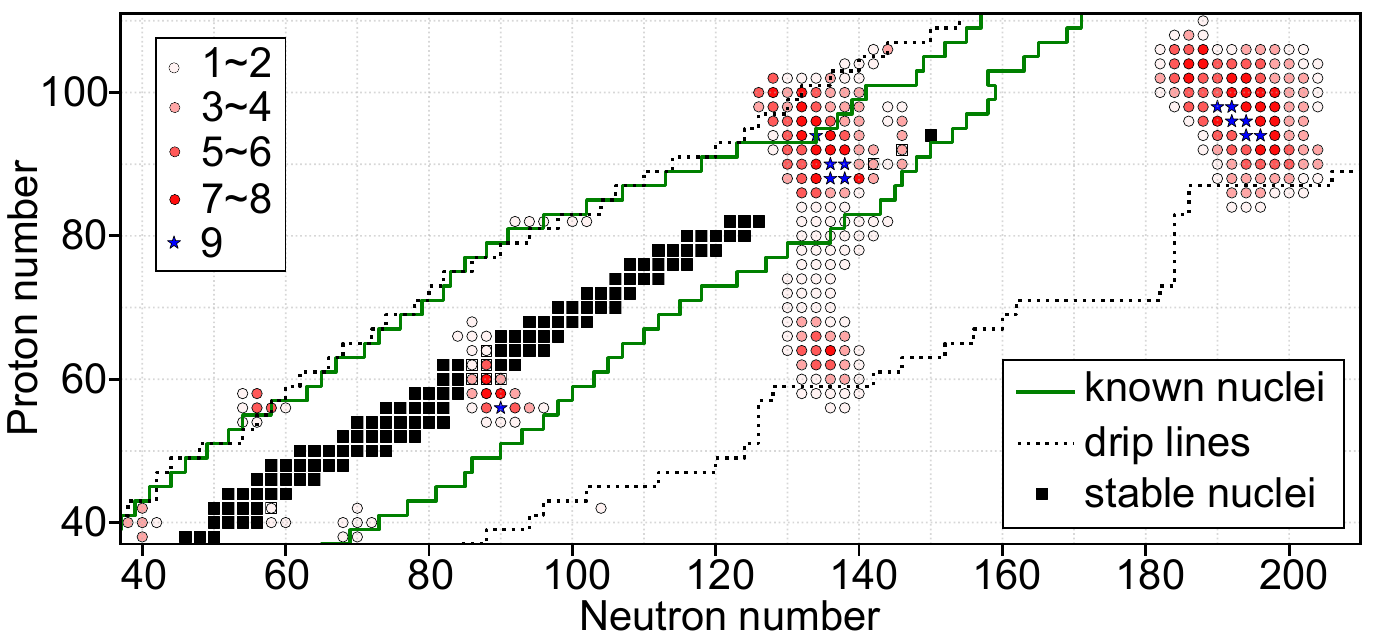}
\caption{\label{multiplicity} The landscape of ground-state octupole deformations in even-even nuclei. Circles and stars represent nuclei predicted to have nonzero octupole deformations. 
The model multiplicity $m(Z,N)$, which is the number of models predicting a nonzero octupole deformation in a nucleus $(Z,N)$,  is indicated by the legend.
 The boundary of known (i.e., experimentally discovered) nuclei is marked by the solid  line. For simplicity, this boundary is defined by the lightest and heaviest isotopes discovered for a given element. The  average two-nucleon drip lines from Bayesian machine learning studies~\cite{Neufcourt2020} are marked  by  dotted lines. Stable nuclides are indicated by  squares.
}
\end{figure*}

According to nuclear DFT, the majority of atomic nuclei have reflection-symmetric ground-states with ellipsoidal  shapes. In rare cases, however, the nucleus can  spontaneously break its intrinsic reflection symmetry and acquire non-zero octupole moments associated with pear-like shapes; for comprehensive reviews see Refs. \cite{Butler1996,Butler2020a}. The necessary condition for  the appearance of localized regions of pear-shaped nuclei in the nuclear landscape is the presence of parity doublets involving  $\Delta\ell=\Delta j=3$ proton or neutron single-particle shells, where $\ell$ and $j$ stand for the orbital and total single-particle angular-momentum quantum numbers, respectively. As discussed in Ref.~\cite{Chen2021} this condition alone is not sufficient to determine whether pear shapes actually appear. The predicted  reflection-asymmetric deformation energies result in fact from dramatic cancellations between even- and odd-multipolarity components of the nuclear binding energy. 

In the context of EDM searches,
the  global DFT surveys \cite{Moller2008,Robledo2011,Agbemava2016,Xia2017,Ebata2017,Cao2020} single-out three specific regions of  octupole collectivity: neutron-deficient  actinides, neutron-rich lanthanides, and neutron-rich heavy and superheavy nuclei that are important for the modeling of heavy-element nucleosynthesis. 
There are 12 even-even nuclei which are systematically predicted  to be octupole-deformed:  $^{146}$Ba, $^{224,226}$Ra, $^{226,228}$Th, $^{228}$Pu, $^{288,290}$Pu, $^{288,290}$Cm, and $^{288,290}$Cf. In the neutron-deficient actinide region, in addition to the Rn, Ra, and Th isotopes, Ref.~\cite{Cao2020}  suggests stable pear-like shapes in  $^{224,226,228}$U, $^{226,228,230}$Pu, and $^{228,230}$Cm. The only stable pear-shaped even-even nuclei expected theoretically are $^{146,148,150}$Nd and $^{150}$Sm.
It is of great interest to carry out systematic DFT studies of pear deformations, parity doublets,  and Schiff moments   
in  odd-mass and odd-odd nuclei. Progress has been made in exploring  particle-odd systems by using projection techniques, primarily in the
systematic computation of Schiff moments \cite{Dobaczewski2018}, but much work still remains to be done.  As discussed in section \ref{sec:bsm}, understanding octupole deformed nuclei is critical for hadronic CPV enhancement via NSMs.

A global DFT and beyond-DFT description of nuclear moments, see,
e.g.,~\cite{(Bon15),(Bor17),(Li18),(Per21)}, has not yet been fully
developed across the nuclear chart, although the first attempts are
promising~\cite{(Sas22b),(Bon22a)}. The challenge here is to replace the
adjustments of interactions, coupling constants, valence spaces, or
effective charges/$g$-factors separately in different regions of the
nuclear chart by a consistent use of the nuclear density functional
applicable to an arbitrary nuclide. For magnetic moments, such a description will require improving the time-odd mean-field
sector of the functional, which has so far been largely neglected
because only the time-even observables have been usually considered in the functional's calibration. The
work in this direction will proceed from the odd near magic
nuclei~\cite{(Sas22b)} through the isotopic and isotonic chains of the
odd neighbours of semi-magic nuclei~\cite{(Ver22b)} and odd-$A$ semimagic
nuclei, and finally to open-shell transitional and deformed nuclei~\cite{(Bon22a)}.

At small deformations, that is, in nearly all semimagic nuclei, one
cannot rely on the standard strong-coupling  approximation as the $K$ quantum number is fragmented,
where $K$ stands for the projection of the intrinsic single-particle angular momentum on the quantisation axis. For a direct
comparison of spectroscopic moments with data, the symmetry
restoration~\cite{(She21)} thus becomes mandatory. This calls for
developing novel functionals based on the density-independent
functional generators, which can guarantee the self-interaction free
approach~\cite{(Tar14a),(She21)}. In addition, since the
polarisation of the charge and spin distributions by odd particles
or holes become essential~\cite{(Sas22b),(Ver22b),(Bon22a)}, one has to properly
consider relative orientations of the intrinsic shapes and currents,
which may significantly affect the results~\cite{(Sat12a)}.
Finally, whenever the single-reference description of nuclear moments
turns out to be inadequate~\cite{(Sas22b)}, a consistent multi-reference
configuration-interaction approach~\cite{(Sat16e)} needs to be systematically
exploited.

\begin{figure*}
\centering
\includegraphics[width=0.7\textwidth]{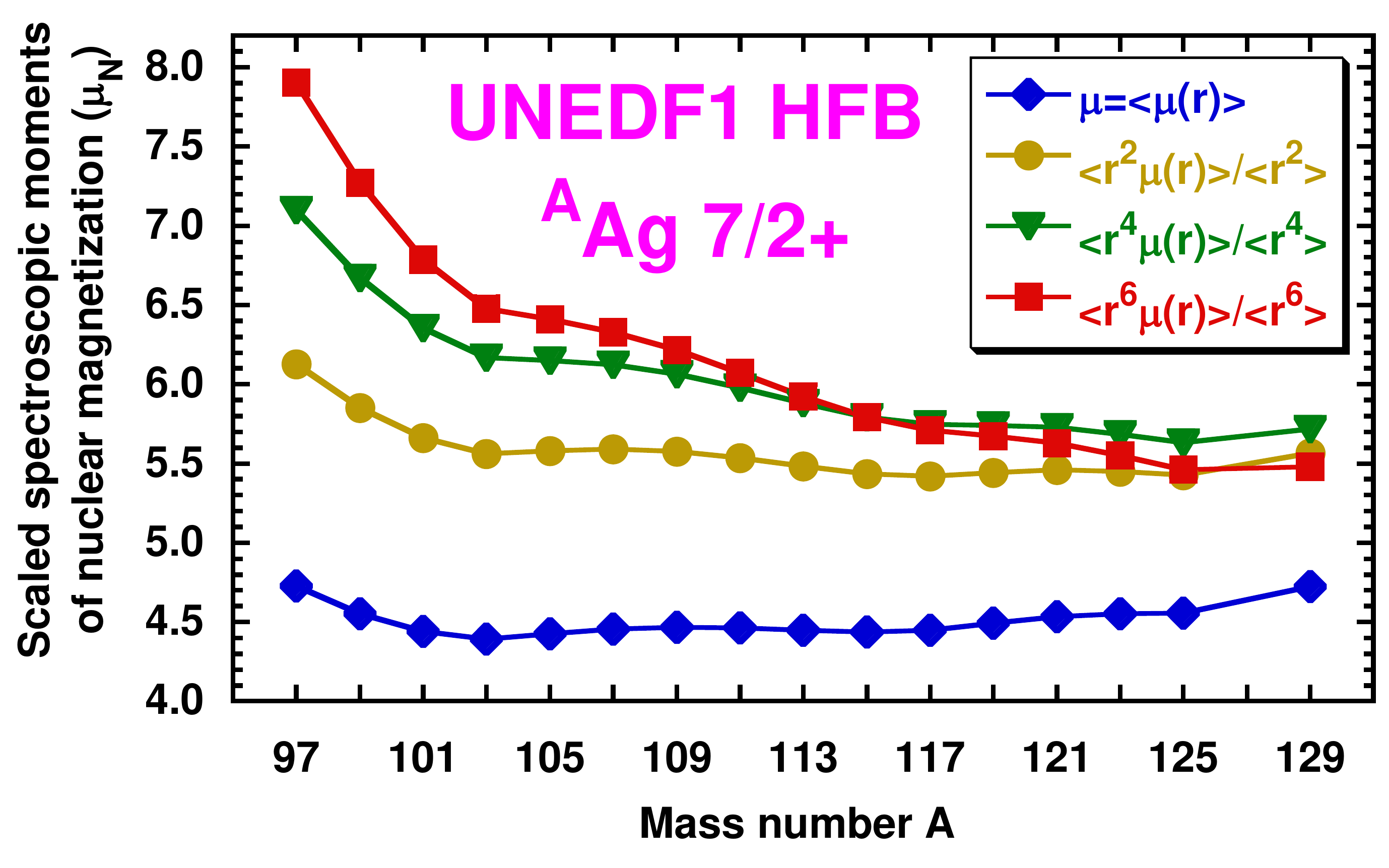}
\caption{\label{ag} Scaled spectroscopic moments of nuclear magnetization calculated
within the Hartree-Fock-Bogolyubov (HFB) DFT with the UNEDF1 Skyrme
functional~\protect\cite{(Kor12b)} for the 7/2+ states in silver isotopes.
}
\end{figure*}

The nuclear DFT can also deliver characteristics of the nuclear
magnetization ${\bm{\mu}}({\bm{r}})$ (magnetic moment density) that
are essential in linking the observed hyperfine atomic or molecular
properties to those of nuclear moments, see
section~\ref{Nuclear_Structure}. In particular, moments of the
nuclear magnetization, $\langle{r}^n{\bm{\mu}}({\bm{r}})\rangle$ which define the BW effect (sec. \ref{Nuclear_Structure}), can be calculated
microscopically without relying on simplifying approximations.
Note that the nuclear magnetic dipole moment $\mu$ corresponds to the $n=0$ moment
of the magnetization. In figure~\ref{ag}, we show the $n=0$, 2, 4, and 6 moments
calculated for the 7/2+ states in silver isotopes. One can see that the moments scaled
by the corresponding radial moments $\langle{r}^n\rangle$ significantly
differ from one another, which indicates that those based on using the single-particle-state
approximation (in which the radial and angular dependencies factorize) may not be
adequate. Further detailed studies of ${\bm{\mu}}({\bm{r}})$ are very much in order.

\begin{figure*}
\centering
\includegraphics[width=0.7\textwidth]{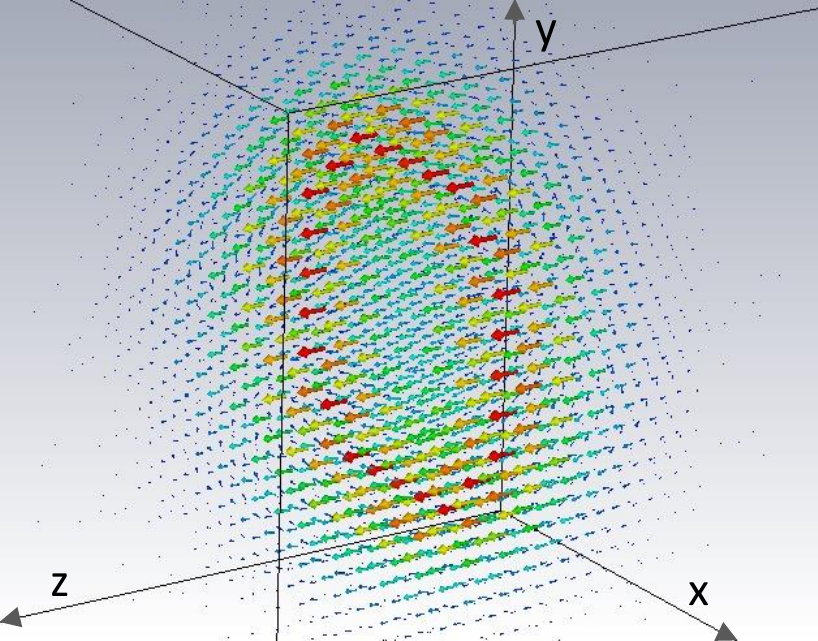}
\caption{\label{sn} A 3D representation of the nuclear total angular-momentum density
(orbital plus spin parts) calculated
within the Hartree-Fock DFT with the UNEDF1 
functional~\protect\cite{(Kor12b)} for the 11/2- state in \iso{Sn}{131}.
Sizes and colours of arrows represent values of the vector field at a given point in space.
Picture courtesy of Nikolay Azaryan, CERN.
}
\end{figure*}

Similarly, anapole moments  (Secs.~\ref{sec:bsm} and \ref{sec:pv}),
which are moments of the nuclear currents
$\langle{r}^n{\bm{j}}({\bm{r}})\rangle$, can be calculated within the nuclear DFT
without relying on the single-particle approximation. In this way, important
contributions coming from the induced orbital and spin currents flowing in the
core can be self-consistently taken into account. An example of such a calculation
is shown in figure~\ref{sn}, where the toroidal structure of the total angular-momentum
density is visible.

\subsubsection{Ab initio}

The last 15 years have seen immense progress in our ability to calculate nuclear properties   \cite{Ekstrom2022}.   \textit{Ab initio} calculations have a crucial advantage over those in lower-resolution approaches such as DFT: the wave functions and operators are determined in a consistent way.  As a result, wave functions are fully correlated. Currently, the chief disadvantage of \textit{ab initio} methods, especially in comparison to DFT, is the difficulty in taking into account the full single-particle strength that is required to describe the absolute values of electromagnetic moments and transitions~\cite{(Ver22b),(Str22)}. Another one is the large amount of computer time and memory needed for heavy nuclei.  Applying the \textit{ab initio} methods systematically will thus require access to our best computing resources.

Two separate achievements have fueled the
growth of \textit{ab initio} theory: the development of nuclear interactions and operators built on the
chiral effective field theory (EFT) of nucleons and pions, and the creation
and/or improvement of several methods for solving the nuclear many-body problem
with controlled approximations.
The effective field theory produces a complete set of operators that contribute
at each order in $p/\Lambda$ or $m_\pi/\Lambda$, where $m_\pi$ is the pion mass, $p$ is a nucleon momentum
and $\Lambda$ is the scale, 500 MeV or so, at which the theory must break down
because other degrees of freedom besides pions and nucleons become manifest.
The nuclear Hamiltonian and other operators can then be written as a sum of
operators in the complete set, with a finite number of numerical coefficients
(at each order) multiplying them.  Although the power counting is guaranteed to
hold only in perturbation theory, in practice it works quite well in
non-perturbative calculations.

Several many-body methods can now exploit the EFT operators. 
Recently, the no-core shell model (NCSM)~\cite{Barrett2013} was applied to calculate anapole moments and electric dipole moments of light nuclei~\cite{Hao2020,Froese:2021civ}.
These calculations were motivated by proposed measurements of the nuclear spin-dependent parity-violating effects in triatomic molecules composed of light elements Be, Mg, N, and C~\cite{Norrgard2019}. The molecules have closely spaced states with opposite parity that may be tuned to degeneracy to enhance parity-violating effects (see sec. \ref{sec:pv}), and {\it ab initio} nuclear calculations are needed for interpretation of these experiments. 

In the NCSM, nuclei are considered to be systems of $A$ nonrelativistic point-like nucleons interacting via realistic two- and three-body interactions. 
The many-body wave function is expanded
in a basis of antisymmetric $A$-nucleon harmonic oscillator (HO) states. 
The only input for the NCSM calculations of parity-violating quantities is the Hamiltonian, consisting of parity-conserving (PC) chiral nucleon-nucleon (NN) and three-nucleon (3N) interaction and parity-violating (PV) NN interactions that admix the unnatural parity states in the nuclear ground state.
NCSM results have been obtained so far in $^9$Be, $^{13}$C, $^{14,15}$N and $^{25}$Mg, which are isotopes of experimental interest ~\cite{Norrgard2019}. 

\begin{figure*}%[t]
\centering
\includegraphics[width=\textwidth]{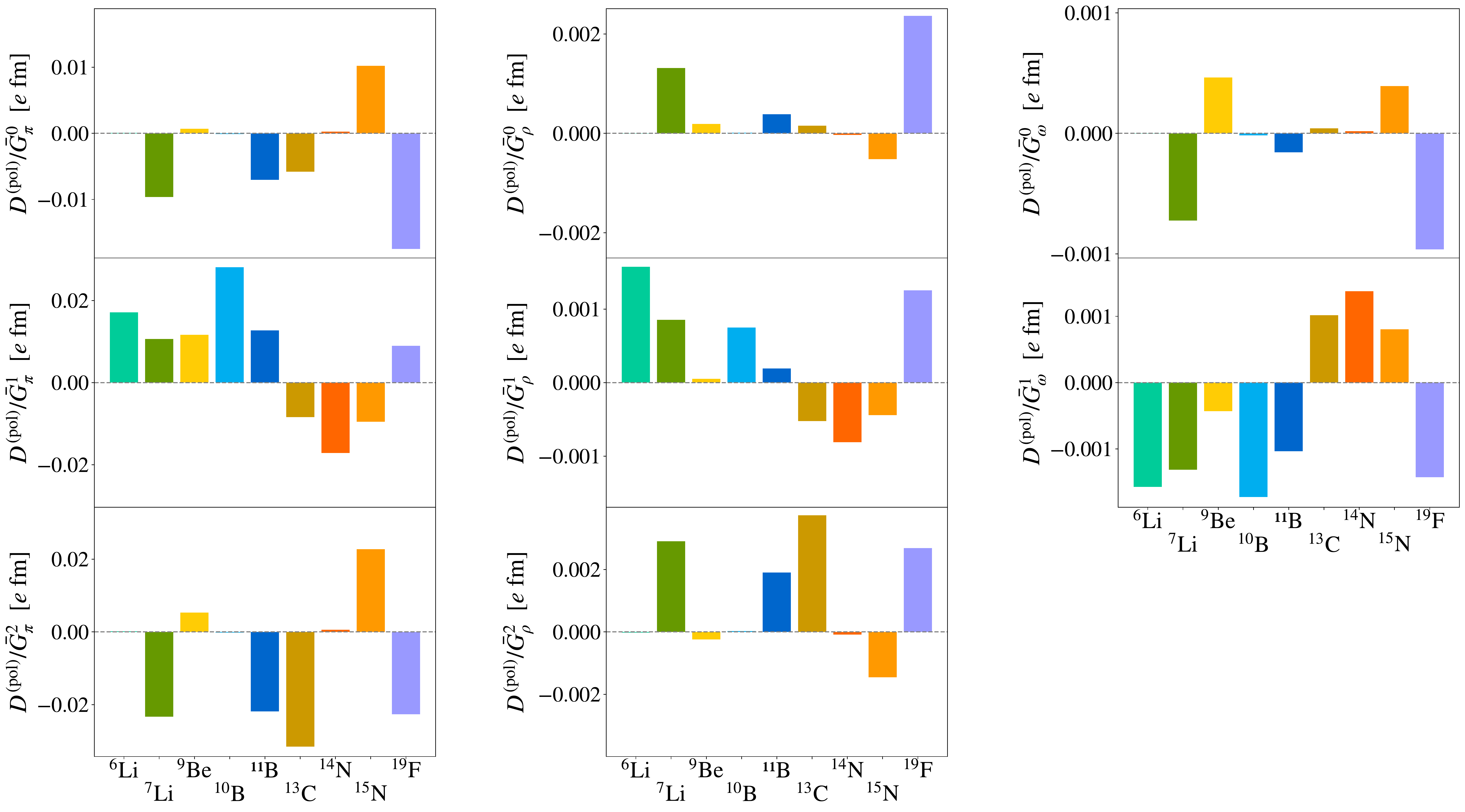}
\caption{\label{fig:Dpolsum} The polarization contribution to EDMs of stable $p$-shell nuclei and $^{19}$F (in $e$~fm) due to the $\chi$-exchange PVTV NN interaction, where $\chi$ stands for $\pi$, $\rho$, or $\omega$. The top (middle, bottom) row represents isoscalar (isovector, isotensor) contribution. The coupling constants $\bar{G}^T_\chi{=}\bar{g}_\chi g_{\chi NN}$ are products of a PVTV $\chi$-meson-nucleon coupling and its associate strong one. Bars of different colours correspond to different isotopes. Figure from ~\cite{Froese:2021civ} where further details are given.}
\end{figure*}

Calculations of light nuclei EDMs are performed in the same way.  The PV NN interaction is simply replaced by the parity- and time-reversal violating (PVTV) NN interaction and the anapole moment operator is replaced by the electric dipole operator. In the NCSM calculations of Ref.~\cite{Froese:2021civ}, one-meson-exchange PVTV NN interaction with the $\pi-$, $\rho-$, and $\omega-$meson exchanges was applied. In that paper, a benchmark calculation for $^3$He was reported, as well as EDM results for the more complex nuclei $^{6,7}$Li, $^9$Be, $^{10,11}$B, $^{13}$C, $^{14,15}$N, and $^{19}$F. The results suggest that different nuclei can be used to probe different terms of the PVTV interaction; see Fig.~\ref{fig:Dpolsum}. EDMs of light nuclei stripped of electrons can be in principle measured at storage rings~\cite{abusaif2019storage}.

NCSM calculations of the anapole moments and EDMs of light nuclei can be improved in several ways. First, higher order terms in the anapole moment operator, including two-body current contributions~\cite{PhysRevC.65.045502} and similar two-body PVTV operators, should be included~\cite{Liu2004}. Second, the very recently developed chiral PV and PVTV interactions~\cite{deVries2020,Gnech2020,deVries21} can be used in place of the one-meson-exchange model forces. Given the satisfactory NCSM basis-size convergence, one can be optimistic that uncertainties in nuclear calculations for light nuclei can be reduced to $\sim$10\% once the above improvements are implemented. The NCSM results for light nuclei can be in turn used to benchmark the methods we discuss next, the in-medium similarity renormalization group (IMSRG) and coupled-cluster approach, and so increase our confidence in the results of those methods in medium mass and heavy nuclei.

The above should make clear that the NCSM (and Quantum Monte Carlo) can produce good results
in light nuclei. They are prevented by exponential scaling, however, from being 
useful in the heavy isotopes used in atomic and molecular EDM experiments.  
In those isotopes, as we've already asserted, one can make use
of two other approaches: coupled-cluster theory and the IMSRG.  

Coupled cluster theory is based on an ansatz for the nuclear ground
state in which one- and two-body operators excite particles from occupied to
empty orbitals in an iterative way. Like all \textit{ab initio} approaches, it 
can be made exact by allowing three-body, four-body, etc., operators to excite particles as well.
The method has been applied primarily to the computation of nuclear
spectra and transitions among low-lying states, but also has been used to compute, e.g., cross sections for photo-excitation 
\cite{Hagen14,Bonaiti21}.  And recently, the  method was applied to the
matrix element for the neutrinoless double-beta decay of $^{48}$Ca \cite{Novario21}, a problem that resembles those addressed in this paper.

The IMSRG focuses on including complicated physics into 
effective Hamiltonians and other operators rather than into wave 
functions.  The procedure is realized through flow equations, which gradually transform the Hamiltonian so that it gives correct results when the ground state is replaced by
a simple ``reference state,'' e.g.\ a Slater determinant. 
The flow
equations accomplish this by driving the part of the Hamiltonian that couples this reference state with others to zero, so that the reference state
becomes the ground state of the transformed Hamiltonian.  The method is in some ways more flexible than coupled cluster because one is free to choose the reference state; it doesn't have to be a Slater determinant. And like coupled cluster theory, it has been applied to many observables --- energies, transitions, nuclear radii, etc. --- in medium mass nuclei \cite{Hergert16,Hergert17,Stroberg19}.  One version of this approach, the In-Medium Generator Coordinate Method (IM-GCM), constructs the reference state from 
a deformed mean-field state or a mixture of such states, projected onto good
angular momentum.  The sophisticated reference state makes the method particularly able to describe collective properties such as E2 transitions \cite{Yao20}. 

The complementary valence-space approach (VS-IMSRG) instead generates an approximate unitary transformation to decouple both the core properties as well as an effective valence-space Hamiltonian from the full $A$-body Hamiltonian. This extends the reach of \textit{ab initio} calculations to that of the standard phenomenological shell model~\cite{Stro17ENO}, essentially all nuclei at least to the $A=100$ region~\cite{Stro21Drip} (at which point diagonalization can become intractable), while generating effective valence-space operators consistent with the Hamiltonian. The physics of parity violation particularly requires valence spaces which include both positive and negative parity orbitals, but a new multi-shell approach has recently been developed to handle such valence-space Hamiltonians~\cite{Miya20lMS}. 

The accuracy of both the IMSRG and coupled cluster theory 
is similar, and the main challenge for both approaches will come from scaling 
efficiently to nuclei in the regions of experimental interest, approximately 200 or so nucleons. The primary factor preventing converged calculations above $A \approx 100$ had been computational limitations on the number of 3N force matrix elements that could be included. However, a recent breakthrough in storing such matrix elements, applicable within methods using spherical references, has now pushed converged calculations to the $^{132}$Sn region~\cite{Miya22Heavy}, where key systems relevant for parity violation searches can now be accessed. More recently this advance has led to converged calculations of even the heaviest doubly magic nucleus $^{208}$Pb~\cite{Hu2021}.

With both IMSRG variants, particular care must be 
taken with effective operators, e.g., the Schiff operator. But, like coupled cluster theory, the IMSRG has already been applied to neutrinoless double-beta 
decay \cite{Yao20,Belley21} as well as spin-dependent dark matter scattering~\cite{Hu22SDDM} up to the region of interest for many parity-violation experiments, where similar issues had to be addressed. Preliminary benchmark calculations of EDMs are indeed ongoing for light nuclei. For example, beginning with the same calculation set up with the NCSM, the VS-IMSRG yields $D^{\rm (pol)} / G^{0}_{\pi}$ about 50\% smaller than the NCSM result for $^{15}$N. The disagreement implies that improving the employed many-body approximation and/or further extension of the valence space is required. The development of the higher-order IMSRG~\cite{Heinz21IMSRG3} and extension to a larger valence space are in progress. Combining with the sophisticated valence space technique~\cite{Shimizu20QVSM} rather than the brute force diagonalization, meaningful predictions of the relevant physics here for heavier experimental candidates are expected to be feasible.

In the field of fundamental physics and radioactive molecules, what can be accomplished right now, and what in the near future?  In light nuclei --- those with fewer than about 20 nucleons --- the NCSM and Quantum Monte Carlo methods can be used to compute almost any static observable.  The same is true, with enough effort, of the IMSRG in any nuclei that are not too deformed.  Thus parity-violating transition rates and anapole moments in nuclei up to the Cs isotopes, for example, should be addressable now.  In the light actinides, which exhibit both strong quadrupole and octupole deformation, DFT is still the best method.  The IM-GCM will be ready to address these nuclei once the necessary codes have been adapted to exploit our best computers.  We discuss their application to the Schiff moments of both  
$^{199}$Hg and $^{225}$Ra in Sections \ref{sec:Hg199} and \ref{sec:Ra}.

\subsubsection{Hadronic theory for CP-violation}
\label{sec:hadronic_CPV}

As EDM, NSM, and MQM experiments are low-energy measurements, different mechanisms of CP violation can be captured through effective field theory (EFT) methods. The Standard Model contains a dimension-four CP-violating operator, the \thetaQCD term, while higher-dimensional operators such as quark EDMs, chromo-EDMs, and CP-odd four-quark operators can arise from beyond-the-Standard-Model physics. A full EFT study of these higher-dimensional operators and their evolution to low-energy scales has been performed in the literature \cite{Dekens:2013zca,Kley:2021yhn}. 

Next, the CP-violating operators must be matched to a new theory that describes the CP-odd interactions among hadronic degrees of freedom, such as pions and nucleons. This matching has been performed using chiral EFT \cite{Hammer:2019poc} that, for each source of CP violation, predicts the form and relative sizes of different hadronic CP-odd interactions \cite{Mereghetti:2010tp,deVries:2012ab}. Important terms in the Lagrangian are CP-odd pion-nucleon and nucleon-nucleon interactions, but also nucleon-photon or pion-nucleon-photon couplings. The chiral interactions can then be used to compute a CP-violating nucleon-nucleon potential and CP-violating nuclear currents ~\cite{deVries2020} order-by-order in the chiral power counting.

Two big outstanding challenges remain. First, each interaction in the chiral Lagrangian comes with a coupling constant, often called a LEC (low-energy constant), that is not predicted from chiral symmetry arguments. For each source of CP violation, e.g. the \thetaQCD term or a quark chromo-EDM, the LECs must be computed with non-perturbative techniques. Lattice QCD has been targeting the LECs; see Ref.~\cite{Shindler:2021bcx} for a review, but many are still unknown. A second problem is the nuclear many-body problem. It has proven complicated to reliably compute nuclear Schiff Moments and magnetic quadrupole moments of heavy nuclei from the CP-odd nuclear potential and currents \cite{Engel2013,Yamanaka2017}. For light nuclei, \textit{ab initio} methods have been used where CP-even and -odd chiral nucleon-nucleon potentials are combined to compute nuclear wave functions and CP-odd moments simultaneously. The associated theoretical uncertainties are small, see e.g. \cite{Bsaisou:2014zwa,Froese:2021civ}. Unfortunately, it is not straightforward to extend these computations to heavier nuclei of experimental interest. 

\subsection{Radioactive atoms}
\label{sec:radioactive atoms}

There are many synergies between work with radioactive molecules and radioactive atoms. Certain radioisotopes are currently being explored for the unique advantages they provide compared to their stable counterparts. For example, $^{133}$Ba$^+$ was recently trapped and laser cooled for its potential for quantum information science \cite{Hucul2017}. Ba$^+$ metastable $D$ states, with a radiative lifetime of approximately one minute, allow for high-fidelity readout, and the long-wavelength transitions enable the use of photonic technologies developed for the visible and near-infrared spectrum. Due to its $I=1/2$ nuclear spin  $^{133}$Ba$^+$ gives the additional advantages of robust state preparation and readout of the
hyperfine qubit, as well as availability of  $m_F = 0$ hyperfine and optical “clock” state qubits, which are relatively insensitive to magnetic fields ($m_F$ is the projection quantum number of the total angular momentum $F$). There are no stable isotopes of Ba with $I=1/2$. 

The perceived radioactivity barrier has been a primary reason radioisotopes haven't been pursued, despite their potential and unique properties, but the opportunities are enough to surpass the obstacles. Several applications of radioactive atoms for fundamental physics are being pursued: parity violation and eEDM studies with Fr \cite{Fr2,Fr1}, EDM searches with \iso{Ra}{225} \cite{2015Ra}, development of a nuclear clock based on \iso{Th}{229} \cite{Peik2021}, the realization of an atomic clock based on Ra$^+$ \cite{Holliman2022}, the development of atomic clocks based on highly-charged Cf$^{15+}$ and Cf$^{17+}$ ions \cite{Cf1,Cf2}, a sterile-neutrino search with $^{131}$Cs \cite{Hunter}, and beta decay asymmetries using laser cooling and optical pumping~\cite{Fenker2018}. In addition, various radioisotopes may be used in fifth-force searches using precision isotope shift measurements to resolve standard-model uncertainties \cite{2020YbIS}. These applications also motivated further improvements in the atomic theory of heavy atoms and atoms with complicated electronic structure \cite{2021Th35,2021Th3,Ginges2015,Dzuba2017}.

\subsubsection{Parity violation and EDM searches.}

As discussed in section \ref{sec:pv}, all previous and ongoing atomic PV experiments rely on
the large enhancement of the observed effect in heavy nuclei. Francium, the heaviest alkali atom ($Z=87$), 
possesses a unique combination of electronic structure  simplicity  and a great
sensitivity to effects such as atomic PV and  permanent electric dipole moments  due to its high
nuclear charge. With up to $10^6$ trapped atoms, the Fr experiment at the Francium
Trapping Facility at TRIUMF  \cite{2016Frtrap} has already reached  the sample size required for a future PV campaign  envisioned to start in the
near future \cite{Fr2} with the M1 strength in the parity-violating transition recently measured \cite{Hucko2022}.

Among the alkali atoms, \iso{Fr}{210} offers the largest enhancement factor to the
electron EDM (eEDM). A  method to measure the eEDM using ultracold entangled Fr atoms trapped in an optical lattice, yielding an uncertainty below the standard quantum limit was proposed in \cite{Fr1}. Estimated statistical and systematic
errors of the proposed measurement scheme, which is based on quantum sensing techniques,  show potential for an eEDM search  at a level below 10$^{-30}$~e~cm. 
A recent theoretical study pointed out advantageous
features for analyzing both the eEDM and scalar-pseudoscalar electron-nucleus interaction contributions from its measurement \cite{2021FrEDMtheory}. 
The Fr EDM experiment can be also used to constrain the P,T-violating effect induced by the exchange of axionlike particles between electrons or between an electron and the Fr nucleus~\cite{maison2022static}.
An eEDM experiment with an atomic fountain at zero magnetic field~\cite{Wundt2012} is being pursued in cesium with a plan to extend it to francium to utilize its much larger tensor Stark shift~\cite{Feinberg2015}.

\subsubsection{Progress towards the search for the atomic EDM of Radium.}
The radioactive \iso{Ra}{225} atom is a favorable case to search for a permanent electric dipole moment because of its strong nuclear octupole deformation and large atomic mass (see section \ref{sec:bsm}). \iso{Ra}{225} is also attractive from an experimental perspective, as the 14.9 day half-life allows this isotope to be obtained as a
radioactive source in sufficient quantities for experiments to
run off-line, away from an accelerator.
The first measurement of its atomic
electric dipole moment, reaching an upper limit of $|d(^{225}\rm{Ra})| < 5.0 \times 10^{-22}$~e~cm (95\% confidence), was reported in \cite{2015Ra}, demonstrating a  cold-atom technique to study the spin precession of \iso{Ra}{225} atoms held in an optical dipole
trap. This was followed up by another measurement with a factor of 36 more sensitivity, and a detailed analysis of systematics indicates that this approach is limited by statistics for \iso{Ra}{225} for the foreseeable future \cite{Bishof2016}. 
Several upgrades to increase the atom number \cite{Booth:2020}, the electric field \cite{ready2021}, and the spin precession readout efficiency \cite{rabgathesis} are underway and, once implemented, will result in orders of magnitude improvement to the statistical sensitivity.
In the long term, isotope harvesting (see section \ref{sec:facilities}) from FRIB \cite{readythesis} will increase the atom number even more, increase the integration time beyond two weeks, and allow for detailed studies of systematic effects.

\subsubsection{Development of optical atomic and nuclear clocks.}

The extraordinary improvement of optical atomic clock precision in the past fifteen years~\cite{2015RMPclocks,Bothwell2022}, now reaching $<10^{-20}$ fractional precision, has enabled testing the constancy of the fundamental constants and local position invariance \cite{2021Ybalpha}, dark matter searches \cite{2020SrcavityDM}, tests of the Lorentz invariance \cite{2019YbLLI}, and tests of general relativity \cite{2020Skytree}. Future clock development will allow for many order of magnitude improvements of these experiments. Deployment of high-precision clocks in space is proposed for many applications, including tests of gravity \cite{2022FOCOS}, search for dark-matter halo bound to the Sun \cite{2020SpaceQ}, and gravitational waves detection in wavelength ranges inaccessible on Earth \cite{Vutha2015,2016clockGW,2022clockGW}.

These advances motivated the development of novel clocks with high sensitivity to the variation of fundamental constants and, therefore, dark matter searches. 
If  the fundamental constants
are space-time dependent, so are atomic and nuclear spectra and the clock frequencies. The variation of the fundamental constants would change the clock tick rate  and make it dependent on the type of the clock since the frequencies of different clocks depend differently on fundamental constants.
Ultralight dark matter can source the oscillatory and transient variation of fundamental constants that can be detected by comparing frequencies of two different clocks or a clock and a cavity \cite{Safronova2018,2020SrcavityDM}. 

An optical clock was recently realized with Ra$^+$ because of its appeal for setting limits on the time variation of fundamental constants, and for the integrated photonic compatible wavelengths for making a transportable optical clock \cite{Holliman2022}. It has the largest positive enhancement to the time variation of the fine structure constant, $K=2.8$, of any demonstrated clock. 

Highly charged ions were demonstrated to have transitions suitable for high-precision clock development  with
large sensitivities   to the variation of the fine-structure constant, $\alpha$, as a consequence of strong relativistic effects and high ionization energies \cite{2018HCIRMP}. A possibility to develop optical clocks using the transitions between the ground and a low-lying excited state
of the highly-charged Cf$^{15+}$ and Cf$^{17+}$ ions was explored in detail in \cite{Cf1}. The dimensionless sensitivity factor $|\Delta K|$ to variation of $\alpha$ for the Cf$^{17+}$ and Cf$^{15+}$  clock pair was predicted to be 107 (see \cite{2018HCIRMP}).
Three out of eight Cf isotopes have a long half-life: $A=249, I=9/2$ (351 y),
$A=250, I=0$ (13.1 y), and $A=251, I=1/2$ (898 y). A  project to develop the optical clocks based on these Cf ions is presently underway \cite{Cf2}.

Nuclear transition frequencies are far outside of the laser-accessible range with the single exception of \iso{Th}{229}, $\tau=7,800$ y.
The current most precise value of the energy difference between its the ground and first excited state is $8.19(12)$ eV ~\cite{Peik2021};  an average of two recent measurements \cite{Sei19,Sik20}. 
 Such a unique feature of this isotope opens up a number of  opportunities, including a design of a super-precise nuclear clock with
very high sensitivity to the variation of the fundamental constants, including the fine structure
constant $\alpha$, the strong interaction, and quark masses~\cite{2020ThalphaFla}. Development of the nuclear clock is presently underway \cite{Peik2021}.

\subsubsection{Fifth-force searches with precision isotope-shift measurements.}
Precision isotope-shift (IS) atomic spectroscopy enables searches for a hypothetical fifth force between the neutrons of the nucleus and the atomic electrons. The method explores the non-linearity of King plots, where the mass-scaled frequency shifts of two optical transitions are plotted against each other for a series of isotopes~\cite{1963King}. Leading Standard Model contributions to the IS, mass and field shifts, lead to a linear relation between two electronic transitions (with respect to different IS measurements). New spin-independent interactions  will cause the non-linearities in the King plots \cite{2016ISproposal1,2018ISproposal}. However, higher-order SM contributions break the linearity of the King plots as well~\cite{2018ISproposal} and must either be calculated with high accuracy~\cite{2020ISNL}, or eliminated with a generalized analysis that uses more transitions and isotopes~\cite{2017GenKPlot,2020GenKP}.
The initial analysis already requires four even-parity isotopes and radioisotopes would be needed for the generalized analysis to remove SM nonlinearities~\cite{2020YbIS,Hur2022}.  Such IS experiments provide a complement to bounds from Big Bang Nucleosynthesis and $K$ meson decays \cite{Knapen2017}.

\subsubsection{The HUNTER experiment: the search for the sterile neutrino.}
While in all the cases above the radioactivity is an unfortunate side effect of using a heavy atom, it is a necessary ingredient in the HUNTER experiment with ultracold  $^{131}$Cs  \cite{Hunter}.
The Heavy Unseen Neutrinos from Total Energy–momentum Reconstruction (HUNTER) is a high-precision laboratory-scale experiment able to search
for very weakly coupled sterile neutrinos that might form
all or part of the galactic dark matter. 
This experiment will fully kinematically reconstruct K-capture events in a population of $^{131}$Cs atoms suspended in
vacuum by a magneto-optical trap. 
The existence of a sterile neutrino of keV/c$^2$ mass that
mixes with the electron-type neutrino produced in the decay would manifest itself in a separated
population of events with non-zero reconstructed missing mass (up to the available
energy of the decay, $Q = 352$~keV).

\section{Opportunities at facilities}
\label{sec:facilities}

The half-lives of the isotopes of interest for research on radioactive molecules may be as short as a few days to weeks, or as in the case of the francium isotope \iso{Fr}{223} for example even on the time scale of minutes. Thus, their radioactive nature requires the artificial production of these radionculides. Over the decades, a world-wide network of dedicated facilities has emerged which provide access to short- and long-lived radioactive atoms for fundamental research and applications in many different fields. For this purpose, the nuclear-reactor based production of rare isotopes is complemented by particle accelerators. On the forefront of research with short-lived, ``exotic'' isotopes there are accelerator-driven radioactive ion beam (RIB) facilities which are each distinct in their respective production mechanisms as well as methods of beam formation and delivery \cite{Blumenfeld_2013}. Broadly speaking, they can be separated into two groups: facilities with in-flight separation and production sites based on Isotope Separation OnLine (ISOL), see Fig. \ref{fig:rib}. In the former, a heavy ion beam is accelerated to typically a few hundred MeV per nucleon and impinges on a thin target to exploit production mechanisms such as fragmentation of the projectile nuclei, fission or fusion-evaporation reactions. The (radioactive)  products carry most of the projectiles' momentum and thus maintain a high-energy ion beam which subsequently passes an in-flight separator to identify and select the ions of interest. While fast-beam experiments directly benefit from these beams, experimental studies at low energies require the beam to be slowed in a degrader followed by an ion catcher or gas cell which dissipates the remaining ion energy and forms the ion beam, transportable via electrostatic ion optics to experimental stations. Major contemporary facilities based on in-flight separation are the Facility for Rare Isotope Beams (FRIB) in the US, the Radioactive Isotope Beam Factory (RIBF) in Japan, GANIL in France or the Facility for Antiproton and Ion Research (FAIR) in Germany.

The Isotope Separation On-Line (ISOL) method of isotope production has been used to create radioactive species since 1951 \cite{Kofoed-Hansen1951}. By coupling an accelerator directly to a mass separator facility, the ISOL method uses an accelerated (light) driver beam to impact a (heavy) target nucleus, generating reaction products through reactions such as fission, fragmentation and spallation. Once produced, the radioactive isotope of interest must diffuse through the target matrix, desorb from surfaces and effuse through pores within the target material. This process, known as ``release'' is often facilitated by heating the target to temperatures above 2000~$^{\circ}$C. The radioactive species must then reach the ion source, where it is ionized using a range of techniques including surface ionization, electron bombardment, and resonant laser ionization. The ionized species is extracted by a voltage difference and can then be post-accelerated, cooled and bunched, or sent to experiments as a charged particle beam. Many target materials \cite{Gottberg2016} and ion sources \cite{Kirchner1996} have been developed, tested and used at ISOL facilities such as ISOLDE \cite{Catherall2017} and TRIUMF \cite{Dilling2014}. The ISOL method continues to grow and develop, along with the rich community of experiments established at various ISOL facilities to take advantage of the available radioactive beams, including those soon available at new ISOL facilities such as the one within SPIRAL2 at GANIL. 

Starting from the low-energy branches of both ISOL and in-flight facilities, the radioactive beams at several online laboratories can be re-accelerated to up to $\sim10$ MeV/nucleon to study and exploit various nuclear reactions. In the context of radioactive molecules, experiments utilising (multistep) Coulomb-excitation at these ion-beam energies provide a central piece of information to identify octupole deformed nuclei which enhance the experimental sensitivity for nuclear Schiff moments, especially when incorporated into polar molecules. 

Low-energy experiments with radioactive molecules themselves will require a translation of experimental methods, well established for molecules assembled from stable nuclides, into the realm of rare-isotope science. In order to master the resulting challenges such as short half-lives, limitations in available sample size or ion-beam temperature, these developments will build upon a wealth of experience in existing RIB programs which are already taking advantage of the high precision and accuracy found in atomic physics techniques. Among others, these include ion and atom-trapping or laser-spectroscopy applications \cite{Blaum_2013,CAMPBELL2016127,annurev-nucl-102711-094939}, which have been advanced over the years to successfully perform high-precision experiments of radionuclides with half-lives of less than 10~ms \cite{PhysRevLett.101.202501} or with production yields lower than one ion per hour \cite{science.1225636}. Moreover, a suite of devices building on buffer-gas \cite{HERFURTH2001254}, laser and sympathetic cooling \cite{PhysRevResearch.4.033229} has been developed to efficiently transform the initially fast and/or hot ion beams into high-quality, cold ion ensembles. The combination of these highly sensitive and fast RIB techniques with state-of-the-art methods in molecular physics will thus usher in a new precision era at RIB facilities.

The formation of the novel field of radioactive molecules coincides favourably with the world-wide emergence of next-generation RIB facilities which have or will soon come online. These will largely enhance the variety as well as availability of radioactive samples. Multi-beam facilities will increase direct online access while isotope harvesting and generator sources are promising options for continuous disposability of radioactive samples, although at an increased complexity in chemistry and radiation protection.  These ongoing facility developments are moreover expected to boost experimental infrastructure beneficial for the science program with radioactive molecules, e.g. in terms of molecular formation processes or laboratory space available for experimenters at the RIB facility. In the following, we will describe present and upcoming opportunities found at those RIB facilities which have or plan to engage in the research on radioactive molecules.

\subsection{FRIB} 

\begin{figure}
    \centering
    \includegraphics[width=0.6\textwidth]{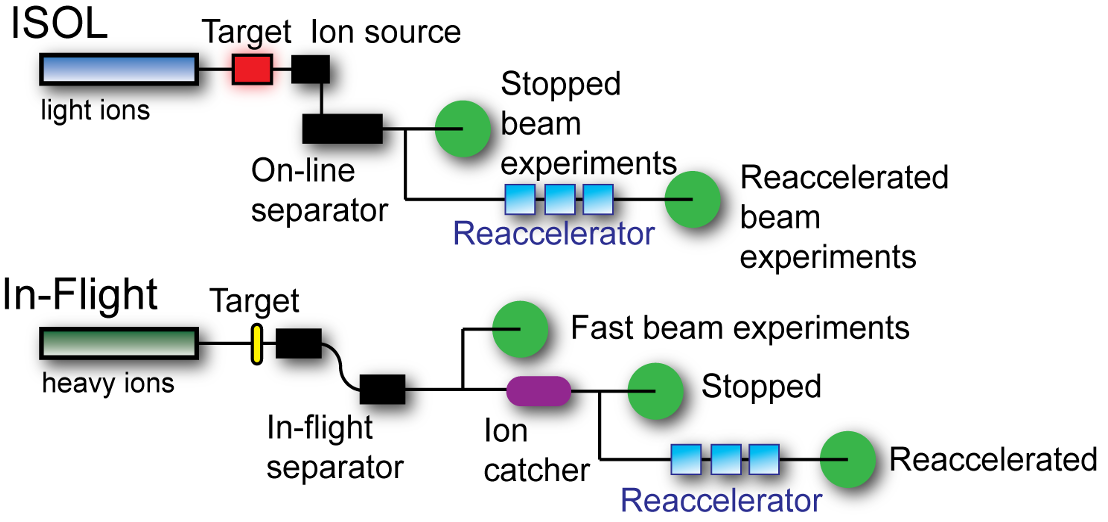}
    \caption{Rare Isotope Beams: ISOL and In-Flight Separation. The ISOL method is used at TRIUMF, ISOLDE, and IGISOL and is discussed in those sections. FRIB uses the In-Flight Separation method and is discussed more in the FRIB section.}
    \label{fig:rib}
\end{figure}

The Facility for Rare Isotope Beams (FRIB) at Michigan State University became operational in June 2022. FRIB will provide unprecedented access to radioactive isotopes across the chart of nuclides into the actinide region. FRIB will enable the study of perhaps up to 80\% of isotopes from hydrogen to uranium \cite{erler2012}. FRIB uses the ``in-flight'' separation method \cite{ms1998}. A heavy primary beam, such as \iso{U}{238}, is directed at carbon (graphite) target. The resulting products, formed by a variety of mechanisms, result in a secondary beam that is separated, filtered, and purified in-flight \cite{aris2012} and then directed towards the appropriate experimental area, see Fig.~(\ref{fig:rib}). These include a fast beam area with secondary-beam energies near the primary-beam energy, a stopped beam area, and a reaccelerated beam area. The latter two are fed by an ion catcher which first stops the secondary beam and then reaccelerates the stopped beam to the appropriate energy. This in-flight technique allows relatively fast access to nearly any isotope of interest.

The heart of FRIB is a three-segment superconducting radiofrequency linear accelerator (LINAC) composed of 46 cryomodules. The LINAC is capable of delivering 400 kW for all heavy ion beams from oxygen to uranium. The third segment of the LINAC has space available for an additional 11 cryomodules for a potential future upgrade that would double the power of the LINAC to 800 kW if implemented. It is expected that the primary beam will be uranium for roughly half the running time, which is crucial for the production of EDM candidate isotopes in the actinide region. The primary beam can then be accelerated up to 200 MeV per nucleon towards a rotating graphite target. This creates secondary beams by projectile fragmentation, nucleon transfer, fission, and Coulomb excitation.  The unused portion of the primary-beam products is directed to a water beam dump where they can be extracted by a process called ``isotope harvesting.'' The purified secondary beam is then directed towards one of three experimental areas.

\begin{figure}
    \centering
    \includegraphics[width=0.85\textwidth]{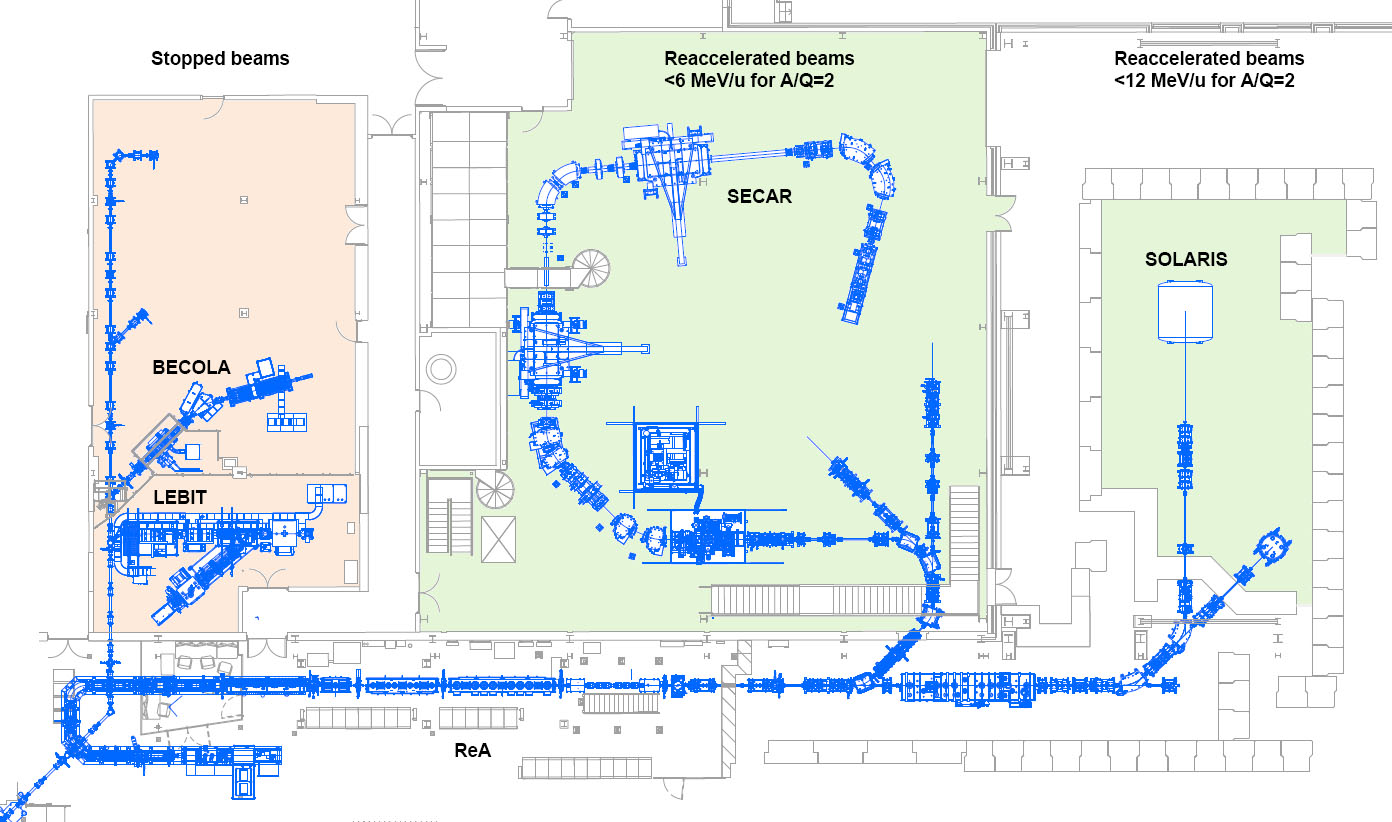}
    \caption{FRIB: Stopped and Reaccelerated Beam Areas.}
    \label{fig:frib1}
\end{figure}

Studies of octupole collectivity in the actinide region will be enabled by the fast and reaccelerated beam areas. The stopped beam area, see Fig.~(\ref{fig:frib1}) will allow for spectroscopy of radioactive atoms and molecules using colinear laser spectroscopy at BECOLA \cite{becola} (BEam COoling and LAser spectroscopy area) and colinear resonant ionization spectroscopy at RISE (Resonant Ionization Spectroscopy Experiments area). A variety of offline ion sources can feed into the stopped beam area which makes laser spectroscopy of radioactive species possible without the need for a dedicated primary beam. This is enabled in great part due to the isotope harvesting program at FRIB \cite{abel2019}.

The infrastructure needed for isotope harvesting at FRIB will be in place by the end of 2023. A shielded cell will be used for the bulk extraction of gaseous and dissolved phase isotopes. They will then be transported to the isotope processing area (IPA), which will contain additional shielded cells, radiochemistry fume hoods, sealed work surfaces, ventilation suitable for open radiochemical work, security, analytical equipment, and short term waste storage. The IPA will allow for the variety of radiochemical work needed to chemically separate the isotopes of interest and make them accessible for offline experiments and other applications. The Chemistry Department at Michigan State University also houses a variety of radiochemistry labs to aid in the conversion of the harvested isotopes into a form usable for offline experiments.

Over the course of about six years starting in 2022, the beam power at FRIB will be increased steadily in phases so that experience with beam tuning can be gained. During this time, the isotope harvesting program will be developed so that it will be robust by the time FRIB reaches its planned operating power of 400 kW around 2028. If the FRIB400 moves forward, then this will increase the yield of secondary beam intensities across the chart of nuclides from a factor of 2 to 100 depending on the isotope. For EDM candidate isotopes in the actinide region, this could triple the harvesting yield \cite{FRIB400wp}. There is also the possibility of dedicated campaigns using a \iso{Th}{232} primary beam which would increase the yield in the actinide region by roughly an order of magnitude compared to the \iso{U}{238} beam. The K500 superconducting cyclotron is expected to still be available for radiation testing of electronic components mimicking the effects of cosmic rays and space radiation. This cyclotron could also be used for dedicated campaigns to produce EDM candidate isotopes in the actinide region using a proton or deuteron beam on a \iso{Th}{232} target. Finally, there are discussions underway to provide experimental space within the facility to host AMO-based EDM search experiments.

\subsection{GSI / FAIR} 
The GSI Helmholtz Centre for Heavy Ion Research (Germany) founded in $1969$ is operating a large accelerator complex consisting of the linear accelerator UNILAC, the heavy-ion synchrotron SIS-18 and the experimental storage and cooler ring ESR. The accelerator facility allows for parallel operation of several experiments, for instance at the UNILAC, at SIS and at the ESR. With the UNILAC, ions of all elements from protons through uranium can be accelerated up to 11~MeV/u, at SIS-18 up to 2~GeV/u, and in the ESR stable or radioactive ion beams can be stored and cooled up to 560~MeV/u (for uranium). 

The Facility for Antiproton and Ion Research FAIR is presently under construction. A part of it, namely the low-energy storage ring CRYRING, is already operational for experiments. It is connected to the ESR beam line and offers cooled primary and secondary beams in the energy range from 4 MeV/u down to 10 keV/u for experiments in atomic and nuclear physics with highly-charged ions. It is also equipped with internal ion sources for stand-alone experiments with stable beams. In the meantime, the existing GSI accelerator facilities are being upgraded towards higher beam intensities, towards even further enhanced parallel operation, and towards design beam operation of FAIR. A beamline is presently being built, which will connect the existing SIS-18 (which will serve as an injector ring) with the new heavy-ion synchrotron SIS-100 and the Super-FRS of FAIR, both of which are under construction. The Super-FRS will be available for first experiments in 2025.

At GSI / FAIR, depending on the primary beam and energy, various types of nuclear reactions, such as fusion, deep inelastic transfer reactions, in-flight fragmentation and fission are used to produce exotic nuclei. The resulting nuclei are separated from the intense primary beams using large-scale in-flight separators, such as for example the velocity Separator for Heavy Ion reaction Products (SHIP) \cite{MUNZENBERG197965}, the FRagment Separator (FRS) \cite{GEISSEL1992286} or the gas-filled TransActinide Separator and Chemistry Apparatus (TASCA) \cite{SEMCHENKOV20084153}. The In-flight production allows for almost pure radioactive ion beams of all elements over the entire nuclear chart to address a variety of scientific questions in nuclear and particle physics as well as in medical, solid state and applied sciences. 

The production of low-energy radioactive ion beams relies on so-called stopping cells \cite{wada2003slow,savard2003development}. In these devices the incoming fast ion beams are fully slowed down and stopped through collisions with helium buffer gas. Once thermalized, they can be extracted from the gas volume using a combination of static and radiofrequency fields into a differentially pumped low-energy beam line, typically relying on the usage of radiofrequency quadrupoles (RFQ) ion traps. Once stopped they can be delivered to high-precision experiments requiring low-energy, high-brilliance ion beams such as for example the future MATS and Laspec experiments \cite{rodriguez2010mats} to be housed at the low-energy branch of the Super-FRS. 
Recently most stopping cells have been built to operate at cryogenic temperatures reaching an improved helium gas purity, which in return enabled the extraction of pure ion beams. At GSI / FAIR several cryogenic stopping cells (CSC) are in operation, such as for example after the FRS \cite{Ranjan_2011,Purushothaman_2013} as part of the FRS Ion Catcher experiment \cite{PLA2013457} and behind SHIP \cite{ELISEEV20084475,DROESE2014126} as part of the SHIPTRAP experiment \cite{Dilling2000}.

\subsubsection{Current and future opportunities for production of radioactive molecules at the FRS Ion Catcher }

\begin{figure}
    \centering
    \includegraphics[width=0.85\textwidth]{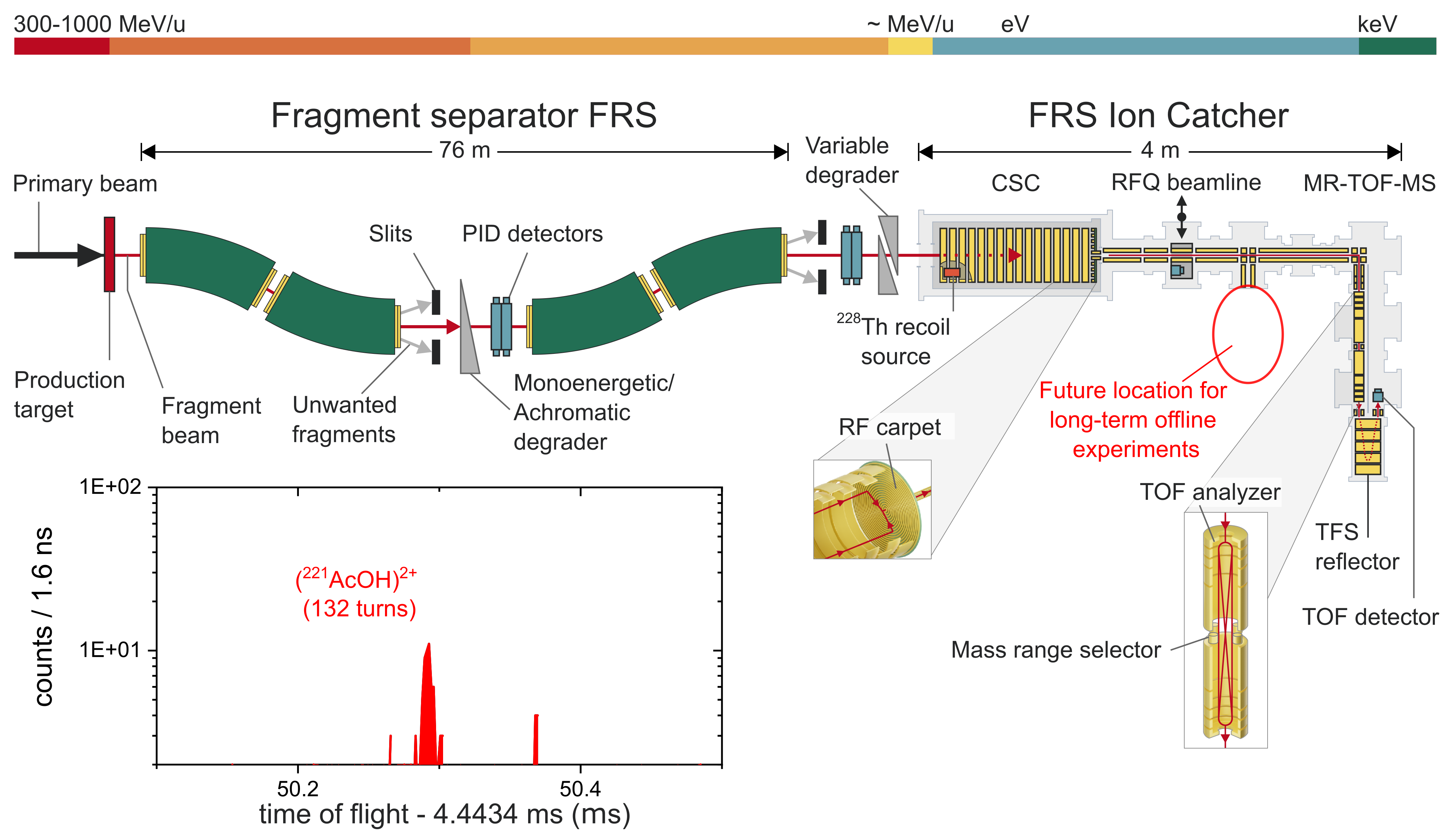}
    \caption{Schematic layout of the FRS Ion Catcher at GSI. Radioactive isotopes are produced In-flight, separated and identified via the FRS at relativistic energies, before being stopped in a cryogenic stopping cell (CSC), transported through a low energy RFQ beamline to a Multiple-Reflection Time-of-Flight Mass Spectrometer (MR-TOF-MS). Trace amounts of reactant species may be used to foster the formation of molecules, such as e.g. $^{221}$AcOH$^{2+}$ shown in the inset. In addition the CSC has been equipped with long-lived recoil sources, such as \iso{Gd}{148}, \iso{Rn}{223} or \iso{Th}{228}, which allow offline experiments of the alpha decay daughter species. A dedicated location is under development which will provide access to radioactive ions and molecules for extended offline experiments.}
    \label{fig:FRSIC}
\end{figure}

The FRS Ion Catcher, as shown in Fig.~\ref{fig:FRSIC}, has specialized in high-precision experiments of stopped fragmentation and fission fragments and serves as a test bench for the future low-energy branch of the Super-FRS at FAIR. It consists of four main parts, (i) the FRS including its particle identification and degrader system, (ii) the cryogenic gas-filled stopping cell (CSC), (iii) an RF-quadrupole-based beam line and (iv) a multiple-reflection time-of-flight mass spectrometer (MR-TOF-MS). Radioactive isotopes are produced in-flight, separated and identified via the FRS at relativistic energies, before being stopped in the CSC. Once thermalized they are extracted from the cell and transported via the low-energy RFQ beamline to the MR-TOF-MS for identification and high-precision mass measurements. Trace amounts of a reactant gas, such as H$_2$O, N$_2$ or CH$_4$ within the ultrapure helium gas of the CSC or the RFQ beam line may foster the formation of radioactive molecules \cite{218d92bc3cbe4328ae05f2a50d4142d9}.

Under the unique and cold conditions many unusual molecules can form and, among others, radioactive molecules of singly charged XeOH and KrOH as well as singly and doubly charged UO, UOH and ThO have been extracted and identified at the FRS Ion Catcher. Recently, the extraction of AcOH, which has applications in parity and time reversal searches  \cite{Oleynichenko:2021AcOH}, was shown. An example mass spectrum of AcOH$^{2+}$ containing the short-lived \iso{Ac}{221} isotope is shown in Fig.~\ref{fig:FRSIC}. 

Further, by using long-lived radioactive sources secondary beams of daughter nuclides can be extracted from the CSC. So far decay daughters from \iso{Gd}{148}, \iso{Rn}{223} and \iso{Th}{228} sources have been extracted and provided to long term experiments. As such, for example a $135$~day long experiment searching for rare decay modes in \iso{Ra}{224} produced from a \iso{Th}{228} sources has been performed. In the future \iso{Cu}{248} and \iso{Np}{237} sources may be anticipated to provide offline beams of \iso{Pu}{244} and \iso{Pa}{233} ions. A dedicated offline location, see Fig.~\ref{fig:FRSIC}, for long term experiments is currently under development which will allow extended experiments harvesting decay daughters from recoil sources. In the future, the production scheme may be expanded by developing dedicated miniature stopping cells which will allow decentralized offline experiments with radioactive molecules at  university laboratories. 

At GSI / FAIR, the advanced high-intensity cryogenic stopping cell \cite{DICKEL2016216,REITER2016240} in combination with the Super-FRS at FAIR will allow the production of radioactive molecules with competitive intensities, which will particularly suit research with radioactive molecules containing reactive elements currently not possible via the ISOL methods.

\subsection{TRIUMF} 

TRIUMF is Canada's particle accelerator centre established in 1968 as the TRI-University Meson Factory. TRIUMF operates the 500 MeV cyclotron accelerating the H$^-$ ions since 1974. The high-intensity 500 MeV proton beam drives the rare ion beam Isotope Separator and ACcelerator (ISAC) facility~\cite{ISAC_Dilling_2014}. The ISAC started its operation in 1995. TRIUMF has recently embarked on the construction of ARIEL, the Advanced Rare Isotope Laboratory, with the goal to significantly expand the rare ion beam program for Nuclear Physics, Life Sciences, and Materials Science~\cite{Dilling2014}. The nuclear physics program will expand its main pillars: nuclear structure, astrophysics, and tests of fundamental symmetries. At the heart of ARIEL there is a 100 kW, 30 MeV electron accelerator (e-linac) for isotope production via photo-production and photo-fission as well as a second proton beam line from TRIUMF’s 500 MeV cyclotron for isotope production via proton-induced spallation, fragmentation, and fission. Also included in ARIEL are two production targets and related infrastructure, mass-separators and ion beam transport to ISAC, and an electron-beam ion source (EBIS) for charge breeding. ARIEL will establish a multi-user capability with up to three simultaneous rare ion beams with more and new isotopes for TRIUMF users. The project completion is planned in 2026 with phased implementation, interleaving science with construction.

The radioactive beams are delivered into the ISAC halls which host a suite of state-of-the-art experimental stations operating in three distinct ranges of ion-beam energies. Post-accelerated beams are available at either medium energies (up to 1.8 MeV/u) or, facilitated by the ISAC-II superconducting linear accelerator, at 5-11 MeV per nucleon. ISAC's low-energy section is dedicated to trap and laser spectroscopy experiments, $\beta$-nmr, radioactive nuclear decay spectroscopy, as well as beamline ports for temporary online access to other initiatives.  

Experimental programs based on atomic-physics techniques include TRIUMF's Ion Trap for Atomic and Nuclear science (TITAN) which is a multi-ion trap facility devoted to high-precision mass measurements, e.g.  \cite{PhysRevLett.101.202501,PhysRevLett.120.062503}, and in-trap decay spectroscopy \cite{PhysRevLett.113.082502,LEACH201591}. Presently, TITAN consists of a linear Paul trap for beam preparation \cite{BRUNNER201232}, a multiple-reflection time-of-flight (MR-ToF) device \cite{TITANmrtof} for mass separation or highly sensitive mass measurements, an electron ion beam trap (EBIT) \cite{LAPIERRE201054, PhysRevLett.107.272501} to charge breed short-lived atoms to high charge states, and a Penning trap for mass measurements at the highest attainable precision \cite{BRODEUR201220}. Moreover, ISAC hosts TRIUMF’s neutral atom trap for beta decay (TRINAT) \cite{Fenker2018} which has trapped radioactive isotopes in a magneto-optical trap (MOT) \cite{PhysRevLett.79.375} following earlier work at Stony Brook  \cite{PhysRevLett.72.3795} and Berkeley \cite{PhysRevLett.72.3791}.
TRINAT’s scientific program is focused on weak-interaction studies and searches for new physics~\cite{Behr_2014} in competition with worldwide atom and ion trap studies~\cite{PhysRevLett.128.202502}. The francium trapping facility \cite{2016Frtrap,PhysRevA.97.042507,Fr2} at TRIUMF is another MOT system which pursues a new, improved atomic parity non-conservation (APNC) experiment. Finally, laser spectroscopy work is performed either directly next to the production target with TRIUMF Resonant Ionization Laser Ion Source (TRILIS) \cite{TRILIS2020,TRILIS2013} or with fast beams in a collinear configuration \cite{VOSS201657}

\subsubsection{Ion Reaction Cell (IRC)}\label{Sec:IRC}

Next-generation experiments require efficient on-line production of exotic radioactive molecules from RIB at high purity and yield. In particular, the demand for new designer or delicate molecules will increase as experiments and quantum chemistry theory identify new species. Some rare molecules including RaF and other robust sidebands do form in ISOL targets using reactive gases~\cite{GarciaRuiz2020}. However, the extreme non-equilibrium conditions inside hot-cavity ion sources and extraction systems tend to fragment molecules and are very unfavorable to forming delicate or complex species by optimized chemical reactions. Difficulties are being addressed as targets improve; however, formidable challenges persist (high temperatures and radiation fields, poor vacuum, significant isobaric interferences, unfavourable energy and target geometry conditions, problems using chemical reactants, etc). New innovative on-line methods are needed to ensure steady progress in this field, to realize the creation of exotic molecules with high purity, yield and flexibility with comprehensive pathways of chemical reactions and reactants.

\begin{figure}
    \centering
    \includegraphics[width=0.95\textwidth]{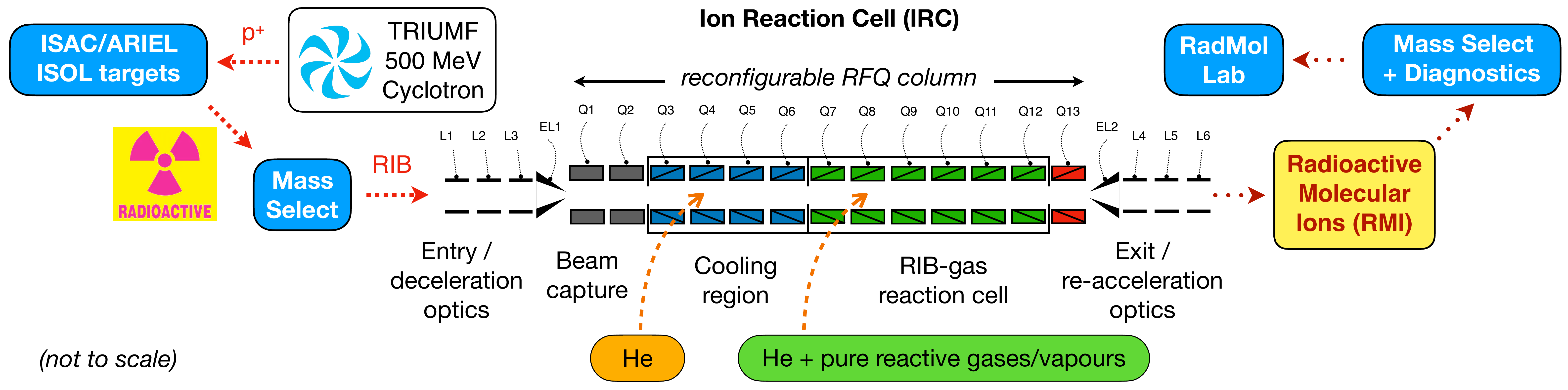}
    \caption{Simplified sketch of the Ion Reaction Cell (IRC) for the on-line creation of exotic radioactive molecules by RIB-gas chemistry at TRIUMF.}
    \label{fig:IRC}
\end{figure}

Figure \ref{fig:IRC} shows the Ion Reaction Cell (IRC). The IRC is a new radiofrequency quadrupole ion guide and gas-reaction cell at TRIUMF having three main functions to (i) accept positive RIB up to 1 $\mu$A at 40 keV (60 keV with future modifications) from the ISAC and ARIEL target stations, (ii) create rare/exotic molecules by controlled, specific gas-phase chemical reactions with RIB at room temperature, and (iii) deliver those newly formed molecules to TRIUMF's upcoming RadMol facility for experiments. The IRC also accepts negative ions that may be useful for future projects.

All RFQ's including the IRC, the ARQB at TRIUMF  and ISCOOL or ISOLTRAP at ISOLDE confine RIB along their longitudinal axis with an oscillating quadrupolar electric field. Inside the IRC, RIB performing quasi-Mathieu motion at low Mathieu $q$-values in-gas interact with helium and mixtures of helium and reactive gases or vapours forming molecular species. The IRC uses dedicated beam deceleration and re-acceleration sections, tunable DC gradient bars on each RFQ segment, and a two-stage differentially pumped re-configurable RFQ column, see Figure \ref{fig:IRC}. This flexibility means the gas-reaction cell can be tailored to selected reactions by changing the cell volume and length, gas pressure and number of collisions, use of multiple reaction gases and mixtures where the collisional cooling and energy distribution of ions along the RFQ column can be fine-tuned. Fine control over the energy transferred to the ions is possible, for example, when forcing endothermic reactions to proceed at $<$1 eV. Efficient optimization of molecule production may then occur at sub-eV energies by highly selective chemical reactions, where formation of other undesired molecules is minimized.

Molecule creation rather than destruction is achieved by providing favourable energy conditions for a suitable reactive gas with an electron affinity close to the RIB ion of interest~\cite{Charles2015}. Fine tuning and matching of reactant energy to the incoming RIB energy promotes chemical bond formation for a large diversity of molecules in high-yields and high-purities from many reagents that could not be used inside hot-cavity on-line targets. Reactive gases include fluorine, chlorine, nitrous oxides, acid vapours and inorganic vapours that are problematic in typical RFQ's at ISOL facilities. A proprietary Altem formulation specific to the IRC ensures chemical resiliency of the RFQ column and vacuum system, together with a semiconductor-grade gas-inlet system. A second inlet adapted from gas-chromatography combustion isotope ratio mass spectrometry (GC-C-IRMS) handles the introduction of complex organic vapours and mixtures of gases and vapours.

Analytical chemists have used RFQ's for years in many iterations for ion-gas reaction mass spectrometry. However, these systems have never been systematically applied to on-line RIB chemistry. As TRIUMF establishes its new RadMol laboratory, the IRC offers a novel and critical on-line technique for ISOL facilities for beam delivery and development of a wide range of rare molecules. Molecules of interest for IRC research include homo- and hetero-nuclear species, many unique di- and tri-atomic species, complex macromolecules, species involving organic compounds and other fragile or designer species. Availability of these exotic and rare radioactive molecules will drive research in BSM physics, nuclear structure, medical molecules and tracers, radiochemistry and astro/circumstellar chemistry.

\subsubsection{RadMol laboratory} 
\begin{figure}
    \centering
    \includegraphics[width=0.75\textwidth]{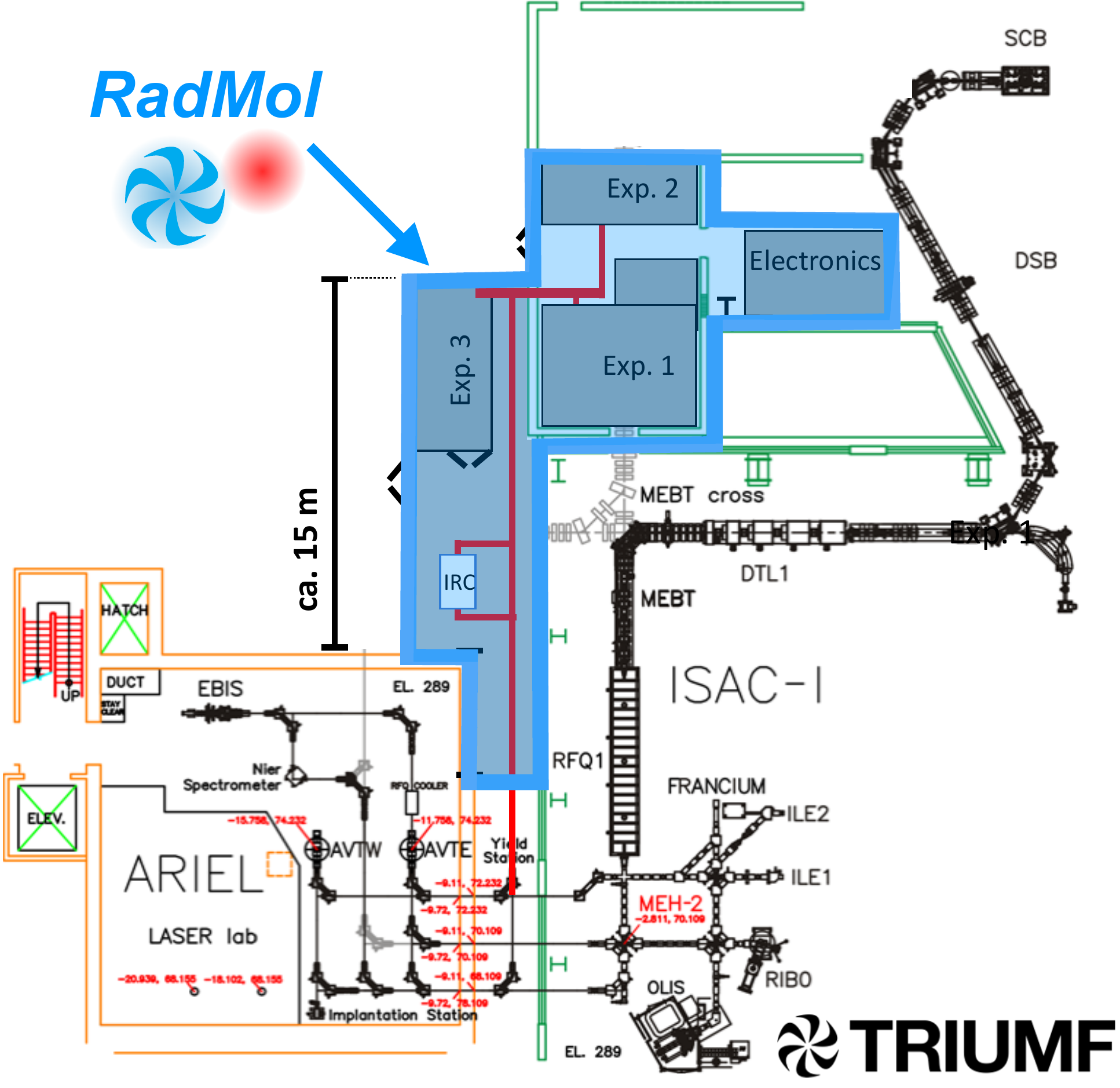}
    \caption{Planned laboratory for radioactive molecules at TRIUMF. A new low-energy beamline (red) delivers radioactive beam from ISAC and ARIEL to three experimental stations, to Exp. 1-3. For molecular formation, an ion reaction cell (IRC; Figure \ref{fig:IRC}) can be employed as part of the beam delivery and formation.  
    An additional, dedicated laser laboratory (not shown) is located in close proximity to Exp. 1 and 2.}
    \label{fig:RadMolLab}
\end{figure}

Building on TRIUMF's capability to produce a large variety of radioactive ion beams, the international RadMol collaboration seeks to establish a dedicated laboratory for radioactive molecules and fundamental physics at TRIUMF.
A schematic overview of the planned RadMol laboratory is shown in Fig.~\ref{fig:RadMolLab}. It will host three independent experimental stations which are directly coupled via a low-energy beamline to TRIUMF's RIB facilities ARIEL and ISAC. The laboratory will also provide space and infrastructure for the previously discussed ion reaction cell (IRC), see Sec.~\ref{Sec:IRC}, as one formation site for radioactive molecules. The facility plans include a dedicated laser laboratory which will be located one floor higher, right next to the experimental stations 1 and 2. 

In February 2022, the project has obtained TRIUMF-internal approval such that the detailed planning of the RadMol laboratory is now underway, including the formulation of its funding strategy. The plans involve a new building for the low-energy beamline, the IRC (Figure \ref{fig:IRC}), and a separate laboratory room for one experiment (Exp. 3 in Fig~\ref{fig:RadMolLab}). The  experimental stations 1 and 2 as well as the laser room are all integrated within already existing laboratory space at TRIUMF and can be readily used once its infrastructure is upgraded to the requirements of RadMol. 

The anticipated early availability of experimental stations 1 and 2 motivates an initial experimental program on longer-lived species without the need for direct online coupling. For this purpose, members of the RadMol collaboration envision the development of a stopping and implantation cell which can be coupled directly to ISAC or ARIEL for the collection of  radioactive samples. Afterwards, these samples can be relocated into the RadMol laboratory to release radioactive ions, either implanted ions or daughter products, for first experiments on radioactive molecules for precision measurement.

\subsection{ISOLDE}

ISOLDE is a radioactive beam facility at CERN \cite{Catherall2017}, where isotopes are produced by 1.4 GeV protons hitting a variety of targets to provide more than 1200 different isotopes and isomers of more than 75 elements \cite{isoldeyields, isoldeemis}. ISOLDE's two on-line target stations are coupled to the two mass separators (HRS and GPS) and provide radioactive ion beams of atomic or molecular species for experiments that address a broad range of scientific questions in nuclear physics, atomic physics, nuclear astrophysics, fundamental interaction physics, and hard and soft condensed matter research.
Isobaric separation of singly-charged atomic or molecular ions is achieved with a magnetic mass separator, from where the ions are sent at a typical energy of 30 to 50 keV into several experimental beam lines. Experiments stationed at ISOLDE use the radioactive ion beams to do very precise mass measurements, ISOLTRAP \cite{Wolf2013, Lunney2017, Mukherjee2008}, study decay properties, IDS \cite{IDSCollaboration2021}, measure hyperfine structure and isotope shifts, CRIS \cite{Cocolios2013CRIS}, and more \cite{HIE-ISOLDE,Butler2005,Borge2014}. 
ISOLDE features also two irradiation points located after the primary target at each target station in which standard ISOLDE targets as well as material samples can be irradiated. Targets irradiated here can be coupled to the CERN-MEDICIS facility for off-line extraction of medical isotopes \cite{MEDICIS} or coupled to the ISOLDE target stations to deliver long-lived isotopes in the so-called winter-physics campaigns, in which ISOLDE is operational, but no protons are available from CERN's accelerator complex. The recent spectroscopy of RaF was achieved by the CRIS experiment at ISOLDE \cite{GarciaRuiz2020} during such an extended winter-physics campaign.

Recent experiments on RaF mark a milestone as the first laser spectroscopy of a short-lived radioactive molecule \cite{GarciaRuiz2020}. In this experiment, radium fluoride was produced at ISOLDE using 1.4 GeV protons from the CERN PSB impinging on a uranium carbide target. CF$_4$ gas was injected into the back of the ion source via a calibrated leak to give a source of fluorine for molecular formation. The radioactive Ra isotope is generated on proton impact by the parent uranium target nucleus. To form the RaF molecule, it must move through the target material by the highly temperature-dependent process of diffusion to find a corresponding fluorine atom. The molecule is then easily ionized using surface ionization in a hot cavity ion source, after which it can be extracted as a beam. For long-lived isotopes, it is possible to irradiate the target without heating, thus limiting diffusion of the produced species and collecting activity of the isotope of interest within the thick target matrix. Such targets can then be reserved as sources of long-lived radioactive molecular beams and stored cold. As was done in the RaF study, the target unit containing the long-lived activity can then be put back online and heated to release the remaining isotopes, allowing operation without the need for proton beam time.

Molecular formation can occur along the path followed by the radioactive species, adding another step to the process of ion beam production. Molecules will have different volatility, vapor pressure and adsorption enthalpy than their constituent radioactive atoms, giving them different release and ionization properties. Additional factors to consider include potential dissociation and breakup behaviour of the molecule, enhancing yield by breaking up larger molecules, or decreasing yield by dissociating the molecule of interest. Previous studies have explored providing reactive gases to different target and ion source combinations via calibrated leaks or vapors from solids, observing the formation of molecules and molecular ions \cite{Eder1992}. Molecular beams have been studied for decades at ISOLDE \cite{Kirchner1997,Koster2000,Koster2007,Koster2008} as a way to enhance volatility of reactive species and provide high-purity beams of isotopes of interest such as carbon \cite{Franberg2008} and boron \cite{Ballof2019}. 

\begin{figure}[hbtp]
 	\centering
 	\includegraphics[width=\textwidth]{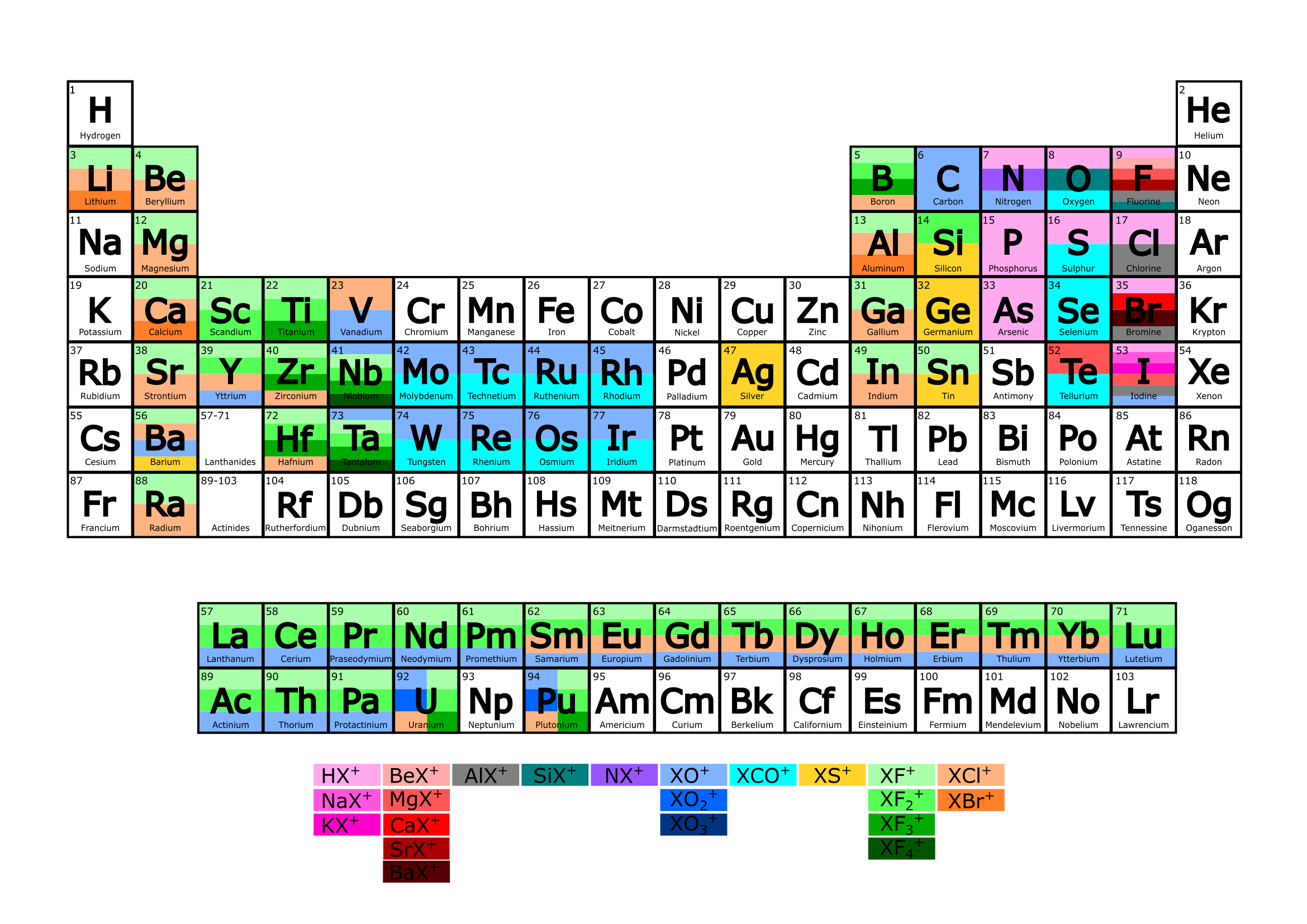}
 	\caption{Summary of potential radioactive molecular ion beams seen and predicted from ISOL targets and ion sources \cite{Au2022_6884293}. Some of the molecules (e.g. the transition metal carbonyl or oxide compounds or COTe) are conceptual and would require ion source developments. The actinides other than U are predicted to form compounds with a variety of valence states and require beam developments. Plasma or electron bombardment ion sources may cause molecular fragmentation.}
 	\label{fig:ISOL_sidebands}
\end{figure}

Fluorine, the most electronegative element, reacts with many species to form molecules that are stable in the environment of the ISOL target and ion source. Radioactive beams of fluoride molecules have been observed as sidebands or contaminants in several experiments (BaF, BF$_\textrm{x}$, RaF, SrF), some even without the presence of reactive fluorine gas. Other such sidebands have been observed and noted over years of operation (Figure \ref{fig:ISOL_sidebands} shows some of these along with others that have yet to be seen), making them promising potential candidates to the study of radioactive molecules. 

While multiple diatomic species have been successfully produced at ISOLDE, for fragile molecules, the high-temperature environment of the various ISOLDE targets and ion sources can cause dissociation and breakup. An alternative route to the production of volatile molecules is the use of ion traps to facilitate molecular formation. Ion traps are a staple tool frequently employed at ISOLDE for experiments with radioactive isotopes and in the preparation of beams for further study \cite{Wolf2013}. The ISOLDE beamline includes a radio-frequency quadrupole cooler-buncher (RFQcb) which uses DC potentials to confine ions axially (along the beam axis) and oscillating potentials for radial confinement. Ions are typically trapped in the RFQcb and cooled by repeated collisions with atoms of a constantly injected buffer gas. Molecular formation from the primary mass-separated beam has already been observed in both the ISCOOL RFQcb and the ISOLTRAP RFQcb at ISOLDE, likely due to collisions with contaminants in the buffer gas. The method of ion preparation with buffer gas lends itself to deliberate injection of reactive gases for molecular formation with the radioactive species. Formation of $^{226}$RaOH$^+$ has already been observed in an ion trap environment \cite{Fan2021}.  By injecting trace amounts of reactive gases into ion traps, ISOLDE could develop the capability to intentionally form and deliver species such as polyatomic symmetric-top and chiral molecules from radioactive beams.
Currently the gas injection of reactive gases to the buffer gas and subsequent molecular formation are studied with homologue elements to Ra using the RFQcb at the ISOLDE Offline 2 facility.

In addition to in-trap molecule formation, the development of an electron-impact ion source that operates at temperatures down to ambient temperature can enable extraction of delicate molecules. In commonly used Versatile Arc Discharge Ion Sources (VADIS) \cite{Penescu2010}, electrons to induce ionization in collision with neutrals are thermionically released from a cathode that is heated to around $2000~^{\circ}\mathrm{C}$. The introduction of a photo-cathode that releases electrons by the photo-electric effect enables operation at ambient temperature. A photo-cathode driven ion source was proposed in \cite{Ballof2022concept}. The feasibility of the approach is supported by a proof-of-concept experiment \cite{Ballof2022cold}, however, further development is required to reach higher ionization efficiencies.

\subsubsection{CRIS}

The collinear resonance ionization spectroscopy (CRIS) experiment is a permanently stationed beamline at ISOLDE, where it is used to perform high-sensitivity and high-resolution laser-spectroscopic studies of radioactive atoms and molecules~\cite{Cocolios2013CRIS}. With the CRIS technique, the incoming radioactive ion beams are first neutralized in an alkali-vapor cell and then step-wise resonantly re-ionized using a series of pulsed lasers. By scanning the frequency of one of the laser and deflecting the ions onto a detector, the hyperfine structure of atomic and molecular systems that contain nuclei with extreme proton-to-neutron ratios can be studied ~\cite{Koszorus2021}. The measurements are then typically used to study the evolution of the nuclear charge radius and electromagnetic moments across long isotopic chains~\cite{deGroote2020b}, and to understand electronic, vibrational, rotational, and hyperfine effects in molecular structure~\cite{GarciaRuiz2020}.

In 2018, the CRIS experiment performed the first laser-spectroscopic study of a molecule without stable or primordial isotopomers, studying the low-lying electronic and vibrational transitions in RaF~\cite{GarciaRuiz2020}. Additionally, by studying the electronic and vibrational structure of RaF molecules containing different isotopes of Ra, small changes in the molecular transition frequencies across different isotopomers were measured and linked to the changes in the charge radii of the radium nuclei as a function of their neutron number~\cite{Udrescu2021}. Demonstrating that the spectra of heavy, polar radioactive molecules can be sensitive to nuclear-size effects further paves the way for utilizing molecular laser spectroscopy for nuclear-structure research.

The experimental campaign on RaF was carried out more than a month after the irradiation of the ISOLDE target with protons, thus profiting from beam time beyond the official CERN running period. As the number of molecules of interest that contain unstable nuclei with sufficiently long half-lives is constantly growing, an extensive research program without the need for concurrent proton irradiation can be envisioned, while experiments on short-lived molecules can be carried out during the proton beam-time periods.

As a next step, the CRIS collaboration is preparing to study, for the first time, the low-lying electronic structure of AcF using laser spectroscopy~\cite{AthanasakisKaklamanakis2021}. The energy levels of AcF have been proposed as being sensitive to the symmetry-violating Schiff moment of the Ac nucleus~\cite{Skripnikov:2020c}, which is predicted to be exceptionally large for $^{227}$Ac and $^{225}$Ac~\cite{Flambaum2020Schiff}. Studying the low-lying electronic structure of AcF with the CRIS technique will benchmark the quantum-chemistry techniques used to predict its molecular properties, the accuracy of which is necessary for linking a future measurement of parity violation in the electronic spectrum to the presence of a Schiff moment in the Ac nucleus.

\subsubsection{In-source spectroscopy}

Applying laser spectroscopy directly in the ion source coupled to the production target~\cite{Alkhazov.1989, Fedosseev.2017}, and using dedicated decay identification-based particle detectors has been shown to reach sensitivity levels down to less than 0.1 produced ions per second~\cite{Marsh.2018}. While the efficiency is unrivalled, limitations arise in potential isotopic contamination ionized by non-laser related mechanisms such as surface ionization, and achievable spectral resolution due to Doppler broadening in the high-temperature environment needed for release. Specialised laser ion sources as ISOLDE's LIST~\cite{Fink.2015} or TRIUMF's IG-LIS~\cite{Raeder.2014} can greatly reduce contamination at the cost of efficiency decline. A recently introduced operation mode featuring crossed laser/atom beam geometry~\cite{Heinke.2017, Heinke.2019} additionally offers a way to enhance spectral resolution approaching the capabilities of collinear spectroscopy setups. This technique was employed off-line already for various nuclear structure studies of long-lived radioactive atomic species~\cite{Heinke.2019, Studer.2020, Kron.2020}.

\subsection{IGISOL}

The Ion Guide Isotope Separation On-Line (IGISOL) technique~\cite{arje_ion_1987, moore_igisol_2014}, conceived in the early 1980s as a novel variation to the helium-jet method, is used to provide exotic low-energy radioactive beams of short-lived nuclei for nuclear structure research, nuclear astrophysics, fundamental studies and applications. Hosted in the Accelerator Laboratory of the University of Jyväskylä (JYFL-ACCLAB), the IGISOL facility is served by a K130 cyclotron equipped with three Electron Cyclotron Resonance (ECR) ion sources (6.4 GHz, 14 GHz and 18 GHz) and a multi-cusp H$^{-}$ light ion source. Currently, primary proton beam energies up to about 65 MeV and intensities of several tens of microamperes can be provided, and acceleration of intense heavy-ion beams up to lead have been realized. A second, high-intensity MCC30/15 light-ion cyclotron is currently being commissioned, and will provide protons (18 to 30 MeV) and deuterons (9 to 15 MeV) for experiments at IGISOL. A second extraction beamline from this cyclotron may offer opportunities at a future target station that will be discussed later.

In the ion guide method of radioactive ion beam production, primary beams interact with a thin target (typically a few mg/cm$^{2}$). Reaction products recoil from the target and are stopped and thermalized in a noble buffer gas, usually helium, at pressures of a few hundred mbar. Charge-state resetting in the gas environment results in the majority of the recoil ions being extracted singly-charged, although if the purity is sufficiently high, extraction of higher charge states is also possible. A variety of gas cell geometries and thin targets, coupled with the wide range of primary beams on offer, provides access to an extensive variety of nuclei via light-ion and heavy-ion induced fusion-evaporation reactions, charged-particle-induced fission of actinide targets and, most recently, multinucleon transfer reactions. This results in access to nuclei far from beta stability, both neutron-deficient and neutron-rich isotopes. This direct on-line mass separation of primary recoil ions from nuclear reactions has achieved similar extraction efficiencies for both volatile and non-volatile elements throughout the periodic table. The universality of the ion guide method combined with extraction times from the gas cell as short as $\sim$1 ms in some cases, provides excellent opportunities for a rich physics program.

As an alternative to the ion guide approach, in-gas laser resonance ionization may be employed. This technique exploits the unique atomic level fingerprint of the desired element, driving electrons in a resonant process with pulsed, tunable laser radiation from the atomic ground state to ionization. One of the challenges to realize this in an on-line environment is the neutralization of the recoiling ions. Usually, to facilitate neutralization, argon gas is employed which supports a faster recombination of atomic ions. Complexities associated with the loss of photo-ions during extraction from the gas cell due to recombination has resulted in the physical separation of the stopping and ionization volumes. Nevertheless, this mode of operation is more difficult to realize and has been employed in only specific experiments, partly due to challenges to obtain comparable laser ionization efficiencies with e.g. hot cavity methods, caused by the competition with collisional de-excitation in the gas cell.

Most recently, the facility has implemented a complementary target-ion source system for the extraction of elements exhibiting a high release efficiency from graphite, an inductively-heated hot cavity catcher laser ion source~\cite{Reponen2015, reponen_evidence_2021}. Similar to the gas cell method, it is a thin-target device whereby reaction products recoil out of the target, implant into a hot graphite catcher and promptly diffuse into a catcher cavity as atoms, before effusing into a transfer tube. There, the atoms are selectively ionized via multi-step laser ionization. 

\subsubsection{Spectroscopic techniques}

Whether the IGISOL is operated with the ion guide or the hot cavity catcher laser ion source, ions are first guided via a radiofrequency sextupole ion guide (SPIG)~\cite{karvonen_sextupole_2008} through a differential pumping region, before being accelerated to a potential of 30 kV towards the mass separator. Mass separation is realized with a dipole sector magnet, with a nominal mass resolving power of $M/\Delta M$ = 500. Downstream of the focal plane of the mass separator, the facility is equipped with a radiofrequency quadrupole~\cite{nieminen_-line_2002} in which continuous beams of ions are cooled and bunched for subsequent experiments. 

A broad range of devices are available to outside users. This includes the double Penning trap JYFLTRAP~\cite{eronen_jyfltrap_2012}, employed for high-precision mass measurements or as a mass purifier for post-trap nuclear decay spectroscopy, a newly-commissioned multi-reflection time-of-flight mass spectrometer, a collinear laser spectroscopy station for high-resolution laser spectroscopy of ions and atoms~\cite{DEGROOTE2020437}, and a decay spectroscopy beam line. Recently, a new low-energy beamline (RAPTOR) has been constructed for collinear resonance ionization spectroscopy (CRIS), designed to accept ion beams with a variable energy of between 1 and 10 keV. This results in less Doppler compression than standard collinear laser spectroscopy experiments operating at several tens of keV, but gives the device a smaller footprint and potentially improved charged exchange efficiency. RAPTOR connects to the JYFLTRAP beamline and thus is available for laser-assisted mass measurements of either isotopically or isomerically pure beams, or for laser-assisted decay spectroscopy. Lastly, the IGISOL facility hosts an atom trap dedicated for trapping of radioactive Cs isotopes and isomers~\cite{GIATZOGLOU2018367}. In 2022 the MORA experiment was installed, a trap-based experiment to search for a signature of CP violation in the nuclear beta decay of radioactive nuclei produced at IGISOL~\cite{delahaye_mora_2019}.

\subsubsection{Production capabilities for actinide elements}

 Neutron-deficient actinide nuclei have traditionally been highlighted for octupole deformation studies, with maximum values of deformation predicted around neutron number $N$=136. Direct experimental information has been reported only for a few isotopes and yet theoretical efforts have indicated several promising candidates in neutron-deficient isotopes of uranium, plutonium and curium. Measurements of relative changes in the nuclear charge radii have been suggested as indicative of the emergence of octupole deformation in nuclei. Recent developments at the IGISOL facility offer a promising roadmap to the production of these isotopes and with the available infrastructure, to potentially explore the nuclear structure in support of efforts at other facilities worldwide.

In recent years, a program of research for the study of the nuclear structure of actinide isotopes has been implemented at the IGISOL facility, motivated by the paucity of ground-state nuclear information that exists above radium, the heaviest element studied online using collinear laser spectroscopy. This lack of data reflects the scarcity of material and the complex atomic structure of these elements. Short-lived isotopes are not available at online isotope separator facilities and require production via fusion reactions in heavy-ion collisions or transfer reactions with radioactive targets. Although high-flux reactors can breed sufficient quantities of transuranium elements, studies are then restricted to long-lived isotopes.

At IGISOL, a three-fold approach has been employed for the production of heavy elements, supporting the development of techniques with which to manipulate and study such isotopes. First, in-gas-cell resonant laser ionization has been applied in combination with thermal desorption of long-lived actinide elements. The actinide region poses special challenges for filament-based sources as the volatility of many actinide elements is relatively poor and the scarcity of some isotopes complicates the filament manufacturing. In collaboration with the Nuclear Chemistry department of the University of Mainz, samples containing $^{238-240,242,244}$Pu were electrolytically deposited onto tantalum substrates. In-gas-cell laser ionization of released atoms offered a window into the gas-phase chemistry exhibited by plutonium~\cite{Pohjalainen2016}. The monoatomic yields were sufficient for high-resolution collinear laser spectroscopy, resulting in measurements of mean-square charge radii and hyperfine structure~\cite{Voss2017}. Similarly, a thorium ion source has been developed using thorium dispensers fabricated at the Institute of Atomic and Subatomic Physics (ATI) of TU Wien \cite{Pohjalainen2020}.

The second approach to RIB production is via the use of alpha-recoil sources. For example, $^{233}$U sources have been used for the production of a thorium ion source to study the low-energy isomer in $^{229}$Th \cite{Wense2015, Pohjalainen2020b}. Extensive target characterization studies have been performed in Jyväskylä using nuclear decay spectroscopy of alpha and gamma radiation, as well as materials-based analysis techniques, for example Rutherford Backscattering Spectrometry, to investigate the elemental composition as a function of target depth. The recoil ion efficiency is very sensitive to target thickness, purity and quality. In addition, and of pertinence to ion manipulation in gas cells, the charge state of the extracted ion can be manipulated through the introduction of trace gases. For example, the addition of xenon into the helium buffer gas has been shown to reduce the charge state of thorium from triply-charged to doubly-charged due to the difference in ionization potential. This can be useful when the production mechanism leads to a distribution of charge states, not all of which may be suitable for subsequent spectroscopy. Other alpha-recoil sources of potential interest are \iso{Ra}{223} ($\tau_{1/2}$ = 11.4 d) and \iso{Pu}{239} ($\tau_{1/2}$ = 24.1 ky). The former is commonly used to determine gas cell efficiencies via a measurement of the mass-separated daughter activity, \iso{Rn}{219} ($\tau_{1/2}$ = 3.96 s). The latter is currently being investigated as a potential route for laser spectroscopy of the alpha-decay daughter, \iso{U}{235}, which hosts the second lowest-lying isomeric state in the nuclear landscape at 76~eV.

Lastly, an exploration of the production of a wider range of short-lived actinide isotopes using proton-induced fusion-evaporation reactions on actinide targets, for example \iso{Th}{232}, has been initiated. This has resulted in the identification, via nuclear decay spectroscopy, of thorium isotopes to \iso{Th}{225} and protactinium isotopes to \iso{Pa}{224}, the latter following an evaporation of 9 neutrons with a mass-separated yield of $\sim$1 pps. These yields were measured with 60 MeV protons on a few mg/cm$^{2}$ metallic \iso{Th}{232} target. According to cross section calculations, these yields are expected to increase for less exotic species with lower primary beam energy, however they cannot be measured using traditional decay spectroscopy due to the long half-lives involved. The use of the Penning trap mass spectrometer for ion identification will be attempted in the near future, informing prospects for future laser spectroscopy experiments depending on the available yields. Important target developments at the University of Mainz have resulted in a novel manufacturing process for actinide targets, a new Drop-on-Demand (DoD) method using an inkjet printer \cite{Haa17}. The first such targets of \iso{Th}{232} have been tested under online conditions at IGISOL, with long-term yield stability monitored under continuous irradiation using a 10 particle microampere, 50 MeV proton beam for 10 hours. Several \iso{U}{233} targets have been manufactured and will offer opportunities for the production of exotic neptunium  through light-ion fusion-evaporation reactions. A clear potential exists to impact our knowledge of other isotope chains by applying these reactions to even heavier target materials in the future.

\subsubsection{Opportunities for molecular beams}

\begin{figure}
\centering
\includegraphics[width=0.9\textwidth]{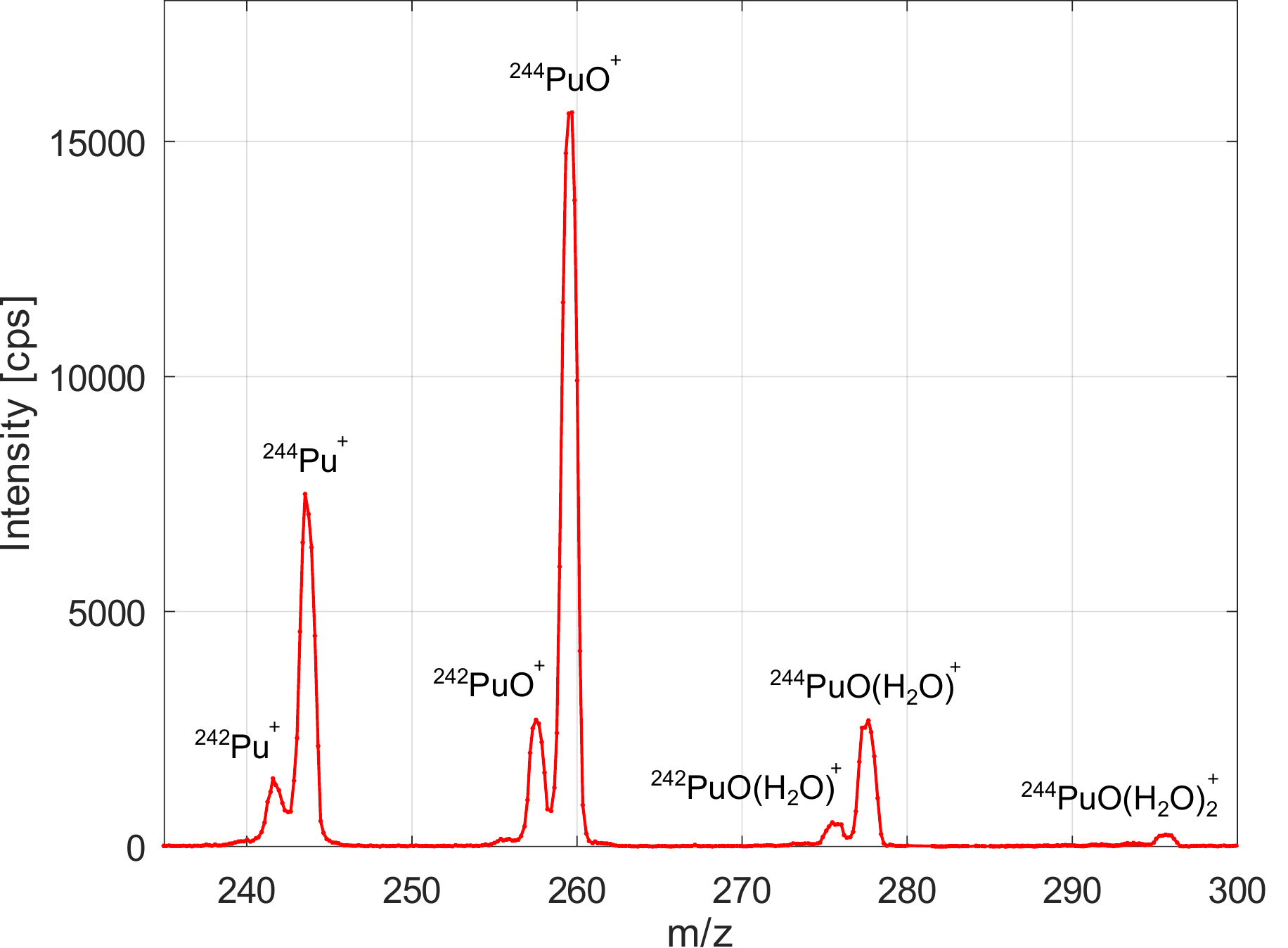}
\caption{Mass spectra in the region of monoatomic plutonium and related molecules measured at the focal plane of the IGISOL mass separator. Isotopic identification as well as molecular identification can easily be made.}
\label{fig_Pu}
\end{figure}

In the ion guide / gas catcher methods of radioactive beam production, gas purity is of critical importance to the survival of atomic ions during extraction from the gas cell, with molecular formation an important loss mechanism that depends on the concentration of impurities. On the other hand, and yet to be exploited at the IGISOL, is the introduction of a controlled bleed of impurities that can offer a simple way to encourage molecular formation of actinide elements, of immediate benefit to potential radioactive molecular beam experiments. Figure \ref{fig_Pu} illustrates a mass spectra in the region of atomic plutonium and related molecules, obtained in our aforementioned filament-based experiments~\cite{Pohjalainen2016}. The ions were directly measured at the focal plane of the IGISOL mass separator. In this work, argon was used as a buffer gas and an unforeseen outgassing from the filament holder led to a source of impurities, mainly water and oxygen. The first molecular isotopic pattern is observed at 16 u above the monoatomic pattern, corresponding to PuO$^{+}$ ions. Further reactions then occur with water molecules, via a thermolecular association reaction, leading to hydrate attachments to the oxide molecule. Very similar chemistry has been observed in thorium~\cite{Pohjalainen2020}, and it would be expected in other reactive actinide elements. It is noteworthy that these molecules do not easily break up during beam transportation, and would therefore be available for study.

It should be noted that the lasers do not selectively ionize neutral molecules in the gas phase, rather molecular formation occurs following the creation of resonant atomic ions which then react with impurities during extraction from the gas cell. By studying the time profiles of the mass-separated species, it is possible to determine the concentration of impurities if the chemical reaction rate coefficients are known. Typical gas purification methods applied at gas cell facilities result in impurity levels below a part-per-billion. In the example shown in Fig. \ref{fig_Pu}, the impurity levels were estimated at a level of a part-per-million.

The potential to form molecules containing radioactive isotopes through the controlled introduction of reactive gases or other impurities could motivate the development of a dedicated gas cell for molecular formation. Combined with the diverse methods available to produce short-lived actinides, described in the preceding section, a wide range of molecules containing heavy, radioactive isotopes could potentially be produced and studied using the techniques presently available at the IGISOL. Thus, the laboratory could provide an excellent testing ground, enabling the exploration of a variety of potentially interesting molecular systems. Opportunities to expand into existing areas of the Accelerator Laboratory, where an IGISOL-like facility, dedicated to the study of molecules, is a potential opportunity for the future. For example, a connection can be made to either the existing K130 cyclotron, or to the second extraction beam line of the MCC30/15 light-ion cyclotron. This latter accelerator presently only serves proton and deuteron beams to the IGISOL laboratory. In doing so, opportunities for regular target irradiations may present themselves, facilitating measurements which require regular access to radioactive isotopes.

\section{Outstanding challenges}
While there have been tremendous advances in the science behind radioactive molecules and their applications, there remain critical questions that must be answered to understand the reach, impact, and potential future opportunities.  Below are selected challenges that are a priority to address.

\subsection{Measuring the $^{229}$Pa parity doublet}\label{sec:pa}

The rare isotope \iso{Pa}{229} was first identified as a promising candidate to search for CP violation in 1983 \cite{Haxton1983}.  
It is expected \cite{Chasman1980} to have an octupole deformed nucleus and a set of opposite parity nuclear states that are separated by a  small amount (50 eV) that would enhance its sensitivity to CP violation by approximately $10^3\times$ compared to the already octupole enhanced sensitivity of \iso{Ra}{225} (55 keV parity doublet separation), as discussed in section \ref{sec:bsm}, and $10^6\times$ that of \iso{Hg}{199}, the atomic physics touchstone \cite{Graner2016}.  
But, unlike the case of \iso{Ra}{225}, the presence of the parity doublet and octupole deformation enhancement of \iso{Pa}{229} have yet to be experimentally confirmed  \cite{Dragoun1993, Lowsch1994,Ahmad2015}, largely due to the fact that this nucleus has historically been quite challenging to work with, in part because the 5/2$^{\pm}$ states, if they exist, are thought to be separated by (60$\pm$50) eV~\cite{Ahmad2015}.  
But, in recent years new technologies have been developed \cite{ponce2018} which are promising for measuring the low nuclear energy level structure of \iso{Pa}{229} including some adapted from attempts to measure the low lying isomeric state in \iso{Th}{229} \cite{Ponce2018a}, and other proposed techniques for measuring the energy of isomeric transitions \cite{Jin2022}.

A measurement of the parity doublet could be performed at FRIB using an intense beam of \iso{U}{229}.  The uranium will need to be filtered in a column from other unwanted radioisotopes before it is placed in a detector.  Then, decays from roughly 80\% of the \iso{U}{229} (58 m half-life) produce \iso{Pa}{229} via electron capture, populating excited states of the \iso{Pa}{229} nucleus.  To measure the ground-state parity doublet, the high-energy gamma and x-rays of \iso{Pa}{229} will need to be filtered out so that the expected 10s of eV photon corresponding to the transition between octupole deformed states may be measured.

\subsection{Lattice QCD for $\bar{g}_0$, $\bar{g}_1$ and quark cEDMs} 

At hadronic scales, BSM CP-violating interactions are captured by effective operators,
containing quarks and gluons as degrees of freedom.
These effective operators contribute to the 
single nucleon, $d_n$ and $d_p$, and nuclear EDMs, and beside the case of the quark-EDM operator, i.e. 
the tensor current, they are all 
affected by large uncertainties, often $\sim 100 \%$~\cite{Engel2013}.
Another big outstanding challenge, as emphasized in Sec.~\ref{sec:hadronic_CPV}, 
is the determination of the CP-violating LECs, $\bar{g}_i$, $i=0,1,2$,
that in principle receive contributions from every CP-violating effective operator.
In both cases non-perturbative techniques are needed and the final goal is 
to provide, using lattice QCD (LQCD), a precise and robust determination of the 
single-nucleon EDMs and of the CP-violating 
pion-nucleon couplings, $\bar{g}_0$ and $\bar{g}_1$,
contributing to the nucleon-nucleon potential 
and the pion loop effects in the nucleon EDM.
The relevance on the precise determination of the single-nucleon EDMs and of the LECs is a direct consequence  
of the fact that they contribute substantially to both atomic and molecular EDMs.

Chiral symmetry considerations provide first estimates for the low-energy couplings,
e.g the \thetaQCD term gives a large contribution to $\bar{g}_0$, but not to 
$\bar{g}_1$, while effective operators like the quark-chromo EDM (qCEDM) contribute equally to 
$\bar{g}_0$ and $\bar{g}_1$. 
For example, in the case of the deuteron, a consequence of these chiral symmetry considerations, 
is that the EDM receives the dominant contribution from the single-nucleon EDMs in the case of the \thetaQCD term,
while in the case of the qCEDM the dominant contribution comes from $\bar{g}_1$.

Lattice QCD is the ideal method to provide a precise determination 
of the nucleon EDM and the low-energy couplings. 
There has been substantial activity in the LQCD community to determine 
the EDMs, $\bar{g}_0$ and $\bar{g}_1$ from the different CP-violating sources, 
for a review see Ref.~\cite{Shindler:2021bcx}.
For the \thetaQCD term, while still dominated by the poor statistical accuracy
of the calculations~\cite{Abramczyk:2017oxr,Dragos:2019oxn,Alexandrou:2020mds,Bhattacharya:2021lol}, 
first LQCD results~\cite{Dragos:2019oxn} seem to confirm the dominance of the chiral 
logarithm~\cite{Crewther:1979pi} and the determination of $\bar{g}_0$, based on hadronic spectroscopic 
calculations~\cite{deVries:2015una}.
For the quark-chromo EDMs there are practically no determinations from LQCD, 
beside some first attempts to measure the bare matrix elements~\cite{Abramczyk:2017oxr,Kim:2018rce,Bhattacharya:2018bkd}.
The main challenge for the determination of hadronic qCEDM matrix elements 
is the rather cumbersome renormalization pattern when using
standard RI-MOM techniques~\cite{Bhattacharya:2015rsa}. In particular the 
largest systematic error is associated with the power-law ultraviolet divergences. 
New techniques based on the gradient 
flow~\cite{Narayanan:2006rf,Luscher:2010iy,Luscher:2011bx,Luscher:2013cpa,Luscher:2013vga} 
show a new promising avenue in the determination of quark-chromo EDMs
and the corresponding $\bar{g}_0$ and 
$\bar{g}_1$~\cite{Shindler:2014oha,Shindler:2015aqa,Dragos:2017wms,Kim:2018rce}. 
First results for a non-perturbative 
determination of the power divergence of the qCEDM~\cite{Kim:2021qae} and 
the perturbative determination of the matching coefficients~\cite{Rizik:2020naq,Mereghetti:2021nkt}
are preparatory for a non-perturbative determination of the nucleon EDM and $\bar{g}_1$ 
stemming from the qCEDM operator. An alternative method to determine $\bar{g}_1$
is based on spectroscopic quantities related to the qCEDM operator~\cite{deVries:2016jox}. 
Furthermore, the use of the gradient flow could be beneficial 
to resolve the challenge of renormalization.

The goal of the next generation of LQCD calculations should be to improve the 
determination of the \thetaQCD term contribution of the nucleon EDM and the corresponding 
$\bar{g}_0$, and to provide first robust determinations of the quark-chromo EDM and the 
correspoding $\bar{g}_0$ and $\bar{g}_1$. These calculations would provide an valuable 
input for heavier nuclei and molecular EDMs.

\subsection{Improved calculations for \iso{Hg}{199} NSM, relationship to CPV $\pi N N$\label{sec:Hg199}}

Even with success in connecting the CP-violating LECs with the underlying 
CP-violating physics, one still needs to connect the LECs directly to molecular EDMs.
An important step in the calculation chain is the computation of the dependence of the Schiff moment on the LECs.  Doing so requires the application of the methods discussed in Section 
\ref{ss:theory}. 

Calculations of the Schiff moment of $^{199}$Hg in nuclear density functional
theory or the shell model still produce values with significant
uncertainty attached.  In DFT, we can use better functionals and a more faithful
reproduction of the soft ground state, which will require a multi-reference
technique known as the generator coordinate method (GCM) that superposes
mean-fields with different shapes.  The shell-model calculations will benefit
from larger spaces that are now possible and a better treatment of excitation to
states outside those spaces, e.g.\ through many-body perturbation theory on top
of the shell model.  Perturbation theory in nuclear physics is challenging,
however, and adapting it will take a lot of work.

Though data on the Schiff strength distribution to excited states or, as a
proxy, the isoscalar-dipole strength distribution to those states would be
extremely useful in constraining the 
%phenomenological 
models used to compute
Schiff moments, the best hope for more reliable moments in the long term are \textit{ab
initio} calculations (see Section \ref{ss:theory}).  One particular variant of
the in-medium similarity renormalization group --- called the valence space
IMSRG --- has the potential to be useful for Schiff-moment calculations in
nearly spherical or softly deformed nuclei such as $^{199}$Hg.  In that
approach, the reference state is driven by the flow to couple only to states
within a valence shell-model space, so that the effective Hamiltonian
$\tilde{H}$ and other effective operators can be used as if they were part of an
ordinary shell model calculation.  The determination of the effective operators
will be more difficult here than in anything done so far because the
time-reversal-violating nucleon-nucleon potential couples shell-model states to
others that are many MeV higher.  With enough effort and computing resources, a
good calculation should be possible.

Lower-resolution methods, which are computationally simpler, can still play
an important role here.  Nuclear DFT, see section~\ref{sec:DFT},  is particularly
useful because of its flexibility, its ability to handle shape deformation and
other kinds of collectivity, and the relatively low demand it places on current
computers.  Analyzing correlations between the Schiff moment and other
observables (such as the intrinsic octupole moment in octupole deformed nuclei)
has proved useful in augmenting the method.  The primary DFT-related task in the
next few years will be to construct wave functions that allow shape mixing to
represent the soft ground state of $^{199}$Hg.  The technique for doing this is
tried and true, but has yet to be applied to the Schiff-moment problem.

\subsection{Improved calculations for radioactive isotopes for CPV sensitivity\label{sec:Ra}}

As discussed in section \ref{sec:bsm}, octupole deformation in radioactive isotopes can enhance Schiff moments.  The
computations in such nuclei are very different than for example in \iso{Hg}{199}.  The
CP-violating potential couples the ground state with many others, but in the
presence of octupole deformation, only one of those is important: the ground
state's parity-doublet partner.  Furthermore, strong deformation of any kind requires an intrinsically deformed
state to build on.  This rules out the shell model and some \textit{ab initio} approaches
that are based on it.    
The DFT methods for treating octupole deformation are now well developed, see Sec.~\ref{sec:DFT}.  They
can still be refined, but again, in the absence of data constraining the matrix
element of the Schiff operator between the ground state and its opposite-parity
partner (see next subsection), \textit{ab initio} calculations offer the best prospects for significant 
improvement.  In nuclei with strong deformation, whether quadrupole or
octupole, the In Medium Generator Coordinate
Method (IM-GCM) or coupled cluster (CC) approach are useful.  Here, the reference state is a
deformed mean-field state or a mixture of such states, projected onto good
angular momentum.  In this way, collective physics is included in the reference state itself.  Though IM-GCM or CC calculations in really heavy nuclei will require
huge amounts of CPU time and computer memory, they are (like their counterparts
in Hg) possible with enough effort \cite{Hu2021}.

\subsection{Nuclei with octupole correlations in EDM measurements}

In a number of elements where modern atomic and molecular techniques could enable EDM searches, nuclei have been identified with low-lying parity doublets that might produce enhancements of Schiff moments from octupole correlations. Challenges for nuclear experimentation include the characterization of octupole effects, ranging from vibrational structures, to octupole-soft systems, to stable octupole deformations; calculations often yield similar Schiff enhancements but appear quite different in the details. To what extent the nuclear wave functions of the parity doublets are affected by octupole correlations  is sometimes in question, as other nuclear structure effects can produce close-lying  doublets of opposite parity states. The experimental and theoretical evidence for most cases is carefully considered in three decades of reviews~\cite{Butler1996,Butler2020,Ahmad1993}. A resource paper pointing to the literature can be found in~\cite{Behr2022}.  Theory challenges are discussed in previous sections.

\subsection{Radiochemistry and molecule production}

In the past few decades, technologies have developed allowing researchers to handle radioactive materials more safely as their availabilities have also seen an increase. Despite these advances, of the actinides that can be made on the milligram scale, protactinium is severely understudied \cite{Wilson2012}. This is due to minimal production of isotopes of protactinium along with protactinium being incredibly insoluble, thus making it difficult to work with. Moreover, protactinium has a propensity to adsorb onto glass surfaces, making it increasingly challenging to manipulate. Protactinium acts similarly to both the transition metals niobium and tantalum, having a stable pentavalent state, and the early actinides, having an accessible tetravalent state. In contrast to uranium, plutonium and neptunium, protactinium forms a mono-oxo ion as opposed to the di-oxo ions of uranyl, neptunyl and plutonyl, making these elements less helpful for modeling protactinium chemistry.

For optimizing experiments with \iso{Pa}{229}, which has a half-life of 1.5 days, it is best to use a non-radioactive analog such as niobium or tantalum, before moving to longer-lived isotopes of protactinium, and eventually \iso{Pa}{229}. \iso{Pa}{231} is the longest-lived isotope of protactinium with a half-life of 32,760 years, which is isolated from uranium ore. Isolating \iso{Pa}{231} from uranium ore is a massive undertaking and has not been repeated since the 1960's. The next longest-lived isotope is \iso{Pa}{233} with a half-life of 27 days. Although \iso{Pa}{233} is produced from the neutron irradiation of \iso{Th}{232} in the nuclear fuel cycle, the separation and purification processes coupled with its short half-life makes its isolation futile for bulk chemistry, especially in comparison to \iso{Pa}{231} \cite{Uribe2018}. The lack of readily available protactinium sources and the fact that protactinium is one of the most understudied actinides makes designing a source for \iso{Pa}{229} studies a great challenge.

Furthermore, once the target has been optimized for experiments, the 1.5 day half-life of \iso{Pa}{229} gives it a high specific activity, resulting in aggressive degradation from merely the radiation and atomic recoil as the sample decays. This can result in material being embedded into the holder and radiation hardening of the sample, making it difficult to extract desired material from the target for experiments. Although the chemistry of making the target can be optimized, the inherent issue of sample degradation endures.

Radioisotopes that can be produced at substantial rates and have long enough half-lives can be collected in quantities that are large enough for the production of samples that can be used in off-line studies. Various methods exist to produce thin layers or radioisotopes that are suitable for this purpose, e.g., manual pipetting or drop-on-demand (DoD) printing of solutions~\cite{Haa17}, or electrochemical deposition from organic solution, referred to as molecular plating~\cite{Par62,Vas12} solution. Laser ablation is a technique that is frequently used to produce (singly) charged ions of these species to make them available for their use in experiments. The technique is rather simple and can lead to atomic ions that may be transformed into molecular ions in a second step~\cite{Hea14,Fan2021} or may directly produce molecular ions~\cite{Eib14}. Which ion species exactly results is sometimes of lesser importance, e.g., in~\cite{Eib14} where atomic masses of actinide isotopes were determined from monoxide ions, simply because this was the most abundant ion species, and the presence of the oxygen was then corrected for. However, many applications demand a specific molecular species. The detailed systematic understanding of how parameters like substrate material, co-deposited chemical species present in the ablation layer, wavelength of the ablation laser, laser pulse duration~\cite{Shi98}, pressure and composition of residual gas, etc. that affect the formation of a specific ionic species in the laser ablation process is incomplete. The importance of details of these parameters vary for different elements, with the chemical reactivity of the ablated element likely being a key factor. A prominent example is that of the very reactive element thorium, the study of which is interesting not least because of the exotic low-excited nuclear isomer in \iso{Th}{229}~\cite{Thi19} located at an optically reachable excitation energy of 8.10(17) eV~\cite{vdW16,Sei19,Sik20,Sei22}. Various experiments focus on studying properties of thorium ions, and we list as one example the TACTICa experiment on Trapped And sympathetically Cooled Thorium Ions with Calcium. In first experiments, atomic $^{232}$Th+ ions were produced by laser ablation from a metallic natural Th foil~\cite{Gro19}.  In the TACTICa experiment, also other Th isotopes will be studied~\cite{Haa20}, which are not available as metallic foils. Th-containing molecules of interest for fundamental symmetry tests~\cite{Flambaum2019Schiff} will also be studied within the TACTICa project. Atomic Th$^+$ ion production from, e.g., samples produced via DoD-printing was so far unsuccessful, likely due to the formation of unidentified molecular species, demonstrating the need for further, systematic studies.

\section{Opportunities}

There is tremendous opportunity for physics with radioactive molecules, and the community is at an exciting threshold.  Experimental advances in the creation, study, and control of complex molecules has enabled rapid progress in precision measurement and quantum science.  The development of theoretical methods can describe and predict increasingly challenging structure of heavy nuclei. Quantum chemistry calculations can reach a percent accuracy for the properties of heavy diatomic molecules.  

Recent and upcoming advanced rare isotope facilities will create exotic elements in quantities which will enable highly sensitive experiments.  The ongoing combination of these efforts will usher in a new era of science with far-reaching impacts in nuclear, high-energy, astrophysical, and quantum sciences.

The outlook for this area is very positive, and the community has many tools available and under development which will help to advance the science.  However, we also list below some opportunities for further advances which should not be missed.

\emph{Molecular structure.}  The structure of even the simplest diatomic molecules is surprisingly complex, and the level of understanding required to perform precision measurement or quantum science experiments often takes years of combined theoretical and experimental effort for a single molecular species.  The complete mapping of energy levels and transitions between states requires a variety of techniques, and the current approach is often a patchwork of different groups with different capabilities.  A critical advance, needed especially for complex and heavy species, would be dedicated efforts to study molecular structure which consolidates and coordinates all of the necessary experimental and theoretical tools.  Such facilities could predict and then measure molecular structure over all energy scales, from optical to microwave, combined with advanced production methods and the ability to work with radioactive species, could shorten these efforts from years to months, resulting in dramatic reduction of time needed to access the science goals.

\emph{Facility space for dedicated precision measurements.}
There are several facilities with complementary radioisotope production capabilities.  A critical addition to these facilities would be dedicated laboratory space suitable for precision measurement experiments based on trace isotopes which are unsuitable for off-site work, due to short half-lives.  As an example, for hadronic CPV experiments a nuclear spin is required which typically results in a shorter half-life compared to spinless isotopes.  Such experiments typically require long run times: weeks or perhaps even months of statistics, and a factor of at least several longer to study and understand systematic uncertainties in detail.  These experiments also require environments that maintain a high degree of temperature stability ($\pm 1$ $^\circ$C) so that lasers do not drift significantly in frequency or amplitude during a measurement.  Magnetic field noise is also a nefarious source of potential systematics, so close proximity to elevators or other switching $B$-fields is non-ideal.  Generally, optics lab spaces require HEPA filtering, and this becomes more important when UV laser light is needed.

\emph{Isotope harvesting at facilities.} 
The time scale for precision measurements precludes direct beam usage.  Therefore, in addition to appropriate laboratory space, it is important to support both on-site and off-site efforts with offline isotopes. FRIB has planned capabilities for harvesting and purifying actinides for offline use.  ISOLDE offers irradiation stations and the possibility to irradiate cold targets, generating an inventory of radioisotopes that can be stored and extracted later as an ion beam without the need for beam time.  It is important to develop similar capabilities at other facilities, as well as provide strong support for the efforts at FRIB, TRIUMF, and ISOLDE.  Longer-lived isotopes, including spin-zero isotopes are very valuable for study and it is important to make these widely available to researchers at universities.  Zero nuclear spin isotopes are very valuable for developing techniques, and harvesting and distributing these isotopes in suitable quantities for research at universities is important for streamlining challenging experiments that must be done onsite. Spinful isotopes with long half-lives such as \iso{Th}{229} ($\tau=7,800$ y) are of particular interest and could be made available to off-site researchers.

\emph{Resources for calculations.}  Molecular, nuclear, and lattice QCD  calculations are critical to these efforts, and require significant computational resources and support. 

Electronic structure calculations of molecules containing heavy elements are computationally demanding. Therefore, in order to conduct reliable investigations of various systems and diverse properties, the combination of availability of powerful high-performance computing (HPC) resources with continuous method development is essential. Specifically equipped compute clusters installed locally at universities and other research institutions are crucial for method development, exploratory studies and short-term projects in this field, whereas large-scale applications that demand the highest computational resources require dedicated HPC facilities on state, national and international levels. Besides availability of high-performance hardware combined with repeated investments in new technology, progress relies strongly on innovative ideas by theoreticians for example to i) propose and explore new schemes for testing fundamental interactions on various levels, ii) identify suitable molecular systems with large enhancements favorable experimental properties, iii) develop efficient methods to predict corresponding molecular properties
and iv) describe molecular systems in external fields.
For instance, recent and future advances in relativistic electronic structure methods for the study of fundamental interactions, in data sorting and compression algorithms and in parallelization and vectorization of computational code will boost the computational efficiency and open the prospects of applying highly sophisticated approaches routinely to heavy many-electron systems. This will in turn allow theorists to provide, on reasonable timescales, even more accurate predictions of a wealth of properties necessary for planning and interpreting experiments with heavy radioactive molecules.  

Nuclear structure research is progressing dramatically as a result of exciting theoretical progress in the nuclear many-body problem and simultaneous experimental developments. The quantum many-body methods used in nuclear physics all have analogues in other fields dealing with complex systems (e.g., condensed matter and atomic and molecular physics), but the unique features of nuclear interactions have particular requirements in the present and future eras of computational science. 
Through the use of advances in applied mathematics, computer science, and computational hardware, \textit{ab initio} methods are now able to address much heavier nuclei than previously thought possible \cite{Hu2021}. These studies can now be extended to complex deformed nuclei, thereby giving access to a dramatically wider array of physics phenomena. DFT addresses the physics of large systems that \textit{ab initio} methods cannot reach today. The DFT results can be tied to \textit{ab initio} approaches \cite{Malbrunot2022} thus bridging light and heavy nuclei. In order to enable these studies, significant advances in large-scale simulations are required, including sparse linear algebra, load balancing, and performance optimization on leadership-class computing systems. Simultaneous advances in numerical optimization and uncertainty quantification are also required to determine the input interactions, interpret simulation results, provide guidance to experiment, and to predict quantities that are difficult, or impossible, to measure.

The goal of lattice QCD calculations is to provide 
theoretically robust predictions for hadronic EDMs and 
CP-violating LECs, from the fundamental degrees of freedom of
quarks and gluons, with fully quantified uncertainties.
In the exascale computing era this target is reachable,
provided that we make very efficient use of 
GPU-accelerated architectures currently deployed 
and planned at national laboratories.
Software developments that would maximize the use
of current and future supercomputers include:
engines that can run on the 
CPU/GPU architectures to build nucleon and pion correlation functions; performant I/O library to share numerically expensive LQCD data; a data analysis suite to handle large sets of highly correlated data with undersampled covariance.

\emph{Dedicated nuclear structure  efforts to measure octupole moments.}
Nuclear structure experiments in the actinide region are needed to calibrate EDM candidate isotopes for their new physics sensitivity. Two distinct types of measurements are needed: (1) low-energy nuclear spectroscopy to precisely measure the energy splitting of the parity doublet involving the ground nuclear state and (2) Coulomb excitation experiments to measure the octupole (E3) nuclear transition strengths. Since parity doublets may be accidental, a measurement of strong nuclear octupole transition strength is needed to verify the presence of octupole collectivity needed to amplify the effects of T- and P- violation within these isotopes. Coulomb excitation experiments using isotopes with non-zero ground state nuclear spin are particularly challenging to analyze because of the large number of gamma ray transitions involved. As a consequence, octupole collectivity for these isotopes is usually inferred from measurements in nearby ground state nuclear spin-0 isotopes \cite{Dobaczewski2018}. FRIB will provide additional access to the actinide region for these types of studies which may prove decisive for the case of \iso{Pa}{229} as a candidate EDM isotope.

\emph{Support for radiochemistry.}
Radiochemistry requires an abundance of consideration. To begin, obtaining radioactive material requires not only the permissions and paperwork, but also the rigorous preparation of a laboratory space equipped to handle the material, i.e. specialized fume hoods and glove boxes, appropriate radiation detection equipment, specialized personal protective equipment, advanced personnel training, etc. The equipment and instrumentation for handling radioactive material may rarely be cross utilized with nonradioactive materials, and often must be isotope specific to avoid cross contamination between isotopes. This necessitates much more laboratory supplies than nonradioactive work, in addition to the constant preventative measures to avoid contamination at every step. The radioactive material itself and individual equipment totals to a costly sum in addition to taking up a vast amount of time, space and labor.

\begin{figure}
    \centering
    \includegraphics[width=\textwidth]{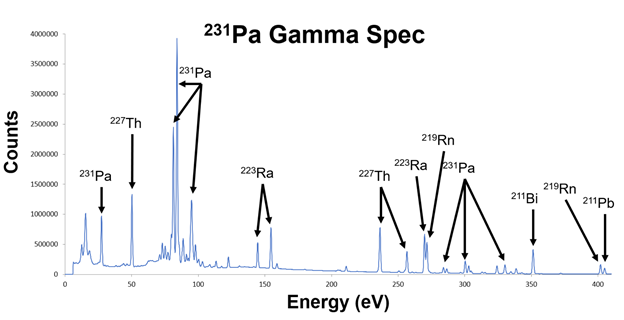}
    \caption{Gamma ray spectroscopy of an aged sample of protactinium-231.}
    \label{fig:PaGammaSpec}
\end{figure}

Furthermore, the isotope being studied is not the only concern, but also the daughter and granddaughter isotopes must be considered for potential containment and shielding issues, and depending on the decay chain, potentially the remainder of the daughter isotopes as well. For instance, \iso{Pa}{231} alone does not have a very high specific activity, but once a sample has aged for several decades, it requires a leaded glovebox to handle safely due to the activity of the daughter ingrowth. Many of the isotopes that are most difficult to handle are consequently understudied, making it even more difficult to efficiently separate daughter isotopes that may interfere with the handling and characterization of the isotope of interest. For example, Figure \ref{fig:PaGammaSpec} shows a gamma spectrum of an aged \iso{Pa}{231} sample. Many of the daughter isotopes within the decay chain are readily identifiable in the spectrum. Each of these isotopes makes the parent isotope more difficult to handle and characterize in further studies without their successful removal.

\section{Outlook}

The outlook for the emerging field of radioactive molecules is promising.  The availability of exotic nuclei in quantities which can be used for a wide range of applications will correspondingly transform how we approach a wide range of science.  This direction has already initiated new lines of inquiry, and brought into contact many diverse areas of science.  Preliminary results achieved for a handful of radioactive molecules are exciting and show tremendous promise, but realizing the potential of radioactive molecules will require sustained and coordinated efforts between experimentalists, theorists, and facilities to tame these complex and challenging species.

\section{Acknowledgements}

We thank John Behr, Vincenzo Cirigliano, and Jordy de Vries for helpful discussions.

This material is based upon work supported by:
the U.S. Department of Energy, Office of Science, Office of Nuclear Physics under award numbers
DE-SC0013365 (Michigan State University),
DE-FG02-97ER41019 (University of North Carolina), 
DE-SC0021179, DE-SC0021176 (MIT),
DE-SC0022034 (University of California, Santa Barbara);
National Science Foundation award numbers
PHY-2146555 (University of California, Santa Barbara),
PHY-1847550 (Caltech),
PHY-2209185 (Michigan State University),
PHY-2012068 (University of Delaware);
Deutsche Forschungsgemeinschaft (DFG) award numbers
328961117 – SFB 1319 ELCH (University of Kassel),
423116110, 390831469: EXC 2118 PRISMA+ Cluster of Excellence (Johannes Gutenberg University);
Science and Technology Facilities Council (STFC) award numbers
ST/P004423/1, ST/V001116/1, ST/X00502X/1 (University of Manchester),
~ST/P003885/1, ~ST/V001035/1;
Polish National Science Centre Contract No.~2018/31/B/ST2/02220;
Academy of Finland Project No. 339245 (University of Jyväskylä);
European’s Union Horizon 2020 Research and Innovation Programme number 861198 project ‘LISA’, Marie Sklodowska-Curie Innovative Training Network (Johannes Gutenberg University);
NIST Precision Measurement Grant numbers 60NANB18D253 (Caltech) and 60NANB21D185 (University of California, Santa Barbara);
Feodor Lynen Fellowship of the Alexander von Humboldt Foundation (MIT);
Heising-Simons Foundation award 2022-3361 (Caltech);
Gordon and Betty Moore Foundation award GBMF7947 (Caltech);
Alfred P. Sloan Foundation award G-2019-12502 (Caltech);
W. M. Keck Foundation (University of California, Santa Barbara);
ONR Grant No. N00014-20-1-2513 (University of Delaware);
NSERC Grant SAPIN-2022-00019 (TRIUMF);
FWO International Research Infrastructure (KU Leuven);
Excellence of Science (EOS) project 40007501 (KU Leuven);
KU Leuven project C14/22/104;
Leverhulme Trust Research Project Grant;
Russian Science Foundation Grant No. 19-72-10019 (Petersburg Nuclear Physics Institute);
Foundation for the Advancement of Theoretical Physics and Mathematics ``BASIS'' Research Project No. 21-1-2-47-1 (Petersburg Nuclear Physics Institute);
National Natural Science Foundation of China No. 12027809.

Computing support is acknowledged from: an INCITE Award on the Summit supercomputer of the OLCF at ORNL; CSC-IT Center for Science Ltd., Finland; The Viking Cluster,
which is a high performance compute facility provided by the
University of York; University of York High Performance Computing service, Viking and the Research Computing team. 

TRIUMF receives federal funding via a contribution agreement with the National Research Council of Canada. 

Support from the Gordon and Betty Moore Foundation Fundamental Physics Innovation Awards (No GBMF6210, APS GBMF-008-2020) is acknowledged for meetings and workshops which inspired this manuscript.

\clearpage

\bibliographystyle{jphysicsB-withTitles}

\small\bibliography{biblio,biblio2}

\end{document}